%% file: cat.tex
%
%
%


\input mndcat
\input mndnote

\input psfig


\def\halve{{\scriptstyle 1 \over 2}}

%

\newcount\eqnumber
\eqnumber=1
\def\step#1{\global\advance#1 by 1}
\def\neweq{{\rm\the\eqnumber}\step{\eqnumber}}
\def\preveq#1){\advance\eqnumber by -#1 \the\eqnumber) \advance\eqnumber by #1}


\pageoffset{-2.5pc}{0pc}


\Autonumber  


\pagerange{000--000}
\pubyear{1996}
\volume{}

\begintopmatter  

\title{A catalogue of galaxy cluster models}
\author{Eelco van Kampen$^{1,2 \star}$ and Peter Katgert$^1$}
\affiliation{$^1$ Sterrewacht Leiden, Postbus 9513, 2300 RA Leiden, The Netherlands}
\affiliation{$^2$ Royal Observatory Edinburgh, Blackford Hill, Edinburgh EH9 3HJ}

\shortauthor{E.\ van Kampen and P.\ Katgert}
\shorttitle{A catalogue of cluster models}


\acceptedline{Accepted ... Received ...; in original form ...}

\input catabs

\keywords galaxies: clustering - galaxies: formation - galaxies: evolution
- cosmology: theory - dark matter - large-scale structure of Universe

\maketitle  

\note{\tx $^\star$ Present address: Theoretical Astrophysics Center,
Juliane Maries Vej 30, DK-2100 Copenhagen, Denmark, {\tt eelco@tac.dk} }



\input cat1

\input cat2
\input cat3

\input cat4
\input cat5

\input cat6
\input cat7

\input cat8

\input catrefs		

\input catapps		

%
%

\end

%% file: mndcat.tex

\catcode `\@=11 

\def\@version{1.3}
\def\@verdate{28.11.1992}


%
%
%
%
%
%

\font\fiverm=cmr5
\font\fivei=cmmi5	\skewchar\fivei='177
\font\fivesy=cmsy5	\skewchar\fivesy='60
\font\fivebf=cmbx5

\font\sevenrm=cmr7
\font\seveni=cmmi7	\skewchar\seveni='177
\font\sevensy=cmsy7	\skewchar\sevensy='60
\font\sevenbf=cmbx7

\font\eightrm=cmr8
\font\eightbf=cmbx8
\font\eightit=cmti8
\font\eighti=cmmi8			\skewchar\eighti='177
\font\eightmib=cmmib10 at 8pt	\skewchar\eightmib='177
\font\eightsy=cmsy8			\skewchar\eightsy='60
\font\eightsyb=cmbsy10 at 8pt	\skewchar\eightsyb='60
\font\eightsl=cmsl8
\font\eighttt=cmtt8			\hyphenchar\eighttt=-1
\font\eightcsc=cmcsc10 at 8pt
\font\eightsf=cmss8

\font\ninerm=cmr9
\font\ninebf=cmbx9
\font\nineit=cmti9
\font\ninei=cmmi9			\skewchar\ninei='177
\font\ninemib=cmmib10 at 9pt	\skewchar\ninemib='177
\font\ninesy=cmsy9			\skewchar\ninesy='60
\font\ninesyb=cmbsy10 at 9pt	\skewchar\ninesyb='60
\font\ninesl=cmsl9
\font\ninett=cmtt9			\hyphenchar\ninett=-1
\font\ninecsc=cmcsc10 at 9pt
\font\ninesf=cmss9

\font\tenrm=cmr10
\font\tenbf=cmbx10
\font\tenit=cmti10
\font\teni=cmmi10		\skewchar\teni='177
\font\tenmib=cmmib10	\skewchar\tenmib='177
\font\tensy=cmsy10		\skewchar\tensy='60
\font\tensyb=cmbsy10	\skewchar\tensyb='60
\font\tenex=cmex10
\font\tensl=cmsl10
\font\tentt=cmtt10		\hyphenchar\tentt=-1
\font\tencsc=cmcsc10
\font\tensf=cmss10

\font\elevenrm=cmr10 scaled \magstephalf
\font\elevenbf=cmbx10 scaled \magstephalf
\font\elevenit=cmti10 scaled \magstephalf
\font\eleveni=cmmi10 scaled \magstephalf	\skewchar\eleveni='177
\font\elevenmib=cmmib10 scaled \magstephalf	\skewchar\elevenmib='177
\font\elevensy=cmsy10 scaled \magstephalf	\skewchar\elevensy='60
\font\elevensyb=cmbsy10 scaled \magstephalf	\skewchar\elevensyb='60
\font\elevensl=cmsl10 scaled \magstephalf
\font\eleventt=cmtt10 scaled \magstephalf	\hyphenchar\eleventt=-1
\font\elevencsc=cmcsc10 scaled \magstephalf
\font\elevensf=cmss10 scaled \magstephalf

\font\fourteenrm=cmr10 scaled \magstep2
\font\fourteenbf=cmbx10 scaled \magstep2
\font\fourteenit=cmti10 scaled \magstep2
\font\fourteeni=cmmi10 scaled \magstep2		\skewchar\fourteeni='177
\font\fourteenmib=cmmib10 scaled \magstep2	\skewchar\fourteenmib='177
\font\fourteensy=cmsy10 scaled \magstep2	\skewchar\fourteensy='60
\font\fourteensyb=cmbsy10 scaled \magstep2	\skewchar\fourteensyb='60
\font\fourteensl=cmsl10 scaled \magstep2
\font\fourteentt=cmtt10 scaled \magstep2	\hyphenchar\fourteentt=-1
\font\fourteencsc=cmcsc10 scaled \magstep2
\font\fourteensf=cmss10 scaled \magstep2

\font\seventeenrm=cmr10 scaled \magstep3
\font\seventeenbf=cmbx10 scaled \magstep3
\font\seventeenit=cmti10 scaled \magstep3
\font\seventeeni=cmmi10 scaled \magstep3	\skewchar\seventeeni='177
\font\seventeenmib=cmmib10 scaled \magstep3	\skewchar\seventeenmib='177
\font\seventeensy=cmsy10 scaled \magstep3	\skewchar\seventeensy='60
\font\seventeensyb=cmbsy10 scaled \magstep3	\skewchar\seventeensyb='60
\font\seventeensl=cmsl10 scaled \magstep3
\font\seventeentt=cmtt10 scaled \magstep3	\hyphenchar\seventeentt=-1
\font\seventeencsc=cmcsc10 scaled \magstep3
\font\seventeensf=cmss10 scaled \magstep3

\def\@typeface{Computer Modern} 

\def\hexnumber@#1{\ifnum#1<10 \number#1\else
 \ifnum#1=10 A\else\ifnum#1=11 B\else\ifnum#1=12 C\else
 \ifnum#1=13 D\else\ifnum#1=14 E\else\ifnum#1=15 F\fi\fi\fi\fi\fi\fi\fi}

\def\mib{\hexnumber@\mibfam}
\def\syb{\hexnumber@\sybfam}

\def\makestrut{%
  \setbox\strutbox=\hbox{%
    \vrule height.7\baselineskip depth.3\baselineskip width 0pt}%
}

\def\bls#1{%
  \normalbaselineskip=#1%
  \normalbaselines%
  \makestrut%
}

%

\newfam\mibfam 
\newfam\sybfam 
\newfam\scfam  
\newfam\sffam  

\def\cal{\fam2}
\def\em{\ifdim\fontdimen1\font>0 \rm\else\it\fi}

\textfont3=\tenex
\scriptfont3=\tenex
\scriptscriptfont3=\tenex

\def\eightpoint{
  \def\rm{\fam0\eightrm}%
  \textfont0=\eightrm \scriptfont0=\sevenrm \scriptscriptfont0=\fiverm%
  \textfont1=\eighti  \scriptfont1=\seveni  \scriptscriptfont1=\fivei%
  \textfont2=\eightsy \scriptfont2=\sevensy \scriptscriptfont2=\fivesy%
  \textfont\itfam=\eightit\def\it{\fam\itfam\eightit}%
  \textfont\bffam=\eightbf%
    \scriptfont\bffam=\sevenbf%
      \scriptscriptfont\bffam=\fivebf%
  \def\bf{\fam\bffam\eightbf}%
  \textfont\slfam=\eightsl\def\sl{\fam\slfam\eightsl}%
  \textfont\ttfam=\eighttt\def\tt{\fam\ttfam\eighttt}%
  \textfont\scfam=\eightcsc\def\sc{\fam\scfam\eightcsc}%
  \textfont\sffam=\eightsf\def\sf{\fam\sffam\eightsf}%
  \textfont\mibfam=\eightmib%
  \textfont\sybfam=\eightsyb%
  \bls{10pt}%
}

\def\ninepoint{
  \def\rm{\fam0\ninerm}%
  \textfont0=\ninerm \scriptfont0=\sevenrm \scriptscriptfont0=\fiverm%
  \textfont1=\ninei  \scriptfont1=\seveni  \scriptscriptfont1=\fivei%
  \textfont2=\ninesy \scriptfont2=\sevensy \scriptscriptfont2=\fivesy%
  \textfont\itfam=\nineit\def\it{\fam\itfam\nineit}%
  \textfont\bffam=\ninebf%
    \scriptfont\bffam=\sevenbf%
      \scriptscriptfont\bffam=\fivebf%
  \def\bf{\fam\bffam\ninebf}%
  \textfont\slfam=\ninesl\def\sl{\fam\slfam\ninesl}%
  \textfont\ttfam=\ninett\def\tt{\fam\ttfam\ninett}%
  \textfont\scfam=\ninecsc\def\sc{\fam\scfam\ninecsc}%
  \textfont\sffam=\ninesf\def\sf{\fam\sffam\ninesf}%
  \textfont\mibfam=\ninemib%
  \textfont\sybfam=\ninesyb%
  \bls{12pt}%
}

\def\tenpoint{
  \def\rm{\fam0\tenrm}%
  \textfont0=\tenrm \scriptfont0=\sevenrm \scriptscriptfont0=\fiverm%
  \textfont1=\teni  \scriptfont1=\seveni  \scriptscriptfont1=\fivei%
  \textfont2=\tensy \scriptfont2=\sevensy \scriptscriptfont2=\fivesy%
  \textfont\itfam=\tenit\def\it{\fam\itfam\tenit}%
  \textfont\bffam=\tenbf%
    \scriptfont\bffam=\sevenbf%
      \scriptscriptfont\bffam=\fivebf%
  \def\bf{\fam\bffam\tenbf}%
  \textfont\slfam=\tensl\def\sl{\fam\slfam\tensl}%
  \textfont\ttfam=\tentt\def\tt{\fam\ttfam\tentt}%
  \textfont\scfam=\tencsc\def\sc{\fam\scfam\tencsc}%
  \textfont\sffam=\tensf\def\sf{\fam\sffam\tensf}%
  \textfont\mibfam=\tenmib%
  \textfont\sybfam=\tensyb%
  \bls{12pt}%
}

\def\elevenpoint{
  \def\rm{\fam0\elevenrm}%
  \textfont0=\elevenrm \scriptfont0=\eightrm \scriptscriptfont0=\fiverm%
  \textfont1=\eleveni  \scriptfont1=\eighti  \scriptscriptfont1=\fivei%
  \textfont2=\elevensy \scriptfont2=\eightsy \scriptscriptfont2=\fivesy%
  \textfont\itfam=\elevenit\def\it{\fam\itfam\elevenit}%
  \textfont\bffam=\elevenbf%
    \scriptfont\bffam=\eightbf%
      \scriptscriptfont\bffam=\fivebf%
  \def\bf{\fam\bffam\elevenbf}%
  \textfont\slfam=\elevensl\def\sl{\fam\slfam\elevensl}%
  \textfont\ttfam=\eleventt\def\tt{\fam\ttfam\eleventt}%
  \textfont\scfam=\elevencsc\def\sc{\fam\scfam\elevencsc}%
  \textfont\sffam=\elevensf\def\sf{\fam\sffam\elevensf}%
  \textfont\mibfam=\elevenmib%
  \textfont\sybfam=\elevensyb%
  \bls{13pt}%
}

\def\fourteenpoint{
  \def\rm{\fam0\fourteenrm}%
  \textfont0\fourteenrm  \scriptfont0\tenrm  \scriptscriptfont0\sevenrm%
  \textfont1\fourteeni   \scriptfont1\teni   \scriptscriptfont1\seveni%
  \textfont2\fourteensy  \scriptfont2\tensy  \scriptscriptfont2\sevensy%
  \textfont\itfam=\fourteenit\def\it{\fam\itfam\fourteenit}%
  \textfont\bffam=\fourteenbf%
    \scriptfont\bffam=\tenbf%
      \scriptscriptfont\bffam=\sevenbf%
  \def\bf{\fam\bffam\fourteenbf}%
  \textfont\slfam=\fourteensl\def\sl{\fam\slfam\fourteensl}%
  \textfont\ttfam=\fourteentt\def\tt{\fam\ttfam\fourteentt}%
  \textfont\scfam=\fourteencsc\def\sc{\fam\scfam\fourteencsc}%
  \textfont\sffam=\fourteensf\def\sf{\fam\sffam\fourteensf}%
  \textfont\mibfam=\fourteenmib%
  \textfont\sybfam=\fourteensyb%
  \bls{17pt}%
}

\def\seventeenpoint{
  \def\rm{\fam0\seventeenrm}%
  \textfont0\seventeenrm  \scriptfont0\elevenrm  \scriptscriptfont0\ninerm%
  \textfont1\seventeeni   \scriptfont1\eleveni   \scriptscriptfont1\ninei%
  \textfont2\seventeensy  \scriptfont2\elevensy  \scriptscriptfont2\ninesy%
  \textfont\itfam=\seventeenit\def\it{\fam\itfam\seventeenit}%
  \textfont\bffam=\seventeenbf%
    \scriptfont\bffam=\elevenbf%
      \scriptscriptfont\bffam=\ninebf%
  \def\bf{\fam\bffam\seventeenbf}%
  \textfont\slfam=\seventeensl\def\sl{\fam\slfam\seventeensl}%
  \textfont\ttfam=\seventeentt\def\tt{\fam\ttfam\seventeentt}%
  \textfont\scfam=\seventeencsc\def\sc{\fam\scfam\seventeencsc}%
  \textfont\sffam=\seventeensf\def\sf{\fam\sffam\seventeensf}%
  \textfont\mibfam=\seventeenmib%
  \textfont\sybfam=\seventeensyb%
  \bls{20pt}%
}

\lineskip=1pt      \normallineskip=\lineskip
\lineskiplimit=0pt \normallineskiplimit=\lineskiplimit




\def\Nulle{0}  
\def\Aue{1}    
\def\Afe{2}    
\def\Ace{3}    
\def\Sue{4}    
\def\Hae{5}    
\def\Hbe{6}    
\def\Hce{7}    
\def\Hde{8}    
\def\Kwe{9}    
\def\Txe{10}   
\def\Lie{11}   
\def\Bbe{12}   


\newdimen\DimenA
\newbox\BoxA

\newcount\LastMac \LastMac=\Nulle
\newcount\HeaderNumber \HeaderNumber=0
\newcount\DefaultHeader \DefaultHeader=\HeaderNumber
\newskip\Indent

\newskip\half      \half=5.5pt plus 1.5pt minus 2.25pt
\newskip\one       \one=11pt plus 3pt minus 5.5pt
\newskip\onehalf   \onehalf=16.5pt plus 5.5pt minus 8.25pt
\newskip\two       \two=22pt plus 5.5pt minus 11pt

\def\Half{\vskip-\lastskip\vskip\half}
\def\One{\vskip-\lastskip\vskip\one}
\def\OneHalf{\vskip-\lastskip\vskip\onehalf}
\def\Two{\vskip-\lastskip\vskip\two}


\def\rTenPT{10pt plus \Feathering}

\def\TenPT{10pt plus \Feathering} 
\def\ElevenPT{11pt plus \Feathering}

\def\Raggedright{
 \rightskip=0pt plus \hsize
}

\def\Fullout{
\rightskip=0pt
}

\def\Hang#1#2{
 \hangindent=#1
 \hangafter=#2
}

\def\EveryMac{
 \Fullout
 \everypar{}
}



\def\title#1{
 \EveryMac
 \LastMac=\Nulle
 \global\HeaderNumber=0
 \global\DefaultHeader=1
 \vbox to 1pc{\vss}
 \seventeenpoint
 \Raggedright
 \noindent \bf #1
}

\def\author#1{
 \EveryMac
 \ifnum\LastMac=\Afe \OneHalf
  \else \Two
 \fi
 \LastMac=\Aue
 \fourteenpoint
 \Raggedright
 \noindent \rm #1\par
 \vskip 3pt\relax
}

\def\affiliation#1{
 \EveryMac
 \LastMac=\Afe
 \eightpoint\bls{\TenPT}
 \Raggedright
 \noindent \it #1\par
}

\def\acceptedline#1{
 \EveryMac
 \Two
 \LastMac=\Ace
 \eightpoint\bls{\TenPT}
 \Raggedright
 \noindent \rm #1
}

\def\abstract{%
 \EveryMac
 \Two
 \LastMac=\Sue
 \everypar{\Hang{11pc}{0}}
 \noindent\ninebf ABSTRACT\par
 \tenpoint\bls{\ElevenPT}
 \Fullout
 \noindent\rm
}

\def\keywords{
 \EveryMac
 \Half
 \LastMac=\Kwe
 \everypar{\Hang{11pc}{0}}
 \tenpoint\bls{\ElevenPT}
 \Fullout
 \noindent\hbox{\bf Key words:\ }
 \rm
}


\def\maketitle{%
  \Two%
  \EndOpening%
  \MakePage%
}


\def\pageoffset#1#2{\hoffset=#1\relax\voffset=#2\relax}


\def\Autonumber{
 \global\AutoNumbertrue  
}

\newif\ifAutoNumber \AutoNumberfalse
\newcount\Sec        
\newcount\SecSec
\newcount\SecSecSec

\Sec=0

\def\:{\let\@sptoken= } \:  
\def\:{\@xifnch} \expandafter\def\: {\futurelet\@tempc\@ifnch}

\def\@ifnextchar#1#2#3{%
  \let\@tempMACe #1%
  \def\@tempMACa{#2}%
  \def\@tempMACb{#3}%
  \futurelet \@tempMACc\@ifnch%
}

\def\@ifnch{%
\ifx \@tempMACc \@sptoken%
  \let\@tempMACd\@xifnch%
\else%
  \ifx \@tempMACc \@tempMACe%
    \let\@tempMACd\@tempMACa%
  \else%
    \let\@tempMACd\@tempMACb%
  \fi%
\fi%
\@tempMACd%
}

\def\@ifstar#1#2{\@ifnextchar *{\def\@tempMACa*{#1}\@tempMACa}{#2}}

\def\section{\@ifstar{\@ssection}{\@section}}

\def\@section#1{
 \EveryMac
 \Two
 \LastMac=\Hae
 \tenpoint\bls{\ElevenPT}
 \bf
 \Raggedright
 \ifAutoNumber
  \advance\Sec by 1
  \noindent\number\Sec\hskip 1pc \uppercase{#1}
  \SecSec=0
 \else
  \noindent \uppercase{#1}
 \fi
 \nobreak
}

\def\@ssection#1{
 \EveryMac
 \ifnum\LastMac=\Hae \Half
  \else \OneHalf
 \fi
 \LastMac=\Hae
 \tenpoint\bls{\ElevenPT}
 \bf
 \Raggedright
 \noindent\uppercase{#1}
}

\def\subsection#1{
 \EveryMac
 \ifnum\LastMac=\Hae \Half
  \else \OneHalf
 \fi
 \LastMac=\Hbe
 \tenpoint\bls{\ElevenPT}
 \bf
 \Raggedright
 \ifAutoNumber
  \advance\SecSec by 1
  \noindent\number\Sec.\number\SecSec
  \hskip 1pc #1
  \SecSecSec=0
 \else
  \noindent #1
 \fi
 \nobreak
}

\def\subsubsection#1{
 \EveryMac
 \ifnum\LastMac=\Hbe \Half
  \else \OneHalf
 \fi
 \LastMac=\Hce
 \ninepoint\bls{\ElevenPT}
 \it
 \Raggedright
 \ifAutoNumber
  \advance\SecSecSec by 1
  \noindent\number\Sec.\number\SecSec.\number\SecSecSec
  \hskip 1pc #1
 \else
  \noindent #1
 \fi
 \nobreak
}

\def\paragraph#1{
 \EveryMac
 \One
 \LastMac=\Hde
 \ninepoint\bls{\ElevenPT}
 \noindent \it #1
 \rm
}


\def\tx{
 \EveryMac
 \ifnum\LastMac=\Lie \Half\fi
 \ifnum\LastMac=\Hae \nobreak\Half\fi
 \ifnum\LastMac=\Hbe \nobreak\Half\fi
 \ifnum\LastMac=\Hce \nobreak\Half\fi
 \ifnum\LastMac=\Lie \else \noindent\fi
 \LastMac=\Txe
 \ninepoint\bls{\ElevenPT}
 \rm
}


\def\item{
 \par
 \EveryMac
 \ifnum\LastMac=\Lie
  \else \Half
 \fi
 \LastMac=\Lie
 \ninepoint\bls{\ElevenPT}
 \rm
}


\def\bibitem{
 \par
 \EveryMac
 \ifnum\LastMac=\Bbe
  \else \Half
 \fi
 \LastMac=\Bbe
 \Hang{1.5em}{1}
 \eightpoint\bls{\TenPT}
 \Raggedright
 \noindent \rm
}


\newtoks\CatchLine

\def\@journal{Mon.\ Not.\ R.\ Astron.\ Soc.\ }  
\def\@pubyear{1994}        
\def\@pagerange{000--000}  
\def\@volume{000}          
\def\@microfiche{}         %

\def\pubyear#1{\gdef\@pubyear{#1}\@makecatchline}
\def\pagerange#1{\gdef\@pagerange{#1}\@makecatchline}
\def\volume#1{\gdef\@volume{#1}\@makecatchline}
\def\microfiche#1{\gdef\@microfiche{and Microfiche\ #1}\@makecatchline}

\def\@makecatchline{%
  \global\CatchLine{%
    {\rm \@journal {\bf \@volume},\ \@pagerange\ (\@pubyear)\ \@microfiche}}%
}

\@makecatchline 

\newtoks\LeftHeader
\def\shortauthor#1{
 \global\LeftHeader{#1}
}

\newtoks\RightHeader
\def\shorttitle#1{
 \global\RightHeader{#1}
}

\def\PageHead{
 \EveryMac
 \ifnum\HeaderNumber=1 \Pagehead
  \else \Catchline
 \fi
}

\def\Catchline{%
 \vbox to 0pt{\vskip-22.5pt
  \hbox to \PageWidth{\vbox to8.5pt{}\noindent
  \eightpoint\the\CatchLine\hfill}\vss}
 \nointerlineskip
}

\def\Pagehead{%
 \ifodd\pageno
   \vbox to 0pt{\vskip-22.5pt
   \hbox to \PageWidth{\vbox to8.5pt{}\elevenpoint\it\noindent
    \hfill\the\RightHeader\hskip1.5em\rm\folio}\vss}
 \else
   \vbox to 0pt{\vskip-22.5pt
   \hbox to \PageWidth{\vbox to8.5pt{}\elevenpoint\rm\noindent
   \folio\hskip1.5em\it\the\LeftHeader\hfill}\vss}
 \fi
 \nointerlineskip
}

\def\PageFoot{} 

\def\authorcomment#1{%
  \gdef\PageFoot{%
    \nointerlineskip%
    \vbox to 22pt{\vfil%
      \hbox to \PageWidth{\elevenpoint\rm\noindent \hfil #1 \hfil}}%
  }%
}

\everydisplay{\displaysetup}

\newif\ifeqno
\newif\ifleqno

\def\displaysetup#1$${%
 \displaytest#1\eqno\eqno\displaytest
}

\def\displaytest#1\eqno#2\eqno#3\displaytest{%
 \if!#3!\ldisplaytest#1\leqno\leqno\ldisplaytest
 \else\eqnotrue\leqnofalse\def\eqn{#2}\def\eq{#1}\fi
 \generaldisplay$$}

\def\ldisplaytest#1\leqno#2\leqno#3\ldisplaytest{%
 \def\eq{#1}%
 \if!#3!\eqnofalse\else\eqnotrue\leqnotrue
  \def\eqn{#2}\fi}

\def\generaldisplay{%
\ifeqno \ifleqno 
   \hbox to \hsize{\noindent
     $\displaystyle\eq$\hfil$\displaystyle\eqn$}
  \else
    \hbox to \hsize{\noindent
     $\displaystyle\eq$\hfil$\displaystyle\eqn$}
  \fi
 \else
 \hbox to \hsize{\vbox{\noindent
  $\displaystyle\eq$\hfil}}
 \fi
}

\def\@notice{%
  \par\Two%
  \bls{12pt}%
  \noindent\tenrm This paper has been produced using the Blackwell
                  Scientific Publications \TeX\ macros.%
}

\outer\def\bye{\@notice\par\vfill\supereject\end}

\everyjob{%
  \Warn{Monthly notices of the RAS journal style (\@typeface)\space
        v\@version,\space \@verdate.}\Warn{}%
}




\newif\if@debug \@debugfalse  

\def\Print#1{\if@debug\immediate\write16{#1}\else \fi}
\def\Warn#1{\immediate\write16{#1}}
\def\wlog#1{}

\newcount\Iteration 

\newif\ifFigureBoxes  
\FigureBoxestrue

\def\Single{0} \def\Double{1}                 
\def\Figure{0} \def\Table{1}                  

\def\InStack{0}  
\def\InZoneA{1}
\def\InZoneB{2}
\def\InZoneC{3}

\newcount\TEMPCOUNT 
\newdimen\TEMPDIMEN 
\newbox\TEMPBOX     
\newbox\VOIDBOX     

\newcount\LengthOfStack 
\newcount\MaxItems      
\newcount\StackPointer
\newcount\Point         
\newcount\NextFigure    
\newcount\NextTable     
\newcount\NextItem      

\newcount\StatusStack   
\newcount\NumStack      
\newcount\TypeStack     
\newcount\SpanStack     
\newcount\BoxStack      

\newcount\ItemSTATUS    
\newcount\ItemNUMBER    
\newcount\ItemTYPE      
\newcount\ItemSPAN      
\newbox\ItemBOX         
\newdimen\ItemSIZE      

\newdimen\PageHeight    
\newdimen\TextLeading   
\newdimen\Feathering    
\newcount\LinesPerPage  
\newdimen\ColumnWidth   
\newdimen\ColumnGap     
\newdimen\PageWidth     
\newdimen\BodgeHeight   
\newcount\Leading       

\newdimen\ZoneBSize  
\newdimen\TextSize   
\newbox\ZoneABOX     
\newbox\ZoneBBOX     
\newbox\ZoneCBOX     

\newif\ifFirstSingleItem
\newif\ifFirstZoneA
\newif\ifMakePageInComplete
\newif\ifMoreFigures \MoreFiguresfalse 
\newif\ifMoreTables  \MoreTablesfalse  

\newif\ifFigInZoneB 
\newif\ifFigInZoneC 
\newif\ifTabInZoneB 
\newif\ifTabInZoneC

\newif\ifZoneAFullPage

\newbox\MidBOX    
\newbox\LeftBOX
\newbox\RightBOX
\newbox\PageBOX   

\newif\ifLeftCOL  
\LeftCOLtrue

\newdimen\ZoneBAdjust

\newcount\ItemFits
\def\Yes{1}
\def\No{2}




\MaxItems=15
\NextFigure=0        
\NextTable=1

\BodgeHeight=6pt
\TextLeading=11pt    
\Leading=11
\Feathering=0pt      
\LinesPerPage=61     
\topskip=\TextLeading
\ColumnWidth=20pc    
\ColumnGap=2pc       

\def\ItemSep{\vskip \TextLeading plus \TextLeading minus 4pt}

\FigureBoxesfalse 

\parskip=0pt
\parindent=18pt
\widowpenalty=0
\clubpenalty=10000
\tolerance=1500
\hbadness=1500
\abovedisplayskip=6pt plus 2pt minus 2pt
\belowdisplayskip=6pt plus 2pt minus 2pt
\abovedisplayshortskip=6pt plus 2pt minus 2pt
\belowdisplayshortskip=6pt plus 2pt minus 2pt

\PageHeight=\TextLeading 
\multiply\PageHeight by \LinesPerPage
\advance\PageHeight by \topskip

\PageWidth=2\ColumnWidth
\advance\PageWidth by \ColumnGap




\newcount\DUMMY \StatusStack=\allocationnumber
\newcount\DUMMY \newcount\DUMMY \newcount\DUMMY 
\newcount\DUMMY \newcount\DUMMY \newcount\DUMMY 
\newcount\DUMMY \newcount\DUMMY \newcount\DUMMY
\newcount\DUMMY \newcount\DUMMY \newcount\DUMMY 
\newcount\DUMMY \newcount\DUMMY \newcount\DUMMY

\newcount\DUMMY \NumStack=\allocationnumber
\newcount\DUMMY \newcount\DUMMY \newcount\DUMMY 
\newcount\DUMMY \newcount\DUMMY \newcount\DUMMY 
\newcount\DUMMY \newcount\DUMMY \newcount\DUMMY 
\newcount\DUMMY \newcount\DUMMY \newcount\DUMMY 
\newcount\DUMMY \newcount\DUMMY \newcount\DUMMY

\newcount\DUMMY \TypeStack=\allocationnumber
\newcount\DUMMY \newcount\DUMMY \newcount\DUMMY 
\newcount\DUMMY \newcount\DUMMY \newcount\DUMMY 
\newcount\DUMMY \newcount\DUMMY \newcount\DUMMY 
\newcount\DUMMY \newcount\DUMMY \newcount\DUMMY 
\newcount\DUMMY \newcount\DUMMY \newcount\DUMMY

\newcount\DUMMY \SpanStack=\allocationnumber
\newcount\DUMMY \newcount\DUMMY \newcount\DUMMY 
\newcount\DUMMY \newcount\DUMMY \newcount\DUMMY 
\newcount\DUMMY \newcount\DUMMY \newcount\DUMMY 
\newcount\DUMMY \newcount\DUMMY \newcount\DUMMY 
\newcount\DUMMY \newcount\DUMMY \newcount\DUMMY

\newbox\DUMMY   \BoxStack=\allocationnumber
\newbox\DUMMY   \newbox\DUMMY \newbox\DUMMY 
\newbox\DUMMY   \newbox\DUMMY \newbox\DUMMY 
\newbox\DUMMY   \newbox\DUMMY \newbox\DUMMY 
\newbox\DUMMY   \newbox\DUMMY \newbox\DUMMY 
\newbox\DUMMY   \newbox\DUMMY \newbox\DUMMY

\def\wlog{\immediate\write-1}


\def\GetItemAll#1{%
 \GetItemSTATUS{#1}
 \GetItemNUMBER{#1}
 \GetItemTYPE{#1}
 \GetItemSPAN{#1}
 \GetItemBOX{#1}
}

\def\GetItemSTATUS#1{%
 \Point=\StatusStack
 \advance\Point by #1
 \global\ItemSTATUS=\count\Point
}

\def\GetItemNUMBER#1{%
 \Point=\NumStack
 \advance\Point by #1
 \global\ItemNUMBER=\count\Point
}

\def\GetItemTYPE#1{%
 \Point=\TypeStack
 \advance\Point by #1
 \global\ItemTYPE=\count\Point
}

\def\GetItemSPAN#1{%
 \Point\SpanStack
 \advance\Point by #1
 \global\ItemSPAN=\count\Point
}

\def\GetItemBOX#1{%
 \Point=\BoxStack
 \advance\Point by #1
 \global\setbox\ItemBOX=\vbox{\copy\Point}
 \global\ItemSIZE=\ht\ItemBOX
 \global\advance\ItemSIZE by \dp\ItemBOX
 \TEMPCOUNT=\ItemSIZE
 \divide\TEMPCOUNT by \Leading
 \divide\TEMPCOUNT by 65536
 \advance\TEMPCOUNT by 1
 \ItemSIZE=\TEMPCOUNT pt
 \global\multiply\ItemSIZE by \Leading
}


\def\JoinStack{%
 \ifnum\LengthOfStack=\MaxItems 
  \Warn{WARNING: Stack is full...some items will be lost!}
 \else
  \Point=\StatusStack
  \advance\Point by \LengthOfStack
  \global\count\Point=\ItemSTATUS
  \Point=\NumStack
  \advance\Point by \LengthOfStack
  \global\count\Point=\ItemNUMBER
  \Point=\TypeStack
  \advance\Point by \LengthOfStack
  \global\count\Point=\ItemTYPE
  \Point\SpanStack
  \advance\Point by \LengthOfStack
  \global\count\Point=\ItemSPAN
  \Point=\BoxStack
  \advance\Point by \LengthOfStack
  \global\setbox\Point=\vbox{\copy\ItemBOX}
  \global\advance\LengthOfStack by 1
  \ifnum\ItemTYPE=\Figure 
   \global\MoreFigurestrue
  \else
   \global\MoreTablestrue
  \fi
 \fi
}


\def\LeaveStack#1{%
 {\Iteration=#1
 \loop
 \ifnum\Iteration<\LengthOfStack
  \advance\Iteration by 1
  \GetItemSTATUS{\Iteration}
   \advance\Point by -1
   \global\count\Point=\ItemSTATUS
  \GetItemNUMBER{\Iteration}
   \advance\Point by -1
   \global\count\Point=\ItemNUMBER
  \GetItemTYPE{\Iteration}
   \advance\Point by -1
   \global\count\Point=\ItemTYPE
  \GetItemSPAN{\Iteration}
   \advance\Point by -1
   \global\count\Point=\ItemSPAN
  \GetItemBOX{\Iteration}
   \advance\Point by -1
   \global\setbox\Point=\vbox{\copy\ItemBOX}
 \repeat}
 \global\advance\LengthOfStack by -1
}


\newif\ifStackNotClean

\def\CleanStack{%
 \StackNotCleantrue
 {\Iteration=0
  \loop
   \ifStackNotClean
    \GetItemSTATUS{\Iteration}
    \ifnum\ItemSTATUS=\InStack
     \advance\Iteration by 1
     \else
      \LeaveStack{\Iteration}
    \fi
   \ifnum\LengthOfStack<\Iteration
    \StackNotCleanfalse
   \fi
 \repeat}
}


\def\FindItem#1#2{%
 \global\StackPointer=-1 
 {\Iteration=0
  \loop
  \ifnum\Iteration<\LengthOfStack
   \GetItemSTATUS{\Iteration}
   \ifnum\ItemSTATUS=\InStack
    \GetItemTYPE{\Iteration}
    \ifnum\ItemTYPE=#1
     \GetItemNUMBER{\Iteration}
     \ifnum\ItemNUMBER=#2
      \global\StackPointer=\Iteration
      \Iteration=\LengthOfStack 
     \fi
    \fi
   \fi
  \advance\Iteration by 1
 \repeat}
}


\def\FindNext{%
 \global\StackPointer=-1 
 {\Iteration=0
  \loop
  \ifnum\Iteration<\LengthOfStack
   \GetItemSTATUS{\Iteration}
   \ifnum\ItemSTATUS=\InStack
    \GetItemTYPE{\Iteration}
   \ifnum\ItemTYPE=\Figure
    \ifMoreFigures
      \global\NextItem=\Figure
      \global\StackPointer=\Iteration
      \Iteration=\LengthOfStack 
    \fi
   \fi
   \ifnum\ItemTYPE=\Table
    \ifMoreTables
      \global\NextItem=\Table
      \global\StackPointer=\Iteration
      \Iteration=\LengthOfStack 
    \fi
   \fi
  \fi
  \advance\Iteration by 1
 \repeat}
}


\def\ChangeStatus#1#2{%
 \Point=\StatusStack
 \advance\Point by #1
 \global\count\Point=#2
}



\def\Zone{\InZoneA}

\ZoneBAdjust=0pt

\def\MakePage{
 \global\ZoneBSize=\PageHeight
 \global\TextSize=\ZoneBSize
 \global\ZoneAFullPagefalse
 \global\topskip=\TextLeading
 \MakePageInCompletetrue
 \MoreFigurestrue
 \MoreTablestrue
 \FigInZoneBfalse
 \FigInZoneCfalse
 \TabInZoneBfalse
 \TabInZoneCfalse
 \global\FirstSingleItemtrue
 \global\FirstZoneAtrue
 \global\setbox\ZoneABOX=\box\VOIDBOX
 \global\setbox\ZoneBBOX=\box\VOIDBOX
 \global\setbox\ZoneCBOX=\box\VOIDBOX
 \loop
  \ifMakePageInComplete
 \FindNext
 \ifnum\StackPointer=-1
  \NextItem=-1
  \MoreFiguresfalse
  \MoreTablesfalse
 \fi
 \ifnum\NextItem=\Figure
   \FindItem{\Figure}{\NextFigure}
   \ifnum\StackPointer=-1 \global\MoreFiguresfalse
   \else
    \GetItemSPAN{\StackPointer}
    \ifnum\ItemSPAN=\Single \def\Zone{\InZoneB}\relax
     \ifFigInZoneC \global\MoreFiguresfalse\fi
    \else
     \def\Zone{\InZoneA}
     \ifFigInZoneB \def\Zone{\InZoneC}\fi
    \fi
   \fi
   \ifMoreFigures\Print{}\FigureItems\fi
 \fi
\ifnum\NextItem=\Table
   \FindItem{\Table}{\NextTable}
   \ifnum\StackPointer=-1 \global\MoreTablesfalse
   \else
    \GetItemSPAN{\StackPointer}
    \ifnum\ItemSPAN=\Single\relax
     \ifTabInZoneC \global\MoreTablesfalse\fi
    \else
     \def\Zone{\InZoneA}
     \ifTabInZoneB \def\Zone{\InZoneC}\fi
    \fi
   \fi
   \ifMoreTables\Print{}\TableItems\fi
 \fi
   \MakePageInCompletefalse 
   \ifMoreFigures\MakePageInCompletetrue\fi
   \ifMoreTables\MakePageInCompletetrue\fi
 \repeat
 \ifZoneAFullPage
  \global\TextSize=0pt
  \global\ZoneBSize=0pt
  \global\vsize=0pt\relax
  \global\topskip=0pt\relax
  \vbox to 0pt{\vss}
  \eject
 \else
 \global\advance\ZoneBSize by -\ZoneBAdjust
 \global\vsize=\ZoneBSize
 \global\hsize=\ColumnWidth
 \global\ZoneBAdjust=0pt
 \ifdim\TextSize<23pt
 \Warn{}
 \Warn{* Making column fall short: TextSize=\the\TextSize *}
 \vskip-\lastskip\eject\fi
 \fi
}

\def\MakeRightCol{
 \global\TextSize=\ZoneBSize
 \MakePageInCompletetrue
 \MoreFigurestrue
 \MoreTablestrue
 \global\FirstSingleItemtrue
 \global\setbox\ZoneBBOX=\box\VOIDBOX
 \def\Zone{\InZoneB}
 \loop
  \ifMakePageInComplete
 \FindNext
 \ifnum\StackPointer=-1
  \NextItem=-1
  \MoreFiguresfalse
  \MoreTablesfalse
 \fi
 \ifnum\NextItem=\Figure
   \FindItem{\Figure}{\NextFigure}
   \ifnum\StackPointer=-1 \MoreFiguresfalse
   \else
    \GetItemSPAN{\StackPointer}
    \ifnum\ItemSPAN=\Double\relax
     \MoreFiguresfalse\fi
   \fi
   \ifMoreFigures\Print{}\FigureItems\fi
 \fi
 \ifnum\NextItem=\Table
   \FindItem{\Table}{\NextTable}
   \ifnum\StackPointer=-1 \MoreTablesfalse
   \else
    \GetItemSPAN{\StackPointer}
    \ifnum\ItemSPAN=\Double\relax
     \MoreTablesfalse\fi
   \fi
   \ifMoreTables\Print{}\TableItems\fi
 \fi
   \MakePageInCompletefalse 
   \ifMoreFigures\MakePageInCompletetrue\fi
   \ifMoreTables\MakePageInCompletetrue\fi
 \repeat
 \ifZoneAFullPage
  \global\TextSize=0pt
  \global\ZoneBSize=0pt
  \global\vsize=0pt\relax
  \global\topskip=0pt\relax
  \vbox to 0pt{\vss}
  \eject
 \else
 \global\vsize=\ZoneBSize
 \global\hsize=\ColumnWidth
 \ifdim\TextSize<23pt
 \Warn{}
 \Warn{* Making column fall short: TextSize=\the\TextSize *}
 \vskip-\lastskip\eject\fi
\fi
}

\def\FigureItems{
 \Print{Considering...}
 \ShowItem{\StackPointer}
 \GetItemBOX{\StackPointer} 
 \GetItemSPAN{\StackPointer}
  \CheckFitInZone 
  \ifnum\ItemFits=\Yes
   \ifnum\ItemSPAN=\Single
     \ChangeStatus{\StackPointer}{\InZoneB} 
     \global\FigInZoneBtrue
     \ifFirstSingleItem
      \hbox{}\vskip-\BodgeHeight
     \global\advance\ItemSIZE by \TextLeading
     \fi
     \unvbox\ItemBOX\ItemSep
     \global\FirstSingleItemfalse
     \global\advance\TextSize by -\ItemSIZE
     \global\advance\TextSize by -\TextLeading
   \else
    \ifFirstZoneA
     \global\advance\ItemSIZE by \TextLeading
     \global\FirstZoneAfalse\fi
    \global\advance\TextSize by -\ItemSIZE
    \global\advance\TextSize by -\TextLeading
    \global\advance\ZoneBSize by -\ItemSIZE
    \global\advance\ZoneBSize by -\TextLeading
    \ifFigInZoneB\relax
     \else
     \ifdim\TextSize<3\TextLeading
     \global\ZoneAFullPagetrue
     \fi
    \fi
    \ChangeStatus{\StackPointer}{\Zone}
    \ifnum\Zone=\InZoneC \global\FigInZoneCtrue\fi
  \fi
   \Print{TextSize=\the\TextSize}
   \Print{ZoneBSize=\the\ZoneBSize}
  \global\advance\NextFigure by 1
   \Print{This figure has been placed.}
  \else
   \Print{No space available for this figure...holding over.}
   \Print{}
   \global\MoreFiguresfalse
  \fi
}

\def\TableItems{
 \Print{Considering...}
 \ShowItem{\StackPointer}
 \GetItemBOX{\StackPointer} 
 \GetItemSPAN{\StackPointer}
  \CheckFitInZone 
  \ifnum\ItemFits=\Yes
   \ifnum\ItemSPAN=\Single
    \ChangeStatus{\StackPointer}{\InZoneB}
     \global\TabInZoneBtrue
     \ifFirstSingleItem
      \hbox{}\vskip-\BodgeHeight
     \global\advance\ItemSIZE by \TextLeading
     \fi
     \unvbox\ItemBOX\ItemSep
     \global\FirstSingleItemfalse
     \global\advance\TextSize by -\ItemSIZE
     \global\advance\TextSize by -\TextLeading
   \else
    \ifFirstZoneA
    \global\advance\ItemSIZE by \TextLeading
    \global\FirstZoneAfalse\fi
    \global\advance\TextSize by -\ItemSIZE
    \global\advance\TextSize by -\TextLeading
    \global\advance\ZoneBSize by -\ItemSIZE
    \global\advance\ZoneBSize by -\TextLeading
    \ifFigInZoneB\relax
     \else
     \ifdim\TextSize<3\TextLeading
     \global\ZoneAFullPagetrue
     \fi
    \fi
    \ChangeStatus{\StackPointer}{\Zone}
    \ifnum\Zone=\InZoneC \global\TabInZoneCtrue\fi
   \fi
  \global\advance\NextTable by 1
   \Print{This table has been placed.}
  \else
  \Print{No space available for this table...holding over.}
   \Print{}
   \global\MoreTablesfalse
  \fi
}


\def\CheckFitInZone{%
{\advance\TextSize by -\ItemSIZE
 \advance\TextSize by -\TextLeading
 \ifFirstSingleItem
  \advance\TextSize by \TextLeading
 \fi
 \ifnum\Zone=\InZoneA\relax
  \else \advance\TextSize by -\ZoneBAdjust
 \fi
 \ifdim\TextSize<3\TextLeading \global\ItemFits=\No
 \else \global\ItemFits=\Yes\fi}
}

\def\BF#1#2{
 \ItemSTATUS=\InStack
 \ItemNUMBER=#1
 \ItemTYPE=\Figure
 \if#2S \ItemSPAN=\Single
  \else \ItemSPAN=\Double
 \fi
 \setbox\ItemBOX=\vbox{}
}

\def\BT#1#2{
 \ItemSTATUS=\InStack
 \ItemNUMBER=#1
 \ItemTYPE=\Table
 \if#2S \ItemSPAN=\Single
  \else \ItemSPAN=\Double
 \fi
 \setbox\ItemBOX=\vbox{}
}

\def\BeginOpening{%
 \hsize=\PageWidth
 \global\setbox\ItemBOX=\vbox\bgroup
}

\let\begintopmatter=\BeginOpening  

\def\EndOpening{%
 \egroup
 \ItemNUMBER=0
 \ItemTYPE=\Figure
 \ItemSPAN=\Double
 \ItemSTATUS=\InStack
 \JoinStack
}


\newbox\tmpbox

\def\FC#1#2#3#4{%
  \ItemSTATUS=\InStack
  \ItemNUMBER=#1
  \ItemTYPE=\Figure
  \if#2S
    \ItemSPAN=\Single \TEMPDIMEN=\ColumnWidth
  \else
    \ItemSPAN=\Double \TEMPDIMEN=\PageWidth
  \fi
  {\hsize=\TEMPDIMEN
   \global\setbox\ItemBOX=\vbox{%
     \ifFigureBoxes
       \B{\TEMPDIMEN}{#3}
     \else
       \vbox to #3{\vfil}%
     \fi%
     \eightpoint\rm\bls{\rTenPT}%
     \vskip 5.5pt plus 6pt%
     \setbox\tmpbox=\vbox{#4\par}%
     \ifdim\ht\tmpbox>10pt 
       \noindent #4\par%
     \else
       \hbox to \hsize{\hfil #4\hfil}%
     \fi%
   }%
  }%
  \JoinStack%
  \Print{Processing source for figure {\the\ItemNUMBER}}%
}


\def\figps#1#2#3#4{%
  \ItemSTATUS=\InStack
  \ItemNUMBER=#1
  \ItemTYPE=\Figure
  \if#2S
    \ItemSPAN=\Single \TEMPDIMEN=\ColumnWidth
  \else
    \ItemSPAN=\Double \TEMPDIMEN=\PageWidth
  \fi
  {\hsize=\TEMPDIMEN
   \global\setbox\ItemBOX=\vbox{#3
     \eightpoint\rm\bls{\rTenPT}%
     \vskip 5.5pt plus 6pt%
     \setbox\tmpbox=\vbox{#4\par}%
     \ifdim\ht\tmpbox>10pt 
       \noindent #4\par%
     \else
       \hbox to \hsize{\hfil #4\hfil}%
     \fi%
   }%
  }%
  \JoinStack%
  \Print{Processing source for figure {\the\ItemNUMBER}}%
}

\def\TH#1#2#3#4{%
 \ItemSTATUS=\InStack
 \ItemNUMBER=#1
 \ItemTYPE=\Table
 \if#2S \ItemSPAN=\Single \TEMPDIMEN=\ColumnWidth
  \else \ItemSPAN=\Double \TEMPDIMEN=\PageWidth
 \fi
 {\hsize=\TEMPDIMEN
 \eightpoint\bls{\rTenPT}\rm
 \global\setbox\ItemBOX=\vbox{\noindent#3\vskip 5.5pt plus5.5pt\noindent#4}}
 \JoinStack
 \Print{Processing source for table {\the\ItemNUMBER}}
}

\let\table=\TH  

\def\UnloadZoneA{%
\FirstZoneAtrue
 \Iteration=0
  \loop
   \ifnum\Iteration<\LengthOfStack
    \GetItemSTATUS{\Iteration}
    \ifnum\ItemSTATUS=\InZoneA
     \GetItemBOX{\Iteration}
     \ifFirstZoneA \vbox to \BodgeHeight{\vfil}%
     \FirstZoneAfalse\fi
     \unvbox\ItemBOX\ItemSep
     \LeaveStack{\Iteration}
     \else
     \advance\Iteration by 1
   \fi
 \repeat
}

\def\UnloadZoneC{%
\Iteration=0
  \loop
   \ifnum\Iteration<\LengthOfStack
    \GetItemSTATUS{\Iteration}
    \ifnum\ItemSTATUS=\InZoneC
     \GetItemBOX{\Iteration}
     \ItemSep\unvbox\ItemBOX
     \LeaveStack{\Iteration}
     \else
     \advance\Iteration by 1
   \fi
 \repeat
}


\def\ShowItem#1{
  {\GetItemAll{#1}
  \Print{\the#1:
  {TYPE=\ifnum\ItemTYPE=\Figure Figure\else Table\fi}
  {NUMBER=\the\ItemNUMBER}
  {SPAN=\ifnum\ItemSPAN=\Single Single\else Double\fi}
  {SIZE=\the\ItemSIZE}}}
}

\def\ShowStack{%
 \Print{}
 \Print{LengthOfStack = \the\LengthOfStack}
 \ifnum\LengthOfStack=0 \Print{Stack is empty}\fi
 \Iteration=0
 \loop
 \ifnum\Iteration<\LengthOfStack
  \ShowItem{\Iteration}
  \advance\Iteration by 1
 \repeat
}

\def\B#1#2{%
\hbox{\vrule\kern-0.4pt\vbox to #2{%
\hrule width #1\vfill\hrule}\kern-0.4pt\vrule}
}

\def\Ref#1{\begingroup\global\setbox\TEMPBOX=\vbox{\hsize=2in\noindent#1}\endgroup
\ht1=0pt\dp1=0pt\wd1=0pt\vadjust{\vtop to 0pt{\advance
\hsize0.5pc\kern-10pt\moveright\hsize\box\TEMPBOX\vss}}}

\def\MarkRef#1{\leavevmode\thinspace\hbox{\vrule\vtop
{\vbox{\hrule\kern1pt\hbox{\vphantom{\rm/}\thinspace{\rm#1}%
\thinspace}}\kern1pt\hrule}\vrule}\thinspace}%


\output{%
 \ifLeftCOL
  \global\setbox\LeftBOX=\vbox to \ZoneBSize{\box255\unvbox\ZoneBBOX
    \ifvoid\footins\else
    \vskip\skip\footins\unvbox\footins\fi}
  \global\LeftCOLfalse
  \MakeRightCol
 \else
  \setbox\RightBOX=\vbox to \ZoneBSize{\box255\unvbox\ZoneBBOX
    \ifvoid\footins\else
    \vskip\skip\footins\unvbox\footins\fi}
  \setbox\MidBOX=\hbox{\box\LeftBOX\hskip\ColumnGap\box\RightBOX}
  \setbox\PageBOX=\vbox to \PageHeight{%
  \UnloadZoneA\box\MidBOX\UnloadZoneC}
  \shipout\vbox{\PageHead\box\PageBOX\PageFoot}
  \global\advance\pageno by 1
  \global\HeaderNumber=\DefaultHeader
  \global\LeftCOLtrue
  \CleanStack
  \MakePage
 \fi
}


\catcode `\@=12 


%% file: mndnote.tex
%


\catcode`@=11
\def\note#1{%
  \let\@sf=\empty \ifhmode\edef\@sf{\spacefactor\the\spacefactor}\/\fi
  \insert\footins\bgroup
    \eightpoint\bls{\rTenPT}\rm
    \interlinepenalty\interfootnotelinepenalty
    \splitmaxdepth\dp\strutbox \floatingpenalty\@MM
    \leftskip\z@skip \rightskip\z@skip \spaceskip\z@skip \xspaceskip\z@skip
    \footstrut #1\strut\par
  \egroup
  \@sf\relax}


\output{%
 \ifLeftCOL
  \global\setbox\LeftBOX=\vbox to \ZoneBSize{\box255\unvbox\ZoneBBOX
    \ifvoid\footins\else
    \vskip\skip\footins\unvbox\footins\fi}
  \global\LeftCOLfalse
  \MakeRightCol
 \else
  \setbox\RightBOX=\vbox to \ZoneBSize{\box255\unvbox\ZoneBBOX
    \ifvoid\footins\else
    \vskip\skip\footins\unvbox\footins\fi}
  \setbox\MidBOX=\hbox{\box\LeftBOX\hskip\ColumnGap\box\RightBOX}
  \setbox\PageBOX=\vbox to \PageHeight{%
  \UnloadZoneA\box\MidBOX\UnloadZoneC}
  \shipout\vbox{\PageHead\box\PageBOX\PageFoot}
  \global\advance\pageno by 1
  \global\HeaderNumber=\DefaultHeader
  \global\LeftCOLtrue
  \CleanStack
  \MakePage
 \fi
}

%% file: catabs.tex
\abstract
We present a technique to construct a fair sample of simulated galaxy
clusters, and build such a sample for a specific cosmological structure
formation scenario. Conventionally one extracts such a sample from a
single low-resolution large-scale simulation. Here we simulate the
clusters individually at high resolution.
This is made possible by the method of constrained random fields, in
which one can put {\it linear}\ constraints on peaks in the initial
smoothed density field.

We assume that these peaks are the progenitors of present-day rich
clusters, and select clusters for a catalogue by
selecting their initial peak parameters. We find that the
final cluster mass can be well approximated by a linear function of
both the amplitude and the curvature of the initial density peak.
Because the probability distributions of these peak parameters are known,
we can construct a model catalogue selected on expected final cluster
mass. Such a catalogue will not have a well-defined richness limit,
because the relation between richness and mass is fairly broad.
However, by applying the appropriate completeness corrections,
the results for the mass-selected catalogue can be compared with
observations for richness-selected cluster catalogues.

Each cluster model is evolved from its constrained initial conditions
by means of an N-body integrator. This includes an algorithm for
galaxy formation, so we produce two-component models consisting of
dark matter background particles and galaxies. The latter allow us to
{\it directly}\ obtain the observable properties of the cluster
models, and match these to the observed properties to define the
present time in the models, and thus derive the amplitude of the
initial density fluctuation spectrum, $\sigma_8$.

We build a model cluster catalogue for the $\Omega_0=1$ CDM scenario
that is designed to mimic the {\it ENACS}\ sample of rich Abell clusters.
We use the distribution of richness, corrected for incompleteness, to fix
the present epoch. We find $\sigma_8=0.4-0.5$, which is consistent with
other determinations. The catalogue is 70 per cent complete for a
richness larger than 50, but we do have a {\it complete}\ subsample
for richnesses larger than 75.

As a first test we compare the cumulative distribution of line-of-sight
velocity dispersions to those found for several observational samples,
and find that they match best for $\sigma_8\approx0.4$.
This means that we find consistent values for $\sigma_8$ for the CDM
$\Omega_0=1$ scenario on cluster scales.


%% file: cat1.tex
\section{Introduction}

\tx Cluster of galaxies are the largest bound structures in the Universe,
but they are dynamically young and still contain signatures of their
initital conditions.  Statistical properties of samples of galaxy
clusters can therefore provide constraints for models of large-scale
structure formation in the universe.  Large catalogues of observed
clusters of galaxies have been compiled, of which the Abell, Corwin \&
Olowin (1990, ACO from here on) catalogue is the best known, but
complete X-ray selected cluster samples have recently become available
(eg.\ Ebeling et al.\ 1996). The ACO clusters have been most
extensively observed in the optical, as that is where they have been
defined.  A renewed interest in their observational properties has
recently led to a flood of new data, and to a better understanding of
their properties. It is therefore desirable to construct a model
catalogue that mimics observed ones, thus allowing a meaningful
statistical comparison between observations and predictions. In this
paper we discuss a technique to construct such a model catalogue, and
we build a catalogue for one particular cosmological scenario, viz.
$\Omega_0$ = 1 standard CDM.

One approach for obtaining a sample of model clusters is to extract
clusters from large-scale N-body simulations. This has been done by
Frenk et al.\ (1990), and more recently by van Haarlem, Frenk \& White
(1996), and Eke, Cole \& Frenk (1996). The advantage of this approach
is that completeness of the resulting catalogue is automatically
guaranteed. However, a disadvantage is that the resolution
of the extracted cluster models is relatively poor since voids and
filaments, which are simulated as well, take up most of the volume.
This means that the fraction of particles in clusters is
only about 10 per cent. We avoid this problem by running an ensemble
of individual, high-resolution cluster simulations. This technique has been
used by various authors in order to study the influence of the choice
of initial conditions on the final cluster properties (Evrard 1989;
Evrard, Mohr, Fabricant \& Geller 1993; van Haarlem \& van de Weygaert
1993). However, because these authors did not aim to construct a fair
sample, a statistical comparison with observed catalogues as performed
by Frenk et al.\ (1990) and others was not feasible. This is exactly
the goal we aim for here: to construct a statistically fair catalogue
of individual cluster models, each of which is simulated at relatively
high resolution. In this context the term `fair sample' is meant to
indicate a sample which is representative of a volume-limited sample.

We add to this a method which includes the formation and merging of galaxies
in dissipationless N-body simulations (van Kampen 1996). The basic idea is
that each group of particles that is roughly in virial equilibrium is replaced
by a single particle with mass, position, velocity and softening corresponding
to that group. This is done several times during the evolution. Since groups
that form later in the evolution can contain one or more earlier formed galaxies,
both growing and merging of galaxies are included in this simulation method,
although somewhat crudely (improvement on this scheme is in progress).
One reason for using this method is that groups of particles that can be
identified as galaxies need to be protected from destruction within the main
cluster by numerical two-body disruption (Carlberg 1994, van Kampen 1995).
But it also is highly desirable to be able to {\it directly}\ compare the
properties of the modelled galaxy distribution to observations, instead
of making assumptions on the relation between galaxies and dark matter.

Both the fluctuations in the galaxy distribution and in the underlying dark
matter distribution grow with time. The relation between the two
(usually described by some bias parameter), may change with time.
In the simulations, the present epoch is identified as the time when the
properties of the simulated galaxy distribution match the observed ones.
We use the normalisation obtained by van Kampen (1996) from a set of field
models run for the same cosmology as adopted here.

In order to construct a sample of individually simulated cluster models, we
need to make a few assumptions.
We assume that clusters of galaxies, as we observe them at the present epoch,
formed from peaks in the initial density field smoothed at the length-scale
appropriate for clusters. These initial peaks have several characteristics
which can be used to predict whether they will evolve into rich Abell
clusters.
We also assume that the initial density fluctuations are Gaussian distributed.
This allows the use of the theory of Gaussian random fields (Rice 1954;
Doroshkevich 1970; Adler 1981; Peacock \& Heavens 1985; Bardeen et al.\ 1986,
BBKS from here on) which provides the probability distributions for the peak
parameters in the early (linear) stages of the evolution of the density field. 

The construction of a sample of cluster models involves the generation of
a set of initial conditions that produce a set of simulated clusters
with the same {\it statistical}\ properties as for an observed sample.
Such realisations of initial matter distributions that produce a cluster
with specific properties can be obtained by means of the Hoffman-Ribak method
of constructing constrained random fields (Hoffman \& Ribak 1991).
We used the implementation of this method by van de Weygaert \& Bertschinger
(1996). An important limitation of this method is that it can only constrain
{\it linear}\ functionals of a field. This means that the defining quantity
of the catalogue has to be a linear functional as well. Given this limitation
on the constrained random field method of generating initial conditions,
we argue (in Section 3.2) that the final cluster mass is the best choice for
the defining quantity.

In order to construct a catalogue, we assume that ${\cal N}$ rich
Abell clusters within a given volume originate from those ${\cal N}$
initial density peaks that have the largest predicted final mass.  We
use the amplitude {\it and}\ the curvature of the initial peak to
predict the final cluster mass (Section 3.3 and Appendix B). This
combination appears to give a better prediction of the actual final
cluster masses than that based on the initial amplitude alone, which
would be sufficient if the evolution would remain linear.  Using the
probability distributions from Gaussian random field theory we then
sample the distribution of expected total masses to construct a fair
statistical sample.

In order to build a model catalogue, we need to choose a cosmological
scenario, and an observational catalogue to compare to.
We adopt the standard CDM $\Omega_0=1$ scenario, because it allows us to
compare with other studies (Frenk et al.\ 1990; White, Efstathiou \&
Frenk 1993; Eke, Cole \& Frenk 1996). It also has the property that the
amplitude of the dark matter fluctuations need not be fixed at the outset.
This is important, since we do not know {\it a priori}\ how the galaxy
distribution will evolve with respect to the dark matter.

The observational catalogue we try to mimic is a complete subset of the Abell
catalogue for which data was gathered in the context of an ESO Key-programme
(Katgert et al.\ 1996; Mazure et al.\ 1996; den Hartog 1995). This sample,
denoted as ENACS from here on, is the largest well-defined catalogue
presently available with extensive information on the internal structure
of clusters of galaxies.

Because there is bound to be significant scatter in the relation
between richness and the expected final cluster mass, our model
catalogue, constructed to be complete in mass, will not be complete in
richness. In Section 5 we discuss the cause of this incompleteness,
and apply a statistical method to correct for it. This gives us
complete distribution functions for cluster properties like richness
and line-of-sight velocity dispersion.

In Section 6 we present the model counterpart of the ESO-Kpgm
catalogue for the $\Omega_0=1$ CDM scenario. We use the distribution of
cluster richness in addition to the galaxy auto-correlation function (van
Kampen 1996) to fix the normalisation (in terms of $\sigma_8$). We
find that both measures provide a consistent timing for the
$\Omega_0=1$ CDM scenario.

We can use the catalogue to test the validity of the chosen scenario by means
of other measures, like the distribution of cluster shapes or line-of-sight
velocity dispersions. We describe several such tests in Section 7. A
preliminary comparison to observations and a more detailed study of the
intrinsic properties of the model clusters can be found in van Kampen (1994).

%% file: cat2.tex
\section{Rich Abell clusters and peaks in Gaussian random fields}

\tx Our aim is to build a sample of model clusters that can be compared to
an observed catalogue.
We therefore need to find a relation between the defining criteria
for observed and model clusters. The former are found within the discrete
distribution of galaxies as observed in projection on the sky, while the
latter are defined in the continuous density field which we sample with
the particle distribution in a numerical simulation. In this section we
discuss the observational and theoretical ways to define clusters of
galaxies, and also the numerical methods to build cluster models.

\subsection{Abell's observational definition of a rich cluster}

\tx We will compare our models to rich Abell (1958) clusters,
so we first list Abell's main criteria for the definition of an
observed richness class $\ge1$ cluster. They read:

{\it Richness criterion} - `A rich cluster must contain at least 50 members
that are not more than 2 magnitudes fainter than the third brightest member'. 

{\it Compactness criterion} - `A rich cluster must be sufficiently compact
that its 50 or more members are within a given (projected) radial distance
$R$ of its centre'.

Abell's choice $R=1.5h^{-1}$ Mpc is usually referred to as the `Abell
radius'.  The richness and compactness criteria jointly set the
length- and mass-scale of Abell clusters. We now know that the Abell
radius is somewhat small compared to more physical scale-lengths like
the turnaround radius, so the Abell radius defines just the central
region of a cluster.  Most observations are restricted to an area
within this radius, or to an even smaller region (the cluster
catalogue extracted from e.g.\ the APM survey uses half the Abell
radius, Dalton et al.\ 1992).

Instead of a richness class we will use the richness measure $C_{\rm ACO}$,
which actually gives the number of galaxies satisfying both of
Abell's criteria.

\subsection{Clusters as peaks in the smoothed density field}

\tx Theoretically, clusters are identified in the density contrast field 
$$\delta({\bf r})\equiv{\rho({\bf r})-\langle\rho({\bf r})\rangle \over
	\langle\rho({\bf r})\rangle} \eqno(\neweq)$$
as fluctuations at a certain mass scale (e.g.\ Kolb \& Turner 1990).
This requires smoothing, or `windowing', of the field at a corresponding
length-scale. The smoothed density contrast field is
$$\delta_W({\bf r},R_W)=4\pi\int\delta(|{\bf r}-{\bf r'}|)W(r',R_W)r'^2
  {\rm d}r'\ , \eqno(\neweq)$$
where $W(r,R_W)$ is a normalised spherically symmetric window function.
Clusters are identified as the peaks in this smoothed density field,
and the properties of these peaks characterise and define them. However,
galaxy clusters exhibit highly non-linear evolution.
This means that the present-day properties can not be linked directly to
the initial conditions from which they formed: we cannot simulate backwards
in time. We are forced to make an educated guess for the initial conditions,
simulate the formation of clusters starting from these conditions,
and compare the resulting models to observations. This means that model
clusters are necessarily defined in the initial density field, preferably
at the epoch before clusters `break away' or `turn around' from the general expansion of the
universe.

To generate a catalogue of cluster models, we must define models by
peak parameters in the initial density field that can be sampled from a
known probability distribution. If we choose peak parameters that are
{\it linear}\ functionals of Gaussian random fields filtered with a
Gaussian filter, we can use the results of BBKS. We therefore choose
the Gaussian filter
\newcount\Wg
\Wg=\eqnumber
$$W_{\rm G}(r,R_{\rm G})=(2\pi R_{\rm G}^2)^{-{3\over 2}}
	e^{-r^2/2R_{\rm G}^2}\ ,\eqno(\neweq)$$ 
and make the implicit assumption that structure forms from small
Gaussian density fluctuations in the early universe. Conveniently,
integrals with a Gaussian kernel behave well and can often be found
analytically.

Note that the probability distributions derived by BBKS are only valid
for the initial, linear regime (i.e.\ $\delta_W\ll 1$). Non-linear evolution
changes these distributions in an unknown manner, which makes the relation
between the initial peak parameters and the final cluster model rather
uncertain. Fortunately, the smoothing scale needed for cluster sized peaks
is large enough that the evolution of clusters in the smoothed density
field is not too non-linear, so that the BBKS distributions can be used.

\subsection{Initial peak parameters}

\subsubsection{Definitions}

\tx It is convenient to express peak parameters in terms of global
parameters for the cosmological scenario in which a cluster is to be modelled.
The general density fluctuation field smoothed with a Guassian with length-scale
$R_{\rm G}$ can be characterised by the spectral moments
\newcount\wmom
\wmom=\eqnumber
$$\sigma_j^2(R_{\rm G})\equiv\int {k^2\over 2\pi^2}|\delta_k|^2
    e^{-R_{\rm G}^2k^2}k^{2j} {\rm d}k\ . \eqno(\neweq)$$
where $\delta_k$ is the Fourier transform of the density field.
Two special spectral parameters are usually defined:
\newcount\gR
\gR=\eqnumber
$$\gamma(R_{\rm G})\equiv{\sigma_1^2(R_{\rm G})\over\sigma_0(R_{\rm G})
  \sigma_2(R_{\rm G})}\ \ {\rm and}\ \ R_*(R_{\rm G})\equiv\sqrt{3}
  {\sigma_1(R_{\rm G})\over\sigma_2(R_{\rm G})}\ .\eqno(\neweq)$$
They relate to the width of the spectrum and its correlation length
respectively. For this paper we adopt the standard adiabatic CDM
scenario for $\Omega_0=1$ with $R_G=4h^{-1}$Mpc, which has
$\gamma(4h^{-1})=0.729$ and $R_*(4h^{-1})=5.16h^{-1}$Mpc.

We define the origin of the coordinate system at the centre of the peak,
and define all peak parameters at this position.
The primary parameter of a peak is its amplitude $\delta_{\rm G}(0,R_{\rm G})$,
expressed in units of the r.m.s.\ fluctuation at the same scale:
$$\delta_{\rm G}(0,R_{\rm G})\equiv\nu\sigma_0(R_{\rm G})\ . \eqno(\neweq)$$
We can further specify the peak using the first and second spatial
derivatives of $\delta_{\rm G}({\bf r},R_{\rm G})$. The gradient
$\nabla\delta_{\rm G}(0,R_{\rm G})$ is zero for an extremum.
This accounts for three parameters. The second derivative tensor
$\nabla_i\nabla_j\delta_{\rm G}$ has six independent elements, resulting in six
additional parameters. They characterise the shape (two parameters), the
orientation (three parameters) and the size (one parameter) of the peak.
The latter is the trace of $\nabla_i\nabla_j\delta_{\rm G}$, and is usually
defined in a dimensionless form as the {\it curvature}:
\newcount\curvature
\curvature=\eqnumber
$$x({\bf r},R_{\rm G})\equiv -{\nabla^2 \delta_{\rm G}({\bf r},R_{\rm G}) \over
  \sigma_2(R_{\rm G})}\ . \eqno(\neweq)$$
This curvature (a measure for
the width of the peak) is positive for a peak, and negative for a dip. The shape of the
peak can be characterised by the two axial ratios of the isodensity contours
around the peak, as found from the small $r$ approximation of the smoothed
density field around the peak using a second order Taylor expansion in
spherical coordinates $(r,\theta,\phi)$ :
\newcount\Taylorexp
\Taylorexp=\eqnumber
$$\eqalign{\delta_{\rm G}(r)&=\delta_{\rm G}(0)+
	{r^2\over 2}\nabla^2\delta_{\rm G}(0)
	{\cal F}_5(a_{12},a_{13},\theta,\phi) \cr 
   =&\,\nu\sigma_0(R_{\rm G})-{r^2\over 2}x(0,R_{\rm G})\sigma_2(R_{\rm G})
	{\cal F}_5(a_{12},a_{13},\theta,\phi)\cr} \eqno(\neweq) $$
(BBKS). Here ${\cal F}_5(a_{12},a_{13},\theta,\phi)$, given by equation
($A7$) in Appendix A, describes the asphericity, and
$$a_{12}\equiv{a\over b}\ ,\ \ a_{13}\equiv{a\over c}\ \eqno(\neweq)$$
are the axial ratios, where $a$ is the major axis, $b$ the intermediate axis,
and $c$ the minor axis.

\subsubsection{Probability distributions}

\tx BBKS give the probability distributions for the peak parameters
(the amplitude $\nu$, the curvature $x$, and the shape parameters
$a_{12}$ and $a_{13}$) in the linear field smoothed with a Gaussian window
of size $R_{\rm G}$.

In order to make dependencies transparent, we use functions ${\cal F}$
(adapted from BBKS) which are listed for reference in Appendix A.
The differential peak amplitude probability density is given in the form of
the expected number of peaks within a volume $V$ with an amplitude in the range
$[\nu, \nu+d\nu]$:
\newcount\Npeaks
\Npeaks=\eqnumber
$$N_{\rm peak}(\nu,R_{\rm G}){\rm d}\nu = V {e^{-\nu^2/2}\over (2\pi)^2 R_*^3} 
	{\cal F}_1(\gamma, \nu){\rm d}\nu\ . \eqno(\neweq)$$
The other probabilities are conditional; given a peak with height $\nu$, the
probability distribution for its curvature to be equal to $x$  is
\newcount\Psize
\Psize=\eqnumber
$$P(x|\nu,R_{\rm G})=
  {e^{-(x-\gamma\nu)^2\over2(1-\gamma^2)}\over\sqrt{2\pi(1-\gamma^2)}}
	{{\cal F}_2(x)\over{\cal F}_1(\gamma, \nu)}\ . \eqno(\neweq)$$
For a peak with curvature $x$, the axial ratios $a_{12}$ and
$a_{13}$ have a joint probability distribution 
\newcount\Pshape
\Pshape=\eqnumber
$$P(a_{12},a_{13}|x) = {225\over\sqrt\pi} {{\cal F}_3(a_{12},a_{13}) x^8
   \over{\cal F}_2(x)}e^{-{5\over 2}x^2/{\cal F}_4(a_{12},a_{13})}\ .
   \eqno(\neweq)$$ 
This distribution does not depend on $\nu$ and $R_{\rm G}$ directly, but only
through $x$. All probabilities do depend on the spectral parameters
$\gamma(R_{\rm G})$ and $R_*(R_{\rm G})$.

\subsection{Building individual cluster models}

\subsubsection{Constrained initial conditions}

\tx It is straightforward to generate initial configurations for an average
patch of universe in a given cosmological scenario (e.g.\ Efstathiou et al.\
1985). However, for our purpose we need initial conditions
that will produce a cluster defined by certain peak parameters.
This is made possible by the method of constrained random
fields, pioneered by Bertschinger (1987). We use a code written by van de
Weygaert \& Bertschinger (1996), based on the Hoffman \& Ribak (1991) method,
which is very well suited for our purpose. It can sample an
unsmoothed density field constrained to form a peak characterised by linear
functionals of the Gaussian smoothed density field. Therefore the only
restriction of this scheme is that we must use {\it linear}\ peak parameters
to define clusters in the initial conditions.
This is an important limitation on the choice of a quantity that defines
a model cluster catalogue (Section 3.2).

\subsubsection{Non-linear evolution and galaxy formation}

\tx The highly non-linear evolution of clusters of galaxies requires the use
of N-body methods. However, N-body simulations suffer from a problem that is
particularly severe for galaxy clusters: small groups of particles that
represent galaxies in such simulations get disrupted by numerical and tidal
effects (Carlberg 1994; van Kampen 1995). This problem can be solved by replacing
those groups of particles by a single 'galaxy particle' just after they have
formed into a virialised system that resembles a galactic halo.
This new galaxy particle is softened according to the size of the group it
replaces, and gets the position and momentum of the centre-of-mass of the
original group. In van Kampen (1995) such galaxy particles were only formed
at a single epoch. Here an extension of this scheme to `continuous' galaxy
formation, which consists simply of applying the algorithm several times
during the evolution, is used (van Kampen 1996). Since groups of particles
can contain already formed galaxies, merging is included as well,
although somewhat crudely.

The actual N-body code we use is the Barnes \& Hut (1986) treecode, slightly
adapted for the modelling of clusters (see van Kampen 1995 for details)
and supplemented with the galaxy formation algorithm.
For all simulations we generated initial conditions in a 64$^3$ cube with
a size of 32 $h^{-1}$Mpc, using an implementation of the Hoffman \& Ribak
(1991) method by van de Weygaert \& Bertschinger (1996).
Because we use non-periodic boundary conditions (the universe outside the
simulation volume is homogeneous), we cut a sphere out of the volume
containing $64^3$ particles. The radius of this sphere is chosen such that
it is sufficiently large to easily include the final turnaround radius,
thus containing the full cluster and its region of influence (see Appendix
C for a discussion on the reliability of the simulations).

\subsubsection{Model cluster richness} 

\tx Abell's richness measure $C_{\rm ACO}$ is defined in a magnitude
interval. This means that we should correctly model the {\it shape and
the amplitude} of the luminosity function in order to obtain a
sensible richness measure from our numerical models.  Frenk et al.\
(1988) found for CDM models that the shape of their mass function
differs from that of the observed luminosity function.  However, such
results strongly depend on how one identifies galactic haloes at the
present epoch (Suginohara \& Suto 1992). The influence of the details
of the modelling of galaxy formation is also present in our
simulations. We expect to get a more realistic luminosity function
since we identify and form galaxies {\it during} the simulation,
rather than only at the end. If we assume a constant mass-to-light
ratio, which should be reasonable for the factor of 30 in mass range,
we find that the resulting luminosity function fits the observed one
quite well. This means that we can obtain a realistic richness measure
for our cluster models.

The assumed constant $M/L$ implies that the mass
interval corresponding to Abell's richness criterion ranges from the
mass of the third most massive galaxy to a mass 6.31 times smaller.
There is one property of our galaxies that plays a r\^ole here:
the galaxy particle masses are multiples of the N-body particle mass,
which is $7\times10^{10}$ M$_\odot$ for our simulations.
Before we determine the richness, we add or subtract a random fraction
of half the N-body particle mass. This removes the artifical
discreteness sufficiently well.

Note that we do not have sufficient foreground and background galaxies
in our simulations to mimic the ACO richness measure. For this reason we
will use a different richness measure, which is statistically
equivalent to Abell's (see Section 5.2).

%% file: cat3.tex
\section{Definition of a model catalogue}

\subsection{From initial density peaks to rich Abell clusters}

\tx In order to build a catalogue of individual cluster models, we
need to assume an one-to-one mapping of the peak parameters $\nu$ and
$x$ in the initial smoothed density to properties of clusters in the
final unsmoothed density field. However, this mapping might be
complicated by initial peaks merging into a single final cluster, a
worry that comes in two flavours.

\subsubsection{Mergers of rich cluster peaks}

\tx In a hierarchical structure formation scenario, like CDM,
merging of clumps into larger clumps is an ongoing process. Because we
assume a one-to-one mapping of peaks in the initial density field to
overdensities in the final density distribution, we need to check if
merging is still important on cluster scales. If evolution would
remain linear, peaks that were separate in the initial smoothed
density field would remain separate for the same comoving smoothing
scale. They can only merge due to non-linear evolution of the density
field. We estimate the fraction of clusters that can merge from the
statistics of the smoothed density field, i.e.\ the correlation
function of Abell cluster peaks, and their peculiar velocities.  The
average peculiar velocity for a cluster-sized peak is about 300 km
s$^{-1}$, with a maximum of about 600 km s$^{-1}$, for a $\Omega_0=1$
CDM cosmology with $\sigma_8=0.6$ (de Theije, van Kampen and Slijkhuis, 1996).
In this scenario, a typical cluster is therefore expected to travel about
2$h^{-1}$ Mpc up to the present epoch.

\newcount\annd
\annd=\eqnumber
The average nearest-neighbour distance $r_{\rm nn}$ is given by
$$ r_{\rm nn}^3 = {3\over 4\pi} <n_{\rm c}>^{-1} -
  3 \int_0^{r_{\rm nn}}\xi_{\rm cc}(r)r^2{\rm d}r\ ,\ \eqno(\neweq)$$
where $n_{\rm c}$ is the cluster number density, and $\xi_{\rm cc}(r)$ the
cluster-cluster correlation function, which is observed to be a power-law,
$$ \xi_{\rm cc} = (r/r_{\rm cc})^{-1.8},\ \eqno(\neweq)$$
where the cluster-cluster correlation length $r_{\rm cc}$ is around
18$h^{-1}$ Mpc (eg.\ Peebles 1993). This means that for a mean cluster
density of $8.6\times10^{-6} h^3$Mpc$^{-3}$ (Mazure et al.\ 1996) the
average nearest-neighbour distance is 20$h^{-1}$.
Therefore one does not expect two Abell sized peaks to merge within a
Hubble time. This is confirmed from an estimate of the probability of
finding, at the present epoch, a cluster-cluster separation of less
than 2$h^{-1}$Mpc (the typical distance travelled
up to the present epoch), using the observed $\xi_{\rm cc}$.
We derive a probability of 2 per cent, i.e.\
about 2 clusters in the catalogue. This means that we expect one
close pair that could merge if moving towards each other.
Note that we derived this probability for the {\it present epoch}, and
that it must have been smaller at earlier times.

\subsubsection{Late cluster formation from subcluster mergers}

\tx A cluster can also form from two or more subclusters, which are more
abundant than rich clusters themselves, through a merger at late times.
One needs somewhat special initial conditions for clusters to form
this way, because the separation of such subclusters in the initial
conditions cannot be too large or they would remain distinct, despite
non-linear clustering.
It is more likely that late merging involves unequal mass
clumps, where the main clump would not have been massive enough to become
a rich cluster were it not for the `secondary infall' of nearby
subclumps. 

Such cases of cluster formation are unavoidable in hierarchical
formation scenarios. The direct way to access their significance is to
look at a large-scale simulation of volume equal to that of the whole
catalogue (see eg.\ van Haarlem, Frenk \& White 1996). We will discuss
this question in Section 5.

\subsection{The catalogue defining quantity}

\subsubsection{Total cluster mass}

\tx As a single cluster model is defined by its local initial conditions,
a model catalogue of clusters is defined by the statistics of peaks in the
initial cosmological density field. We will construct a model catalogue
under the assumption that a certain subset of peaks in the initial smoothed
density field will each produce a cluster at the present epoch.
In order to build a catalogue that mimics a sample of rich Abell clusters,
we need to establish a relation between peak characteristics in the initial
density field and the criteria used by Abell at the present epoch.
Unfortunately, his criteria are not easily translated into the formalism of
Gaussian random fields. Therefore we are left with a choice for a sensible
physical or observable quantity that does provide a relation between initial
and final conditions.

Physical parameters are very easily obtained for model clusters since
the data are clean and unprojected. Far less choice is available in the
observations, which provide projected, biased and generally less clean data.
We will focus mainly on galaxies, not on dark matter, although it is probably
the dominant mass component. The dark matter distribution can only be
obtained through indirect measurements like the X-ray brightness
distribution (e.g.\ Sarazin 1986; Forman \& Jones 1990) or the effects
of gravitational lensing (e.g.\ Fort \& Mellier 1994).

Which physically significant quantities can we derive from the galaxy
distribution?  The quantities that are best observable are the two
spatial coordinates on the sky, which can be obtained very
accurately. Nevertheless, the position of the {\it centre} of the
projected galaxy distribution, required for most of the derived
properties, is relatively difficult to obtain.  Global properties
derived from the projected distribution are the total number of
galaxies within a certain radius, which was used by Abell for his
richness criterion, the shape (ellipticity) and orientation of the
distribution (e.g.\ Binggeli 1982; Rhee, van Haarlem \& Katgert 1989;
de Theije, Katgert \& van Kampen 1995), measures of substructure
(reviewed by Beers 1992), the azimuthally averaged surface density
profile (e.g.\ Beers \& Tonry 1986; Rhee et al., ibid), and the
projected mean harmonic radius, which is often used for mass estimates.

With regard to velocity information, only line-of-sight components are
readily observable. Apart from the distance to the cluster, the most
useful global quantity derived from these is the central line-of-sight
velocity dispersion. This would be a good candidate for the catalogue
defining quantity, except that it is a non-linear quantity, and can
therefore not be used in the the Hoffman-Ribak method of constrained
random fields. The total velocity dispersion is correlated with the
total mass of the cluster, which is also one of its most basic
physical properties. We demonstrate below that the final cluster mass
can be estimated fairly reliably from the initial peak parameters, and
we therefore choose {\it the expected final mass}\ to be the defining
quantity of a catalogue of cluster models.

Although in principle mass estimates can be obtained from the
combination of the projected galaxy distribution and the central
line-of-sight velocity dispersion, reliable total masses are not known
for most Abell clusters.  All we can do is to hope that richness
correlates sufficiently well with mass, and that those peaks in the
Gaussian smoothed initial density field that have the largest expected
final mass are the progenitors of the rich Abell clusters.

\subsubsection{Smoothing scales}

\tx Before we discuss the relation between the final mass of a cluster
and the parameters of the initial density peak, we need
to choose the smoothing scale that is appropriate for clusters. We base our
choice on an argument involving the cluster mass. In general, the total mass
for a given spherical window function $W(r,R_W)$ is given by
$$M_W = V_W \bar\rho [1+\delta_W(0,R_W)]\ , \eqno(\neweq)$$
where the volume $V_W(R_W)$ of the window function is
$$V_W(R_W) = 4 \pi \int W(r,R_W) r^2 {\rm d}r\ .\eqno(\neweq)$$
The (observational) compactness criterion implies that Abell clusters are
defined using a Top-Hat window
\newcount\TopHat
\TopHat=\eqnumber
$$W_{\rm TH}(r,R_{\rm TH})=\cases { 1 & for $r\le R_{\rm TH}$\cr
                    	    	    0 & for $r >  R_{\rm TH}$\ ,\cr}\eqno(\neweq) $$
with $R_{\rm TH}=1.5 h^{-1}$Mpc, the Abell radius. The mass
enclosed by a sphere this size is about $4\times10^{12}h^{-1}$M$_\odot$
for the background universe (for $\Omega_0=1$).
A `typical' cluster has a mass of about $3\times10^{14}\Omega_0h^{-1}$M$_\odot$
(e.g.\ Peebles 1993), so the cluster mass originates from a volume 70 times
larger. This corresponds to a Top-Hat filter of about $6h^{-1}$Mpc in the
initial density field. The corresponding Gaussian smoothing radius $R_{\rm G}$
is found by setting the enclosed volumes equal for both windows (BBKS).
This gives $R_{\rm G}\approx4h^{-1}$Mpc, as the ratio between the two radii is
$(4/18\pi)^{1/6}\approx2/3$.

\newcount\Fcumuden
\Fcumuden=1
\figps{1}{S}{\psfig{file=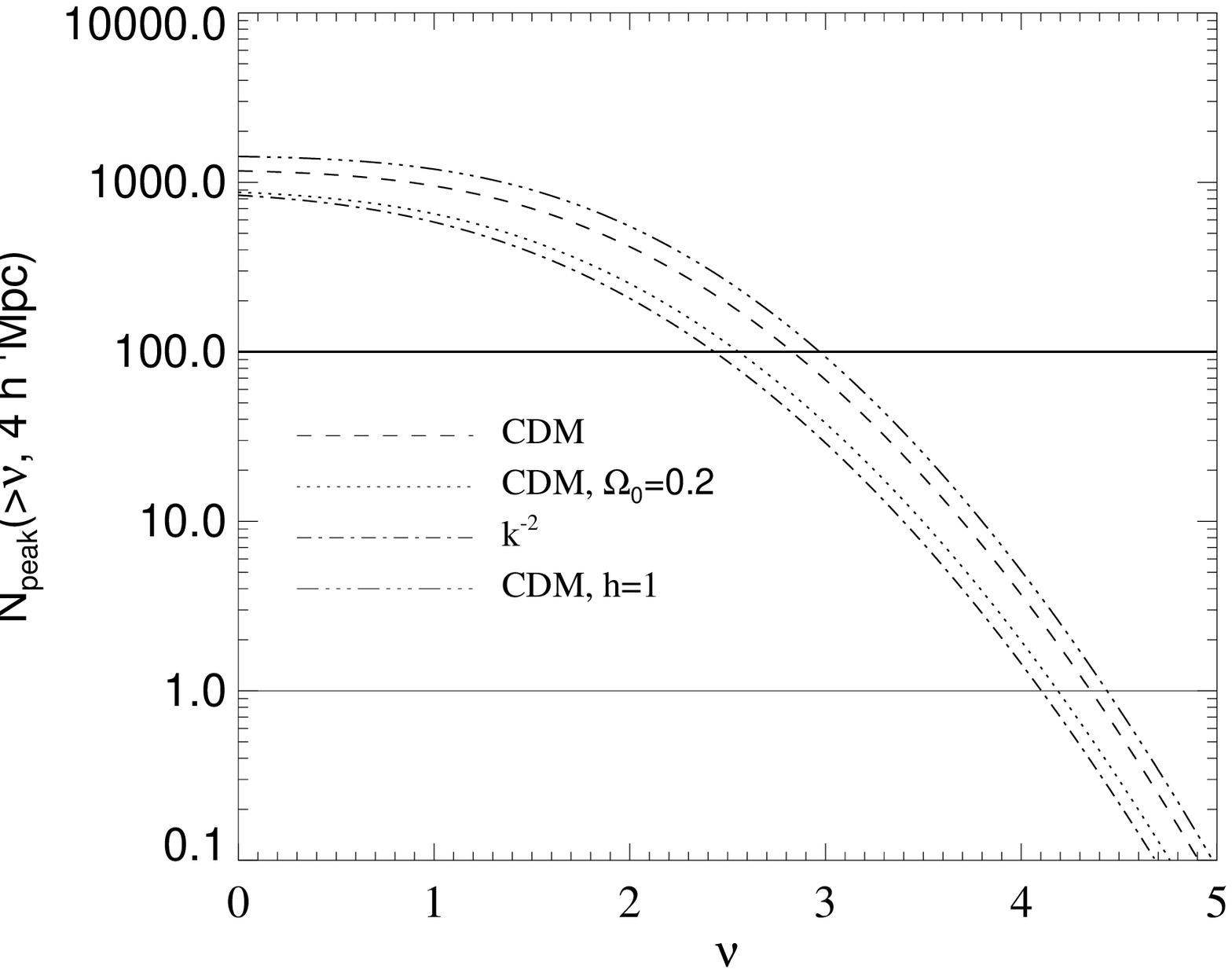,width=8.5cm,silent=}}
{{\bf Figure 1.} Expected range in $\nu$ for 100 clusters in a volume of
$10^7 h^{-3}{\rm Mpc}^3$. All scenarios are for $\Omega_0=1$ and
$h=\halve$ unless stated otherwise.}

\subsection{Cluster mass threshold}

\tx In this section we estimate the total mass of a cluster given the
properties of the peak in the initial smoothed density field it
originates from. 

\subsubsection{Peak amplitude threshold: `linear' mass}

\tx In the linear approximation the final total mass is
proportional to the peak amplitude in the initial smoothed density
field. We define model clusters using the Gaussian window function with
radius $R_{\rm G}=4h^{-1}$Mpc. If we take the same value for the present
density field as for the initial field, and assume the field to evolve
linearly, the total mass is determined exactly by the peak amplitude $\nu$
alone and given by 
\newcount\Mgauss
\Mgauss=\eqnumber
$$M_{\rm G}(\nu)=(2\pi R_{\rm G}^2)^{3/2}\bar\rho[1+\nu\sigma_0]\ , \eqno(\neweq)$$
because the relation $\delta_{\rm G}=\nu\sigma_0$ remains valid throughout
the evolution.
In that case the linear relation $\sigma_0(t)=a\sigma_0(t_0)$ describes the
time dependence, where $t_0$ is the present epoch and $a$ is the cosmological
expansion factor (with $a(t_0)\equiv 1$), and $\nu$ remains constant.

We could then define a model catalogue by setting a threshold on $\nu$, which
is that value of $\nu$ for which $N_{\rm peak}(>\nu,R_{\rm G})={\cal N}$,
where ${\cal N}$ is again the total number of clusters in the catalogue
with volume $V$.
The threshold $\nu_{\rm min}$, which is formally an expectation value,
should then correspond to the threshold in the defining quantity of the
observational catalogue. In the case of rich Abell cluster catalogues this
is the value of 50 for the Abell richness measure.

We can calculate the expected maximum peak amplitude
$\nu_{\rm max}$ in that volume from $N_{\rm peak}(>\nu,R_{\rm G})=1$.
We illustrate this in Fig.\ \the\Fcumuden\ for several cosmological
parameters and fluctuation spectra, smoothed on the cluster scale of
$4h^{-1}$Mpc. The heavy horizontal line determines the threshold
$\nu_{\rm min}$ for the choices ${\cal N}=100$ and $V=10^7 h^{-3}{\rm Mpc}^3$,
which are typcial values for a cluster catalogue. The thin line indicates
the expected $\nu_{\rm max}$, which is supposed to represent the largest
cluster in $V$. Of course, larger peak amplitudes will occur in larger
volumes.

Note that it is not sufficient to just threshold the 
`linear' mass in the initial smoothed density field in order to guarantee
the presence of a rich Abell cluster in the final density field.
Because richness and mass are not likely to be perfectly correlated,
there are peaks with a small `linear' mass which produce rich
clusters (and vice versa), making the catalogue selected on mass
incomplete in richness.
Since we cannot use the constrained random
field method to constrain Abell richness, we have to tighten the
correlation of initial peak parameters and final richness as much as
possible. One way of doing that is to go from `linear' mass, as given
by the peak parameter $\nu$, to a `non-linear' mass.

Several non-linear effects come into play which determine the final
mass of a galaxy cluster. The collapse time of a cluster
not only depends on the peak amplitude of the initial overdensity,
but also on its curvature $x$ (van Haarlem \& van de Weygaert 1993),
on tidal effects from the environment, on the presence of substructure
(Cavaliere et al.\ 1986) which can result in pre-virialisation (Peebles 1990)
and thus a slower collapse, and on the amount of shear in the initial
velocity field (Bertschinger \& Jain 1994). 
If the non-linear evolution of these properties is significant, the peak in
the final density field may not coincide with the peak in the initial
density field, not even for the field smoothed on the scale of 4$h^{-1}$Mpc.

We show below that the effect of the peak parameter $x$, which is
related to the slope of the initial density profile, can be taken into
account, and can be used for a better approximation of the expected
final cluster mass.

\subsubsection{Expected `non-linear' cluster mass}

\tx Cluster peaks in the density field smoothed at
4$h^{-1}$ become non-linear. We find that $\delta_{\rm G}$ can
grow up to about 10 for the structure formation scenario that we have adopted,
viz.\ standard CDM for $\Omega_0=1$. Therefore $\nu$ becomes a function of
time, and $\sigma_0(t)$, the {\it r.m.s.}\ fluctuation on scale $R_{\rm G}$,
will become quasi-nonlinear. Note that $\nu(t)$ is a {\it locally}\
evolving quantity whereas $\sigma_0(t)$ evolves {\it globally}, so the
latter will grow less rapidly than the former which describes the
evolution of an individual peak.
Furthermore, the effect of the initial curvature of the peak, as set
by $x$, will become important. Van Haarlem \& van de Weygaert (1993)
found that the final cluster mass is larger for smaller $x$, i.e.\ more
extended initial configurations.
We can try to incorporate these effects by looking at the possible evolution
of the initial smoothed radial density constrast profile around a peak
with given $\nu$ and $x$. The initial density profile is given
by BBKS:
\newcount\BBKSprofile
\BBKSprofile=\eqnumber
$$\delta_{\rm G}(r)={\nu-\gamma x\over\sigma_0(1-\gamma^2)}\xi_{\rm G}(r)
   +{\nu-x/\gamma\over3\sigma_0(1-\gamma^2)}\nabla^2\xi_{\rm G}(r)\ , \eqno(\neweq)$$
where $\xi_{\rm G}(r)$ is the smoothed density autocorrelation function. Remember
that all quantities depend on the smoothing radius $R_{\rm G}$. In the linear
regime $\delta_{\rm G}(0)=\nu\sigma_0$, and $\xi_{\rm G}(0)=\sigma_0^2$ by
definition, so we can derive that initially
\newcount\initnabla
\initnabla=\eqnumber
$$\nabla^2\xi_{\rm G}(0)=-3\gamma^2\sigma_0^2\ . \eqno(\neweq)$$
For small $r$ both terms in eq.\ (\the\BBKSprofile) initially contribute
roughly equally to the density profile for most cosmological scenarios.
For large $r$ the $\nabla^2\xi_{\rm G}(r)$ term quickly vanishes and the
profile is well approximated by just the first term. For small $r$ the profile
is well approximated by a second order Taylor expansion around the peak,
as in equation (\the\Taylorexp). Here we will just consider the spherically
averaged profile, which means that the contribution from ${\cal F}_5$ cancels.

We now substitute non-linear quantities in eq.\ (\the\BBKSprofile) in
order to get an evolution equation for the (smoothed) density profile.
If the evolution would remain linear, $\nu$, $x$ and $\gamma$ would remain
constant, and $\sigma_0$ would evolve linearly with the cosmological expansion
factor $a$. The autocorrelation function scales with $\sigma_0^2$, and so do
its derivatives because the slope of $ \xi_{\rm G}(r,t)$ remains constant,
so that the $x(t)$-terms in eq.\ (\the\BBKSprofile) cancel.
How does non-linear evolution change this behaviour ? The {\it r.m.s.}\
fluctuation $\sigma_0$ at 4$h^{-1}$Mpc Gaussian smoothing is about unity
at the present epoch for most cosmological scenarios, and its time 
evolution therefore becomes non-linear. Yet, $\nu$ and $x$ become 
non-linear more quickly because they are defined for {\it peaks}.
Now, if $\xi_{\rm G}(0,t)$ would grow at the same rate as
$\nabla^2\xi_{\rm G}(0,t)$, both $x(t)$-terms in eq.\ (\the\BBKSprofile)
still cancel and we would only see non-linear behaviour in 
$\delta_{\rm G}(0,t)$ due to $\nu(t)\sigma_0(t)$.

However, non-linear evolution causes the slope of $\xi_{\rm G}(t)$ to become
steeper, which changes the evolution of $\nabla^2\xi_{\rm G}(0,t)$ with
respect to $\xi_{\rm G}(0,t)$.
So the second term of eq.\ (\the\BBKSprofile) evolves somewhat faster
than the first because of the increasing slope of $\xi_{\rm G}(t)$.
This might seem unimportant, but the $x$-terms, which grow relatively fast,
do not cancel anymore.
This is demonstrated in more detail in Appendix B. If we use a first order
approximation for $\nu(t)$ and $x(t)$, we find a relation of the form
\newcount\Mtheory
\Mtheory=\eqnumber
$$M_{\rm G}(\nu_{\rm i},x_{\rm i},t_0) = (2\pi R_{\rm G}^2)^{3/2}
	\bar\rho\bigl[1 + (c_0 + c_1\nu_{\rm i} + c_2 x_{\rm i})
	\sigma_0\bigl]\ , \eqno(\neweq)$$
where the $c_j$ are constants at epoch $t_0$, and $\nu_{\rm i}$ and $x_{\rm i}$
are the initial peak parameters.
If the field and its derivatives would just evolve linearly, $c_0=0$,
$c_1=1$ and $c_2=0$. Non-linear evolution results in a non-vanishing $c_0$,
where the sign depends on the exact balance of the constants in the linear
approximations for $\nu(t)$ and $x(t)$, $c_1>1$ and $c_2<0$. Again, see
Appendix B for more details.

\subsection{Sampling constraints above the expected final mass threshold}

\tx We have defined a model catalogue of ${\cal N}$ clusters as follows.
We first draw a large number of $(\nu,x)$ sets for a
sufficiently small $\nu_{\rm min}$ found from (\the\Npeaks),
and then apply the $M(\nu,x)$ threshold so that ${\cal N}$ models remain.
However, the values for the parameters $c_j$ in (\the\Mtheory)
are unknown {\it a priori}\ because no model has actually been built yet.
So one first needs to choose a few representative models from the ${\cal N}$
models selected for small deviations of the linear values $c_0=0$, $c_1=1$
and $c_2=0$, for example $c_0=0$, $c_1=1.5$ and $c_2=-0.5$ (see Appendix B). 
After simulating these models, one can fit
the $c_j$ to the measured total mass in these simulations within a
comoving sphere of radius $6h^{-1}$Mpc around the centre.
This provides a new total mass threshold, which can then be used to adjust
the selection of the ${\cal N}$ models from the large ensemble.
If the initial sets of constraints are chosen sensibly,
along the lines described in Appendix B, and not too close to the
initial threshold, most of the trial models will remain in the catalogue.
This iterative procedure is repeated until all ${\cal N}$ models are
run and the $c_j$ have converged to values that define the final catalogue.

Each model is defined not only by $\nu$ and $x$, but also by its shape
parameters. These do not form part of the predicted mass relation that defines
the catalogue, but are still chosen so that we can also study the dependence
of the final cluster models on initial shape. Given $\nu$ and $x$, the
parameters $a_{12}$ and $a_{13}$ are drawn simultaneously from the probability
distribution in eq.\ (\the\Pshape).
The orientation is drawn from an isotropic distribution, as the phases of the
Gaussian density field are random. The actual orientation is important only to
determine the projected shape of the object on the sky.

Deviates are generated using the rejection method (e.g.\ Press et al.\ 1988).
To draw $\nu$, a convenient comparison function is found by substituting
$\nu^3$ for ${\cal F}_1(\gamma, \nu)$ in (\the\Npeaks).
The comparison function needed for drawing $x$ is obtained by replacing
${\cal F}_2(x)$ in (\the\Psize) with $x^3$ (see also Appendix A).
For the joint sampling of the shape parameters the comparison function
is simply a constant that is larger than the maximum of (\the\Pshape).
Deviates of $a_{12}$ and $a_{13}$ are (uniformly) generated between
1 (by definition) and some maximum for which the probability is really small
already, for which we took 10. This may not be the most efficient algorithm,
but it works sufficiently well for our purposes given that (\the\Pshape) is
rather complicated.

%% file: cat4.tex
\section{Construction of a model catalogue}

\newcount\Fspectra
\Fspectra=2
\figps{2}{S}{\psfig{file=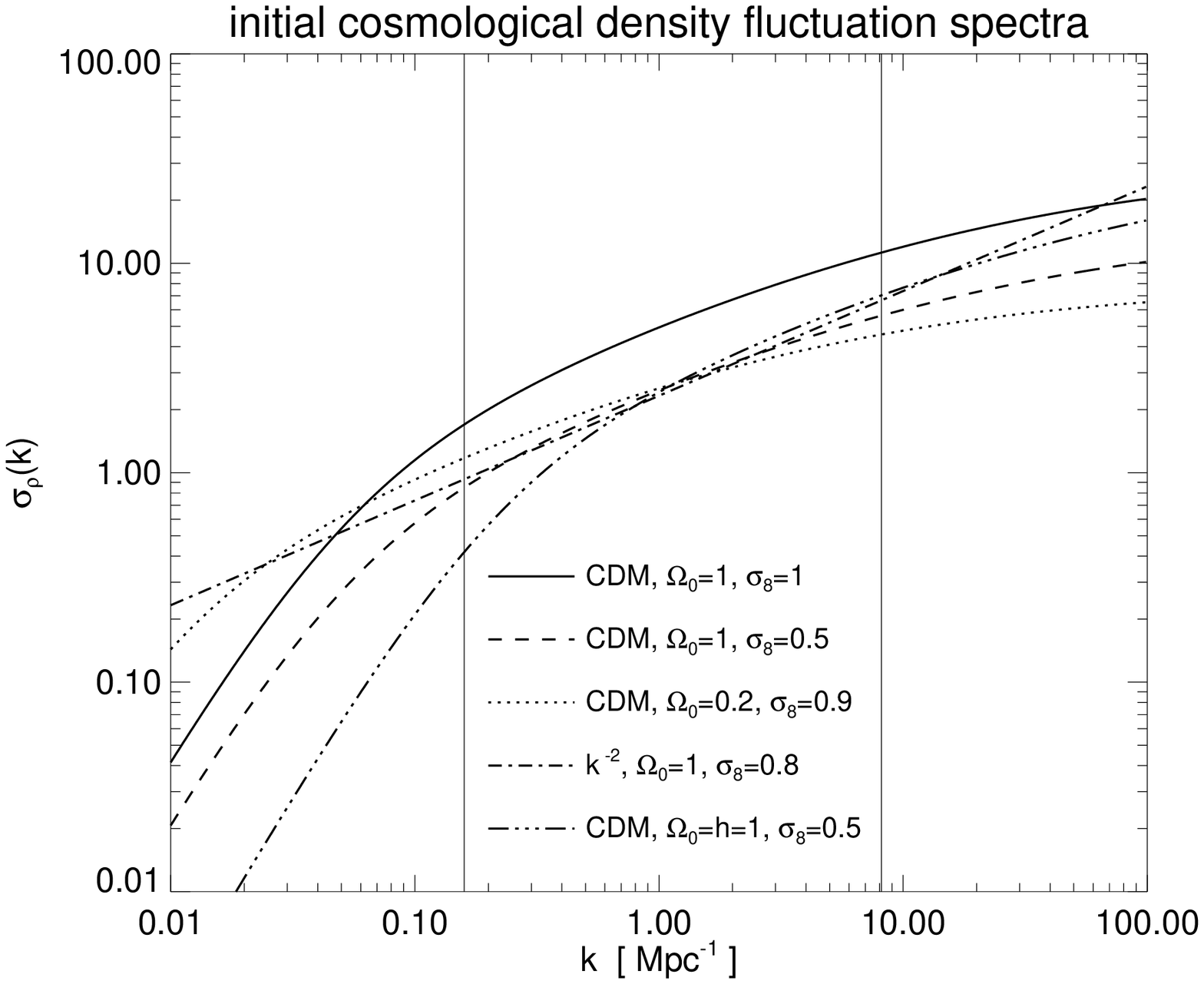,width=8.5cm,silent=}}
{{\bf Figure 2.} Selected csomological density fluctuation spectra,
plotted in the form of $\sigma_\rho(k)\equiv (4\pi k^3)^{1/2}|\delta_k|$.}

\subsection{Choice of the cosmological scenario}

\tx We build a catalogue of model clusters within the standard CDM
scenario, i.e.\ we adopt the $\Omega_0=1$ adiabatic CDM initial
fluctuation spectrum (Davis et al.\ 1985). This scenario has the
advantage that the amplitude of the initial density fluctuation spectrum,
denoted by $\sigma_8$, can be fixed {\it after}\ having run the models, by
comparing the properties of the galaxy distribution to observational data.
Fig.\ \the\Fspectra\ shows the unbiased standard $\Omega_0$=1 adiabatic
CDM spectrum as given by Davis et al.\ (1985) in the form of
$\sigma_\rho(k)\equiv (4\pi k^3)^{1/2} |\delta_k|$,
which is the r.m.s.\ contribution to the density fluctuation field.
If $\sigma_8$ is less than the observed value of unity, this decreases the 
amplitude of the spectrum by a factor of $\sigma_8$.
for all wavenumbers. The vertical lines indicate the range of wavenumbers
included in our simulations. The dashed line shows the $\Omega_0=1$ CDM
scenario, which we adopted here, for $\sigma_8=0.5$, a value that should
not be far off from the one that best fits the observational data.
We also show a few other scenarios, with values for $\sigma_8$ such
that the spectrum is not too different from the $\Omega_0$=1 CDM spectrum
within the range of wavenumbers present in the simulations.
Of course they will be significantly different at larger and/or smaller
wavenumbers: eg.\ the featureless power-law spectrum with index
$-2$ (dashed-dotted line) has substantially more power on both larger
and smaller scales.

Another important reason for choosing the $\Omega_0=1$ CDM scenario is that
it allows comparison to published results, notably those of
Frenk et al.\ (1990) and White et al.\ (1993).
Furthermore, CDM does not have many more problems than most of
the competing scenarios (e.g.\ Davis et al.\ 1993; Ostriker 1993;
Bertschinger 1993), and explains many observations, although not in its
original unbiased form.

\newcount\Fcatalog
\Fcatalog=3
\figps{3}{S}{\psfig{file=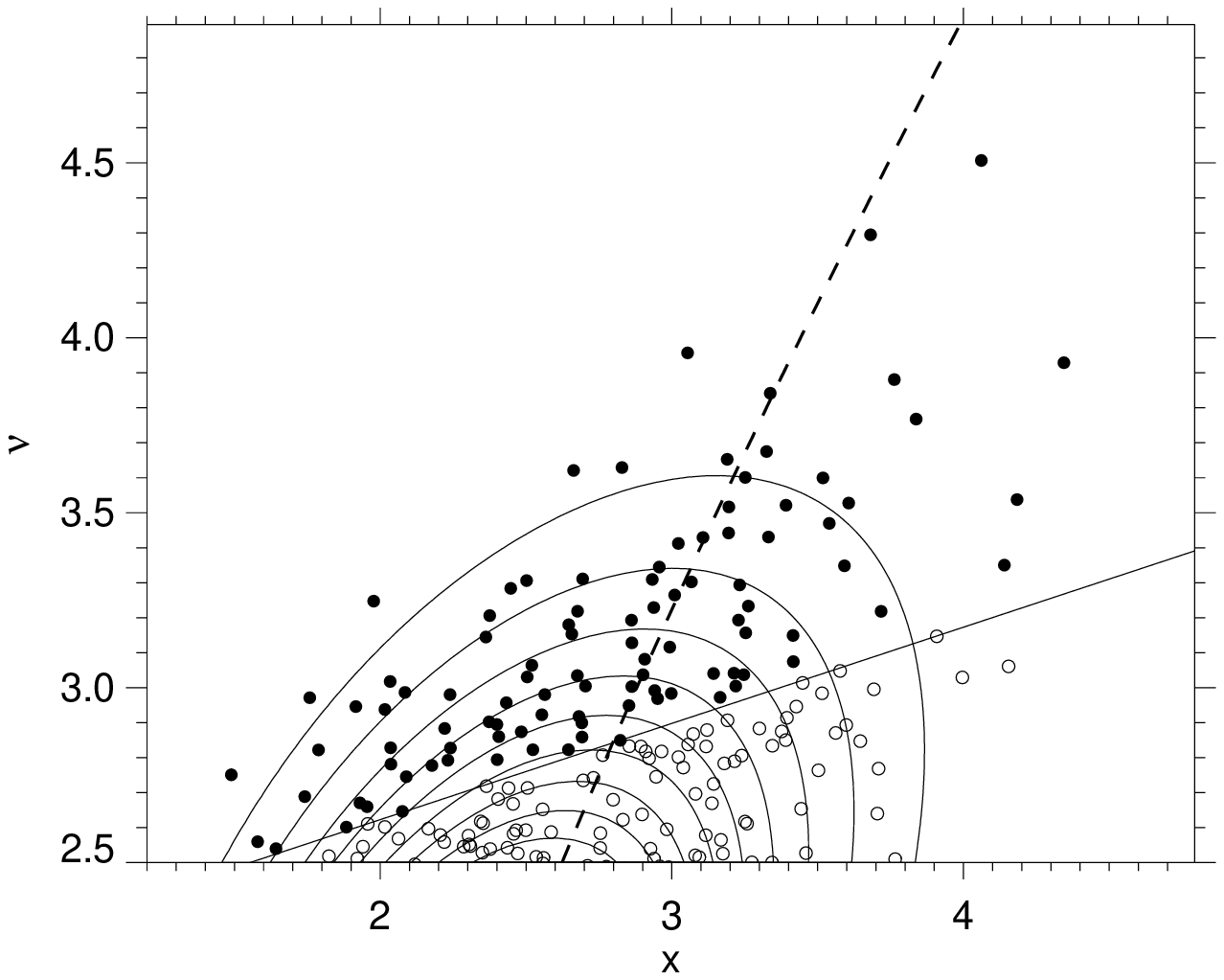,width=9.0cm,silent=}}
{{\bf Figure 3.} Selection of models on expected total mass within
4$h^{-1}$Mpc at $\sigma_8=0.46$ from 200 generated $(\nu,x)$ sets.
The thick solid line selects the 99 cluster models that have been simulated
(filled symbols) and form the model catalogue. The dashed line denotes
the expectation value of $x$ for a given $\nu$ (eq.\ $A10$).
The thin solid lines show probability density contours of $P(x|\nu)$
(see eq.\ (\the\Psize) in Section 2.3, and Section 4.4.3).}

\newcount\Fmasscomp
\Fmasscomp=4
\figps{4}{S}{\psfig{file=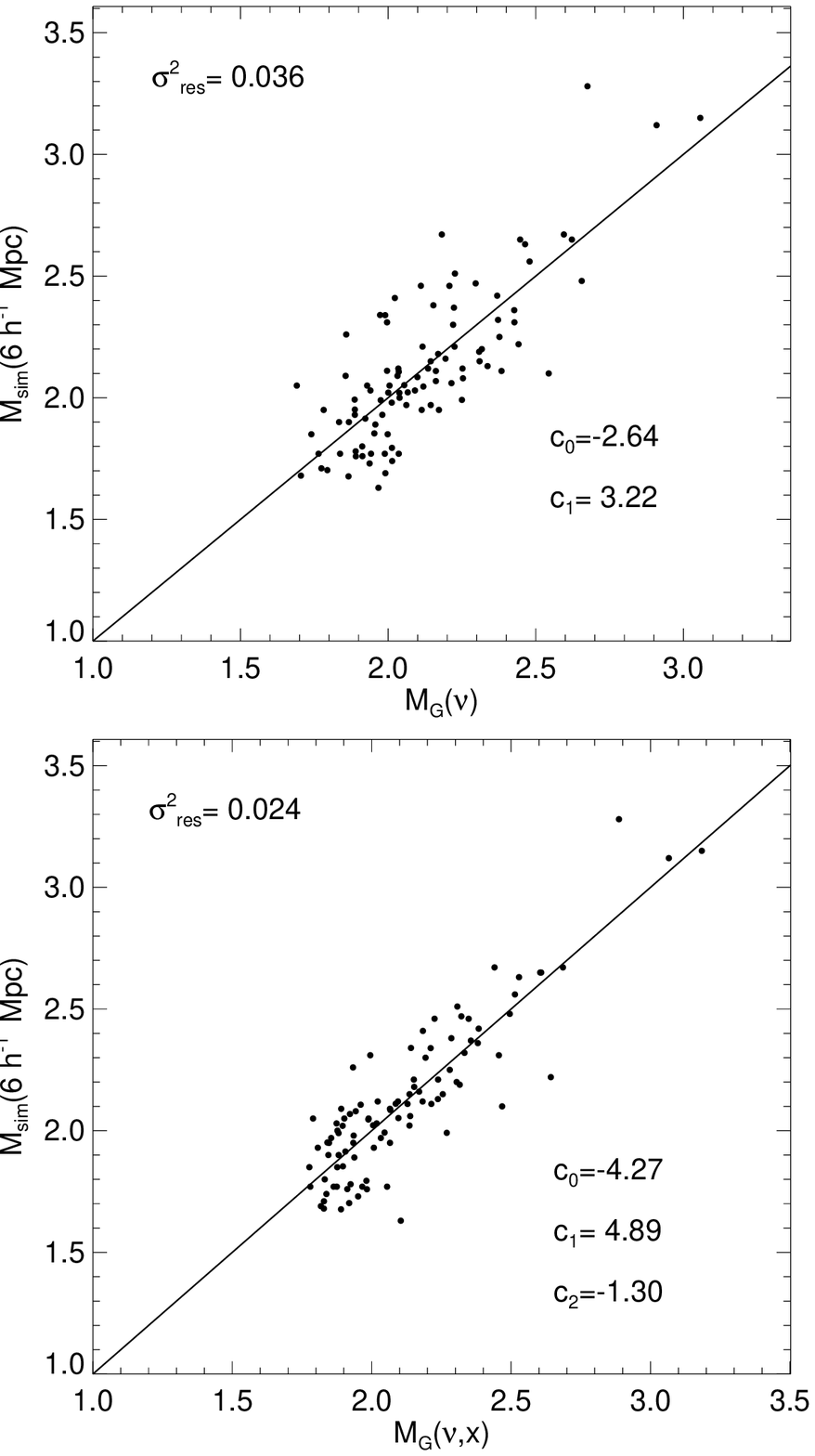,width=8.5cm,silent=}}
{{\bf Figure 4.} Fits for the expected masses $M_{\rm G}(\nu)$ (top panel)
and $M_{\rm G}(\nu,x)$ (bottom panel) to the total mass found within
6$h^{-1}$Mpc in the simulations for $\sigma_8=0.46$.}

\subsection{The target observational catalogue}

\tx As we want our sample of model clusters to be fair, i.e.\
statistically complete for a specific volume, its observational
counterpart also needs to be as complete as possible.
A sample that was constructed with this requirement in mind is one
for which data has been obtained in the context of an ESO Key-programme.
This sample of rich Abell clusters, the ENACS sample for short,
is based on a subsample of 128 $R\ge 1$ clusters selected from both
the Abell and ACO catalogues in a cone of about 2.6 steradian out to
$z=0.1$ (Katgert et al.\ 1996). It occupies a single
closed volume that is well away from the Zone of Avoidance.

We choose a $z<0.08$ subsample of the ENACS sample (which goes out to
$z=0.10$) as the observational catalogue to be mimicked, because it is
the most complete subsample selectable which is large enough for
statistical purposes (Mazure et al.\ 1996).  It contains 76 clusters
in a volume of $9.2\times 10^6 h^{-3}$Mpc$^3$.  After corrections for
completeness, Mazure et al.\ ({\it ibid.})  found that there should
have been 79 in that volume, implying a comoving number density of
$8.6\pm0.6\times 10^{-6} h^3$Mpc$^{-3}$ for $R\ge1$ Abell clusters.

\subsection{The $\Omega_0=1$ CDM model catalogue}

\tx We now define a catalogue of cluster models that mimics the ENACS sample.
For the volume of the catalogue we take $10^7 h^{-3}$Mpc$^3$. This means
that according to the number density found by Mazure et al.\ (1996) there
should be 86 rich Abell clusters in this volume, of which 83 define
an ACO-like observational sample.
To get close to the target of 83 cluster models, we built a simulation
set that consists of 99 cluster models.
This redundancy somewhat alleviates completeness problems associated
with thresholding on the predicted final mass, which arises from
scatter in the relation of richness versus final mass (Sections 5 and 6). 

We generated the simulation set for the CDM $\Omega_0=1$ scenario
using the technique described in Section 3.4.
We used 200 sets of $(\nu,x)$ from which the desired 99 will be selected.
The 200 sets are shown in Fig.\ \the\Fcatalog\ in the form of a
scatter plot of $\nu$ versus $x$. The solid line depicts
the fitted $M(\nu,x)$ threshold found after the last iteration that
separates the 200 sets in 99 candidate catalogue entries (filled circles),
and 101 models that are below the expected final mass threshold
(open circles).

The present epoch for the adopted scenario, as obtained from comparison
to the galaxy autocorrelation function, is most likely to be
$\sigma_8=0.4-0.5$ (Davis et al.\ 1985, van Kampen 1994).
Note that the {\it COBE}\ measurements require a rather different range
of values, as is discussed in Section 8. Our simulations have data saved
for $\sigma_8=\{0.25,0.31,0.36,0.41,0.46,0.51,0.56,0.61,...\}$,
so we took $\sigma_8=0.46$ for the purpose of fitting the $c_j$ in
eq.\ (\the\Mtheory).
In Fig.\ \the\Fmasscomp\ we show two fits: one for the linear mass
estimate based on a linear function of $\nu$ alone, and one for
$M_{\rm G}(\nu,x)$, given by (\the\Mtheory). The first clearly has a
larger spread than $M_{\rm G}(\nu,x)$; the variance of the residuals of
the fit, $\sigma^2_{\rm res}$, is twice as large.
We found $c_0=-4.3$, $c_1=4.9 \pm 0.3$ and $c_2=-1.3 \pm 0.2$, with no
correlation between the residual scatter in the $M_{\rm G}(\nu,x)$
relation and the shape parameters $a_{12}$ and $a_{13}$. We suspect that
the remaining scatter is mostly due to the non-linear evolution and
the difference between Gaussian and Top-Hat smoothing, as well as to
differences in the initial {\it velocity}\ field, notably
the shear (Bertschinger \& Jain 1994; van de Weygaert \& Babul 1994).
This latter effect will be investigated in another paper.

\subsection{Testing the construction technique}

\tx With the aid of a conventional large-scale, low-resolution
simulation we have tested the assumptions made in the method we use to
construct a catalogue of galaxy cluster models. 
What we test here is whether the final galaxy clusters (i.e.\ the
final peaks) indeed have a one-to-one correspondence to initial density
peaks, and how well the expected mass estimator $M_{\rm G}$ performs.

\subsubsection{Description of the large-scale simulation}

\tx We evolved a patch of universe in a cubic volume of (256$h^{-1}$Mpc)$^3$
using a P$^3$M code (Bertschinger \& Gelb 1991), and saved particle
positions and velocities at the same epochs as for the individual
cluster models. A more detailed description of this simulation can
be found in de Theije, van Kampen \& Slijkhuis (1996).
Given a cluster density for rich Abell clusters of $8.6\times 10^{-6}
h^3$Mpc$^{-3}$, we expect 144 of such clusters in the chosen volume.
We employed 128$^3$ particles, so that each particle
has a mass of $4.4\times 10^{12}$ M$_\odot$.
This provides sufficient resolution to identify rich clusters and trace
them back to the initial density field. Note that the resolution is
significantly lower than for the individual simulations, so that
properties obtained for the clusters in the single large-scale simulation 
will be much noisier than for their individually simulated counterparts.

After running the simulation, the 4$h^{-1}$Mpc Gaussian smoothed density
field and its second derivative where obtained on a 128$^3$ lattice for
each output time. Peaks were identified on these lattices by searching
for those cells which have a larger density than their 26 neighbouring cells.
We then use second order polynomial fitting to find the off-grid positions
of the peaks, and the actual peak parameters $\nu$ and $x$ at those positions.
At the final epoch these peaks correspond to N-body particle groups which
should be the final-epoch rich galaxy clusters. 

\subsubsection{Tracing cluster peaks}

\tx We first look at the link between initial and final density
peaks by linking the former to the latter using our single large-scale
simulation. We link the peaks at different epochs by finding the smallest
spatial separation between the respective positions in consecutive time-slices.
 
We found that none of the initial peaks have merged during the evolution up
to $\sigma_8=0.6$, well above the value of $\sigma_8=0.4-0.5$ found 
previously (see Section 4.3).
This justifies our use of initial peaks to construct a cluster catalogue, 
and confirms the estimate in Section 3.1.1.

\newcount\Fifrel
\Fifrel=5
\newcount\Fnux
\Fnux=6
\newcount\Fexpmasstest
\Fexpmasstest=7

\figps{5}{S}{\psfig{file=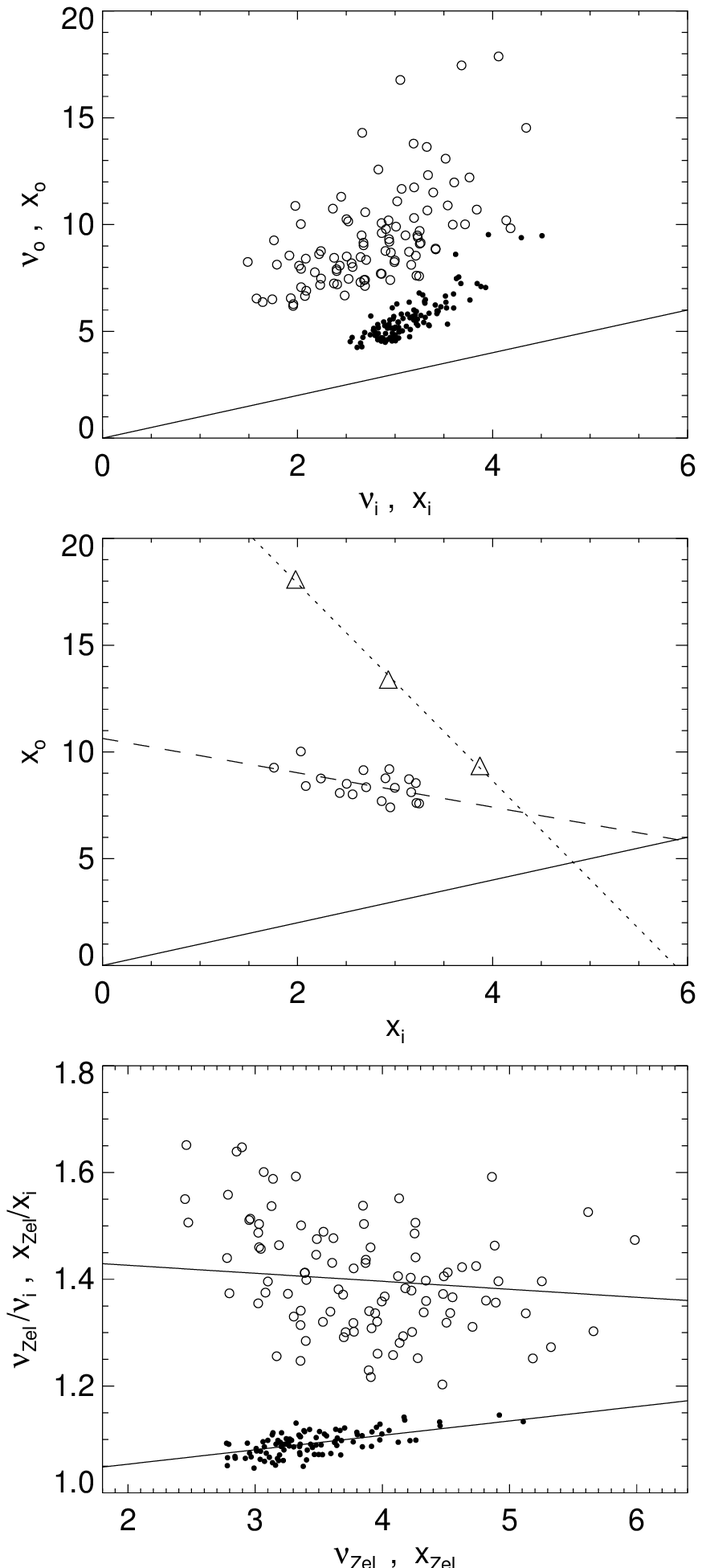,width=8.45cm,silent=}}
{{\bf Figure 5.} (a) Relations between linear and final $\nu$
(closed symbols) and $x$ (open symbols) from the model catalogue,
for $\sigma_8=0.46$.
(b) same as (a), but for those $(x_{\rm i},x_0)$ that have
$|\nu_{\rm i}-3|<0.05$. The dashed line is a fit to these points.
Open triangles are for simulations by van Haarlem \& van de Weygaert (1993),
for $\sigma_8=1.0$. A fit to these three points, which all have
$\nu_{\rm i}=3$, is shown as a dotted line.
Notice the evolution from solid line ($\sigma_8=0$), via dashed line
($\sigma_8=0.46$), to dotted line ($\sigma_8=1.0$).
(c) $\nu_{\rm Zel}$ and $x_{\rm Zel}$ versus the ratios
$\nu_{\rm Zel}/\nu_{\rm i}$ and $x_{\rm Zel}/x_{\rm i}$),
with solid lines showing fits to these, which were used for
Figs.\ \the\Fnux(b)\ and \the\Fexpmasstest. See text for more details.}

\figps{6}{S}{\psfig{file=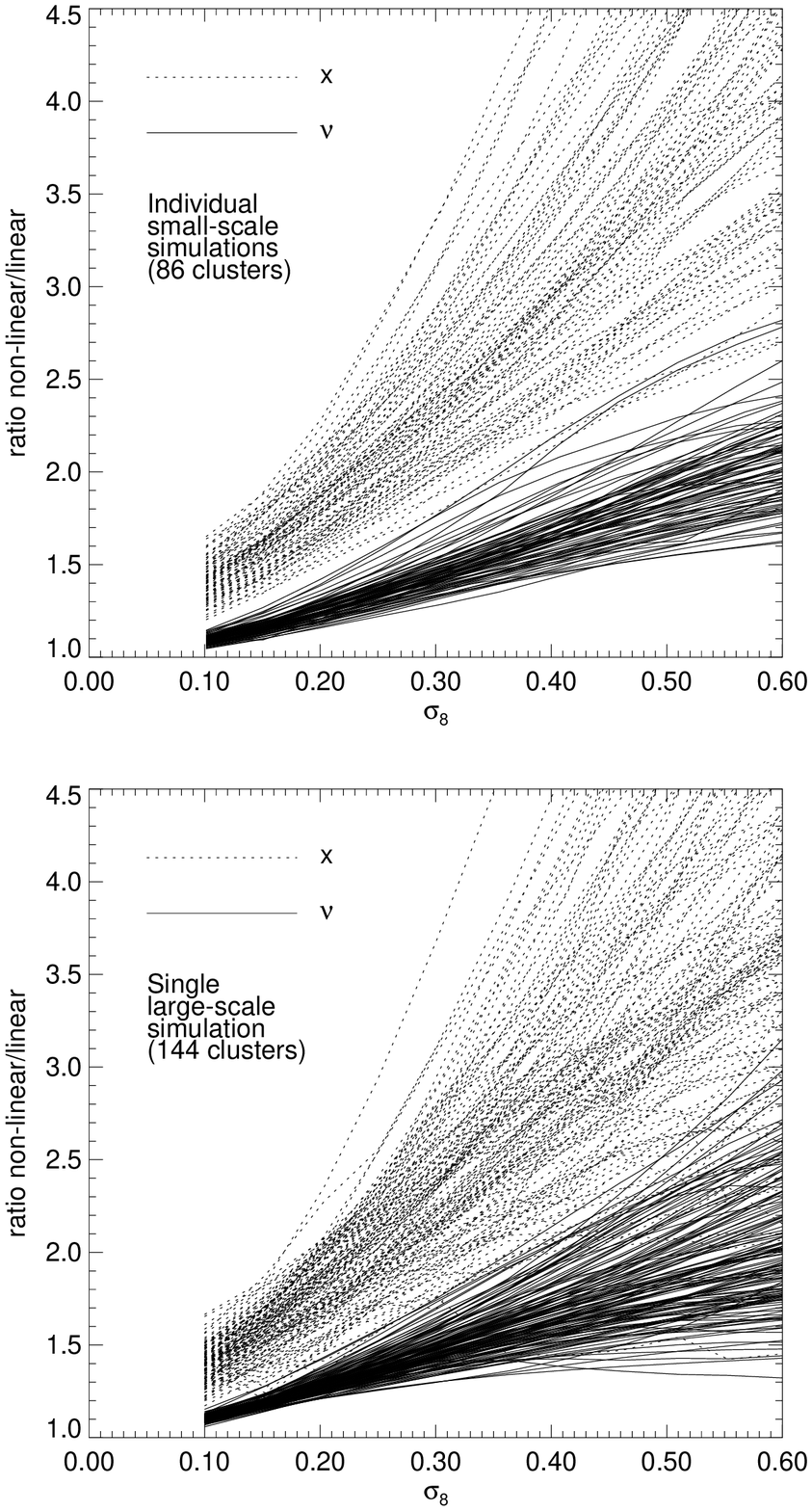,width=8.5cm,silent=}}
{{\bf Figure 6.} (a) Evolution of the non-linear to linear ratio for the
peak amplitude, $\nu$, and the peak curvature, $x$, for all catalogue models.
(b) Same as (a), but for the peaks found in the single large-scale
simulation.}

\figps{7}{S}{\psfig{file=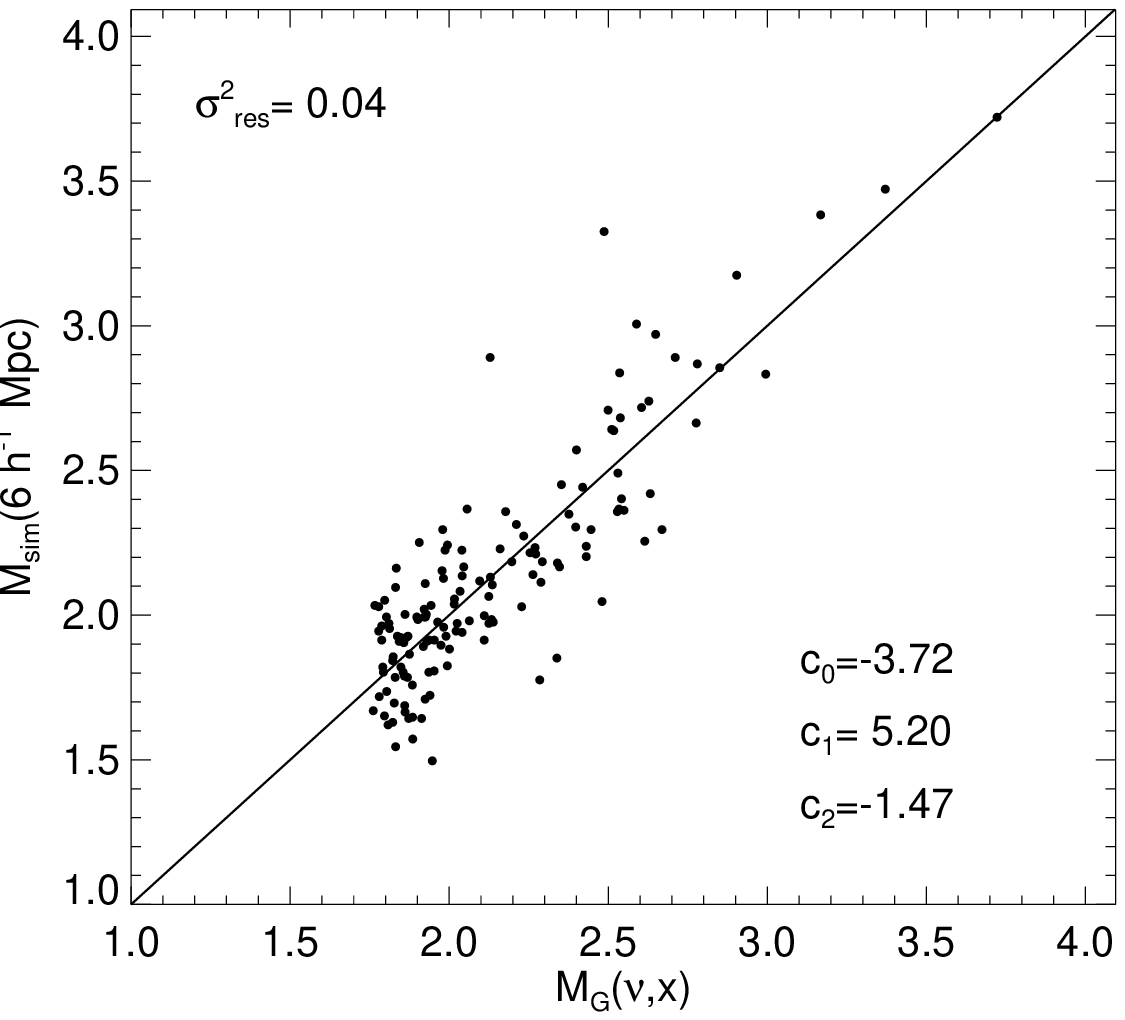,width=8.5cm,silent=}}
{{\bf Figure 7.} Fit for the expected mass $M_{\rm G}$ of peaks in
the single large-scale simulation to the total mass found within
$6h^{-1}$Mpc around these peaks.}

\subsubsection{Distribution and evolution of $\nu$ and $x$}

\tx We examined the $(\nu,x)$ distribution found in the single
large-scale simulation, which is complete by definition, in order to
test the generation of deviates for the peak parameters that define
the catalogue of high-resolution models. We indeed find a similar
distribution for $(\nu,x)$ as for the models making up the model
catalogue (see Fig.\ \the\Fcatalog), so the sampling of the latter is
representative (see also van Kampen 1994).

For the catalogue of high-resolution models we show, in Fig.\
\the\Fifrel(a), for each simulation, the relation between the initial
peak parameters, $\nu_{\rm i}$ and $x_{\rm i}$, and their final values
$\nu_0$ and $x_0$ at the present epoch, $t_0$. Filled symbols are
for $\nu$, whereas open ones are for $x$.
The solid line indicates linear evolution, for which the peak parameters
remain constant (i.e.\ $\nu_0 = \nu_{\rm i}$ and $x_0 = x_{\rm i}$).
Whereas the evolution of $\nu$ is mildly non-linear, the curvature $x$
evolves well into the non-linear regime, but with a large scatter.

In Fig.\ \the\Fifrel(b) we show the results of three simulations by
van Haarlem \& van de Weygaert (1993) (open triangles), who give
initial-final relations for $x$ for peaks with $\nu_{\rm i}=3$ (note that
they have their final epoch at $\sigma_8=1$, whereas we have
$\sigma_8=0.46$). The dotted line is a fit to
those points. We also include those models in Fig.\ \the\Fifrel(a)
with $|\nu_{\rm i}-3|<0.05$, as those are closest to the $\nu_{\rm i}=3$
peaks of van Haarlem \& van de Weygaert (ibid.). A fit to their
data is shown as a dashed line, and it can be seen that the slope of
the initial-final relationship for $x$ at a fixed $\nu_{\rm i}$ evolves
from unity for the initial epoch (solid line), through zero near
$\sigma_8\approx 0.4$, to a steepening negative slope for larger $\sigma_8$.

The evolution of the peak parameters of the high-resolution models
might not be similar to that in the low-resolution simulation if the
size of the high-resolution simulation spheres was not chosen large
enough, as that would artificially
reduce large-scale tidal fields.  We show the evolution with
$\sigma_8$ of the ratio of final to initial values of $\nu$ and
$x$ for each high-resolution cluster model in Fig.\ \the\Fnux(a).

A similar plot for the low-resolution simulation cannot be readily
made, as we do not know the initial values of the peak parameters, but
only $\nu_{\rm Zel}$ and $x_{\rm Zel}$, the values at the end of the
Zeldovich' approximation which was used to calculate the linear
evolution and the first part of the non-linear evolution. 
In Fig.\ \the\Fifrel(c) we show the relation between the value of the
peak parameters at the start of the N-body integration ($\nu_{\rm Zel}$
and $x_{\rm Zel}$, and the ratios $\nu_{\rm Zel}/\nu_{\rm i}$ and
$x_{\rm Zel}/x_{\rm i}$). This figure shows that at the start of the N-body
integration the amplitude $\nu$ of the peak is on, average, 1.1 times
larger than the initial value, while the curvature is on average 1.4
times larger than its initial counterpart.  Using Fig.\
\the\Fifrel(c), we have produced Fig.\ \the\Fnux(b), which shows the
evolution of the peak parameters for the low-resolution
simulation. The initial scatter in Fig.\ \the\Fnux(b)\ was put in by
hand; it reflects the scatter in the fits from Fig.\
\the\Fifrel(c)\ and was sampled from a Gaussian distribution. It 
appears that the individual simulations perform well; the peak
parameters from the single large-scale simulation show a very similar
evolution with time as those for the collection of high-resolution
models, although they are somewhat noisier.

\subsubsection{Fairness of the expected mass relation}

\tx Using the centre-of-mass of the actual N-body particle group near the
density peaks in the low-resolution simulation, we obtain the mass
$M_{\rm sim}$ within a radius of 6.0 $h^{-1}$Mpc (i.e.\ the Top-Hat
mass-scale that was used to define the catalogue). We expect
144 rich Abell clusters in this volume (see above), so we select a
threshold for $M_{\rm G}$ that produces that number.

To test the $M_{\rm G}-M_{\rm sim}$ catalogue defining relation,
we trace back clusters found at the present epoch to peaks in the
smoothed initial density field. In Fig.\ \the\Fexpmasstest\ we
plotted the final mass $M_{\rm sim}$ against the predicted
final mass $M_{\rm G}$ that is calculated from the initial peak
parameters according to eq.\ (\the\Mtheory), where we again have
used the fits of Fig.\ \the\Fifrel(b)\ to get from the `Zeldovich'
peak parameters to the initial ones.

The relation in Fig.\ \the\Fexpmasstest\ compares quite well to the
relation for the single, high-resolution, cluster models, as shown in
Fig.\ \the\Fmasscomp(b). Although the constants $c_j$ are slightly
different, the ratio $c_1/c_2$, which determines the slope of the
final mass threshold, is almost the same (3.54 as compared to 3.76).

%% file: cat5.tex
\section{Completeness}

\newcount\Frichaco
\Frichaco=8
\figps{8}{S}{\psfig{file=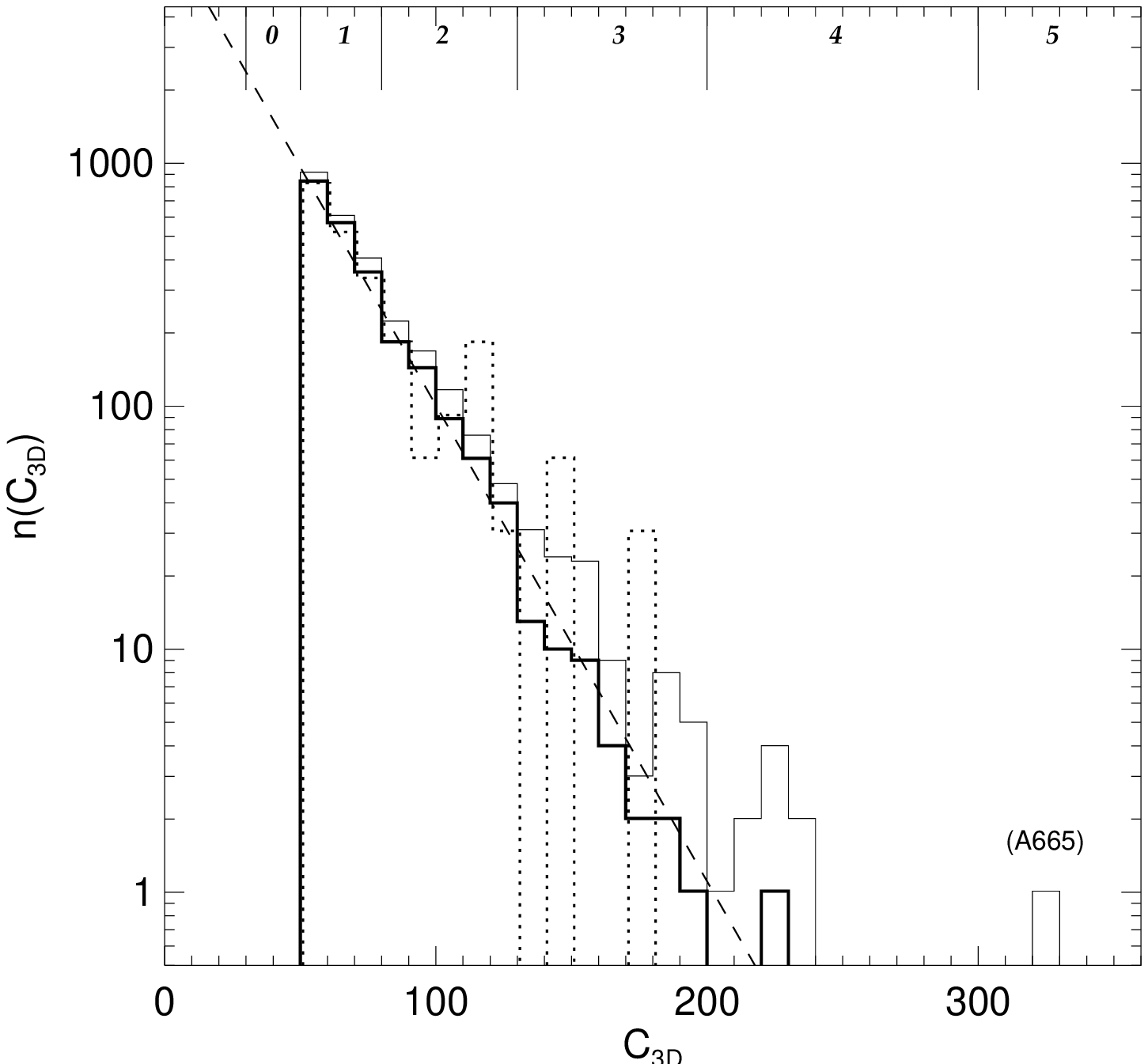,width=8.5cm,silent=}}
{{\bf Figure 8.} Richness distribution for the ACO catalogue (thin line),
its $z<0.08$ subsample (thick line), and the rescaled ENACS sample
(dotted line), also for $z<0.08$. Numbers at the top indicate the
traditional Abell richness classes.}

\subsection{Completeness of the ENACS sample}

\tx The ENACS sample was constructed to be complete in richness
(Katgert et al.\ 1996) by selecting a subsample from the ACO catalogue
within a particular volume (see Section 4.3).
However, from comparison to the Edinburgh-Durham Cluster Catalogue
(EDCC from here on), Mazure et al.\ (1996) estimated that the full 
$R\ge1$ $(C_{\rm ACO}\ge50)$ ACO catalogue is 94 per cent complete out
to $z=0.08$. We have to accept this incompleteness, since the ACO
catalogue is still the largest observational sample available.
In the following we therefore assess completeness with respect to
the full ACO cluster catalogue, keeping in mind that this catalogue
is only 94 per cent complete.

We see from Fig.\ \the\Frichaco\ that the ENACS sample, as a subsample
of the ACO catalogue, is 100 per cent complete out to $z=0.08$.
Plotted is the distribution of richness $C_{\rm ACO}$ of all ACO clusters
(thin solid line), ACO clusters out to $z=0.08$ (thick solid line),
and ENACS clusters out to the same redshift (dotted line), where
the latter is rescaled to the ACO volume. The distribution of the
latter two is clearly statistically the same, indicating that the
ENACS catalogue is a representative subsample of the ACO catalogue.
An exponential function of the form $n(R)=a\ e^{-bR}$ has been fitted
to the richness distribution of the $z<0.08$ ACO clusters for use in
Section 6, and is shown in Fig.\ \the\Frichaco\ as a dashed line.

\subsubsection{ACO and 3D richness}

\tx Both Abell (1958) and ACO (1990) have corrected their richness
determinations for foreground and background galaxies, by subtracting
a `field correction' term from the raw number count.
Since we model clusters individually, our simulation volume does not
contain any obvious foreground or background galaxies. Common observational
practice is to use a `gapper' criterion to remove such galaxies,
i.e.\ eliminate all galaxies that are on the non-cluster side of a gap
in the line-of-sight velocity distribution for all galaxies observed
towards the cluster. Note that we will remove `interlopers',
non cluster members near the turn-around radius which are not
eliminated by the `gapper' criterion, when we calculate line-of-sight
velocity dispersions in Section 7.1.

For our model clusters, with their limited
simulation volume, we did not find any gaps of 1000 km s$^{-1}$, which
is the value used by Katgert et al.\ (1996) for the ENACS sample.
So we indeed do model just what an observer would call the `main' system,
and we can simply apply Abell's (1958) criteria to each of our simulations,
without making any corrections. In fact, this means that we can only
obtain the {\it 3D richness} of a cluster.

We define the 3D richness measure $C_{\rm 3D}$ as the number of
galaxies within the magnitude interval $[m_3, m_3+2]$ within a
cylinder of radius $1.5h^{-1}$ Mpc (Abell's compactness criterion)
and depth $10h^{-1}$ Mpc,
which is the typical cluster turn-around radius for our catalogue.
We chose a constant depth because the radius of the simulation sphere
was adjusted according to the expected final cluster mass in order
to include into the modelling a sufficient volume around the cluster
turn-around radius.

Mazure et al.\ (1996) found that {\it on average} $C_{\rm ACO}$ is
equal to $C_{\rm 3D}$, but there exists a significant scatter in
their relation, because for $C_{\rm ACO}$ a constant field correction
term was, while for $C_{\rm 3D}$ the field correction term was estimated
for each individual cluster. We used their data as plotted in their
Fig.\ 1(c) to find that scatter. A linear fit gives
$\sigma_{C_{\rm ACO}-C_{\rm 3D}}(C_{\rm 3D})=0.19C_{\rm 3D}$.

\subsection{Completeness of the model catalogue}

\tx The main purpose of our catalogue of galaxy cluster models is the
comparison of its statistical properties to those of the ENACS
catalogue, so we need both samples to be complete in the same (observable)
quantity. The ENACS sample is complete in the ACO richness measure
$C_{\rm ACO}$, so we desire that the model catalogue is complete
in $C_{\rm ACO}$ as well. However, we can only obtain $C_{\rm 3D}$
for our models, so we can only strive for completeness in $C_{\rm 3D}$.
However, as $C_{\rm ACO}$ is statistically equivalent to $C_{\rm 3D}$
(Mazure et al.\ 1996), we can still compare the statistical properties
of both catalogues.

Unfortunately, completeness in $C_{\rm 3D}$ is not readily achieved.
As discussed in Section 3, we had to construct the catalogue using the
final mass estimator $M_{\rm G}$. Only if there exists a noiseless
relationship between $M_{\rm G}$ and $C_{\rm 3D}$
will completeness in 3D richness immediately be guaranteed.
However, richness is determined from the projected galaxy distribution,
while $M_{\rm G}$ is related to the total matter distribution within
a sphere. For this reason, and others which are discussed in Section 5.3
below, there is a rather large scatter in the correlation between
$M_{\rm G}$ and $C_{\rm 3D}$, so near the mass threshold some clusters
drop out of the richness selected catalogue, while others can enter
from below the expected mass threshold. In other words, low-mass clusters
with a relatively high richness are, by construction, not included in
our catalogue, and a catalogue that is complete in mass will not be
complete in richness. As this comprises a significant fraction of all
clusters, we need to correct for this.

\newcount\Fmassrich
\Fmassrich=9
\figps{9}{S}{\psfig{file=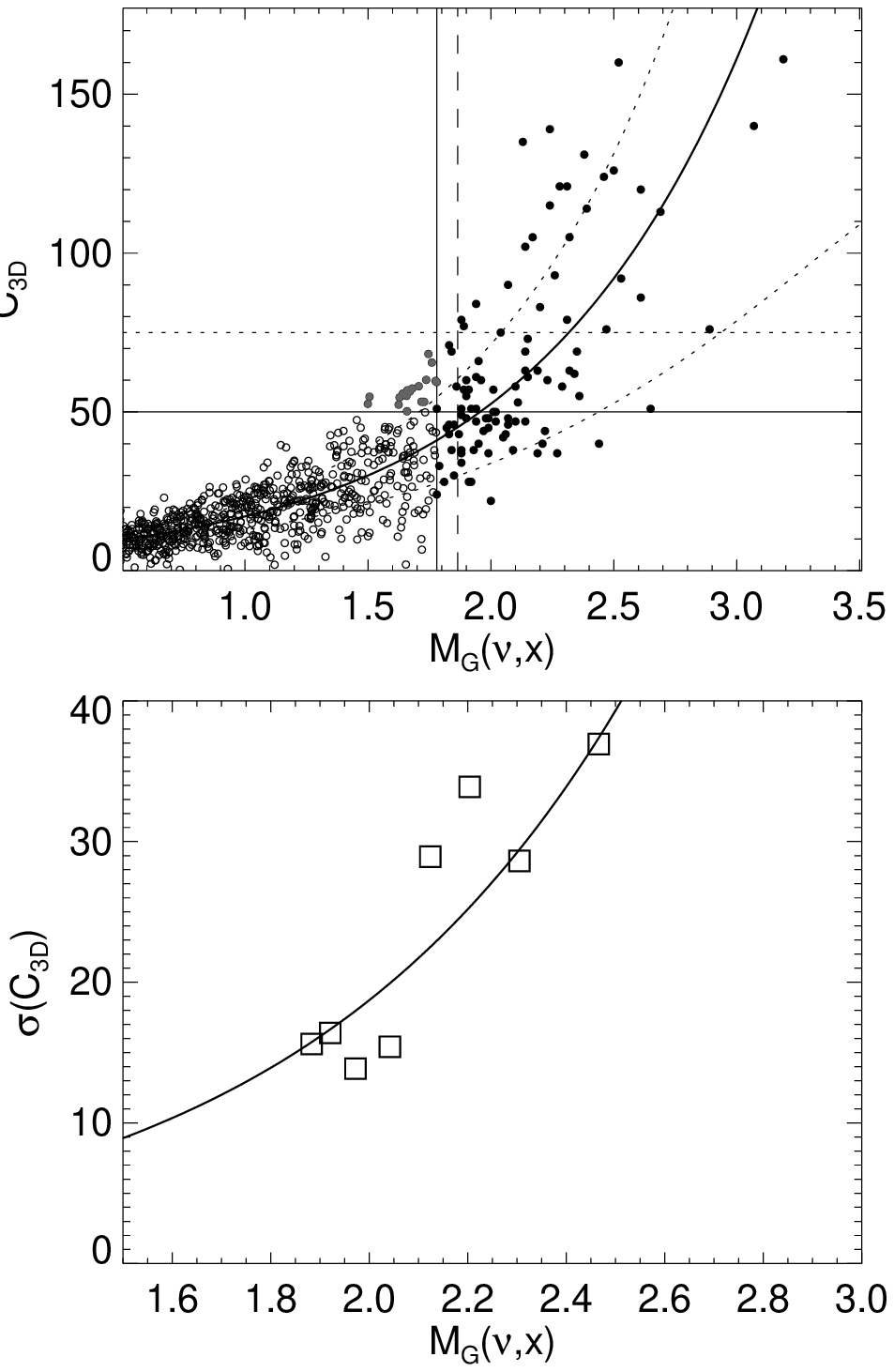,width=8.6cm,silent=}}
{{\bf Figure 9.} (a) 3D richness $C_{\rm 3D}$ versus predicted final
cluster mass $M_{\rm G}$ for our simulated clusters at $\sigma_8=0.46$
(filled symbols), and peaks sampled below the $M_{\rm G}$ threshold for 99
clusters (open and grey symbols). The dashed line indicates the
$M_{\rm G}$ threshold for 86 clusters. The horizontal lines indicate
the $C_{\rm 3D}=50$ and $C_{\rm 3D}=75$ richness limits. The solid curve
is an exponential fit to the points above the $M_{\rm G}$-threshold (the
vertical solid line), with the dotted lines showing the $1\sigma$ deviation
around that fit, as obtained from (b).
(b) Variance in richness as a function of 3D richness, as obtained from
the {\it simulated}\ models, i.e.\ the filled symbols displayed in (a).
The solid curve is again an expontial fit to these points, and was used
to plot the dotted lines in (a).}

\subsection{Correcting for incompleteness}

\tx We correct for incompleteness in richness {\it directly}\ from the relation
between richness $C_{\rm 3D}$ and predicted final mass $M_{\rm G}(\nu,x)$
(the catalogue defining quantity). The method consists of fitting an
appropriate function to the ($M_{\rm G}$,$C_{\rm 3D}$) pairs
obtained from the simulations, calculate the scatter of the residuals
for several bins in $M_{\rm G}$, and fit a function to those as well.
In order to find a richness for the $M_{\rm G}$ that we did not select for
simulation, we sample these richnesses from a
normal distributed with mean and variance given by the two fits.

In Fig.\ \the\Fmassrich (a) we show the relation between
$C_{\rm 3D}$ and $M_{\rm G}$. The filled symbols indicate the
{\it simulated} clusters at $\sigma_8=0.46$.
The vertical dashed line shows the $M_{\rm G}$ threshold that selects 86
clusters, the number appropriate to the chosen volume (see Section 4.3),
and the vertical solid line the $M_{\rm G}$ threshold that selects the
99 cluster models that we have actually simulated. 
The horizontal line shows the richness limit used to compile
observational cluster catalogues (on average, $C_{\rm 3D}\ge50$ gives the
same number of clusters as $C_{\rm ACO}\ge50$).

To estimate the distribution of rich Abell clusters with masses below the
$M_{\rm G}(\nu,x)$ threshold, we assume a functional form for the mass-richness
relation, and measure its total scatter. This functional form should
follow the general trend of the points, and stay positive for all $M_{\rm G}$.
We find that an exponential of the form $C_{\rm 3D}=a\ e^{b M_{\rm G}}$
achieves all this. A fit of this form is shown in Fig.\ \the\Fmassrich (a)
as a solid line. We found $a=5.6$ and $b=1.12$.
The scatter around this fit as a function of $M_{\rm G}$ is shown
in Fig.\ \the\Fmassrich (b), along with an exponential fit to it:
$\sigma(C_{\rm 3D})=0.96\ e^{1.48 M_{\rm G}}$. This fit provides
the $1\sigma$ deviations from the $M_{\rm G}$-$C_{\rm 3D}$ relation
that are plotted as dotted lines in Fig.\ \the\Fmassrich (a).
Combining both fits we derive that $\sigma(C_{\rm 3D}) = 0.35C_{\rm 3D}$.

We now assume that low-mass/high-richness clusters are still corresponding
to initial density peaks, and should therefore be found amongst peaks
{\it below} the $M_{\rm G}$ threshold. For the standard CDM scenario
chosen here, the adopted volume of $10^7h^{-3}$Mpc is expected to contain
$N_{\rm peak}(>0,R_{\rm G})=1164$ peaks (see also Section 3.3.1).
We then use both fits to estimate $C_{\rm 3D}$ for
each $M_{\rm G}$ of the remaining 1078 peaks {\it below} the $M_{\rm G}$
threshold. This is done by sampling them from a normal distribution with
mean $5.6\ e^{1.12 M_{\rm G}}$ and variance $0.35C_{\rm 3D}$, i.e.\ the
fits from Fig.\ \the\Fmassrich.
The resulting data points are plotted as grey dots and open circles in
Fig.\ \the\Fmassrich (a). All points below the $M_{\rm G}$ threshold that
have $C_{\rm 3D}>50$ should be the clusters that we miss in our catalogue;
these points are plotted in grey. Note that most of them are near the
$M_{\rm G}$ threshold, so it did make sense to run 13 more cluster models
below that threshold. In fact, in turns out that four of those are
showing up as rich clusters (i.e.\ $C_{\rm 3D}\ge50$) at $\sigma_8=0.46$,
as can be seen in Fig.\ \the\Fmassrich (a). According to the
exponential fit, and its scatter, this is a plausible number.

The distribution of all points plotted in Fig.\ \the\Fmassrich (a) is 
the {\it complete} richness distribution (in a statistical sense).
Since, on average, $C_{\rm 3D}$ is equal to $C_{\rm ACO}$,
we found the complete distribution function for $C_{\rm ACO}$ as well.
Note that the number of $C_{\rm 3D}\ge50 $ rich Abell clusters we find
from this method is not necessarily the amount we need (i.e.\ 86 in
our case). This will depend on the value of $\sigma_8$, and we
actually use this to constrain its value in Section 6.1.2.

For the choice of $\sigma_8=0.46$ that we used to illustrate the method,
we find that we have 78 clusters with $C_{\rm 3D}\ge 50$, of which 60
were actually simulated. For the other 18 clusters we just sampled
the richness from an estimated probability distribution, i.e.\ we did not
simulate them, and therefore any other properties of these models are unknown.
In Section 7.1.1 we use a similar method to obtain line-of-sight velocity
dispersions.

\subsection{A complete subsample}

\tx If we increase the richness threshold used to define a sample of
galaxy clusters, we can obviously make the sample complete since the
richest clusters are all simulated. Indeed Mazure et al.\ (1996) found
that they could construct a truly {\it complete}\ subsample of ACO
clusters for a $C_{\rm 3D}$ richness limit of 75, in a cone out
to $z=0.1$. After inspection of
Fig.\ \the\Fmassrich(a) we find that our simulation set at $\sigma_8$=0.46
is also complete for $C_{\rm 3D}\ge75$ (i.e.\ above the horizontal dotted line
in the Figure). This allows a direct comparison of statistical properties
of these complete observational and model samples, but the sample size
is obviously rather small: 33 and 29 clusters respectively (for slightly
different volumes, and for $\sigma_8=0.46$).

\newcount\Frichkpgm
\Frichkpgm=10
\figps{10}{S}{\psfig{file=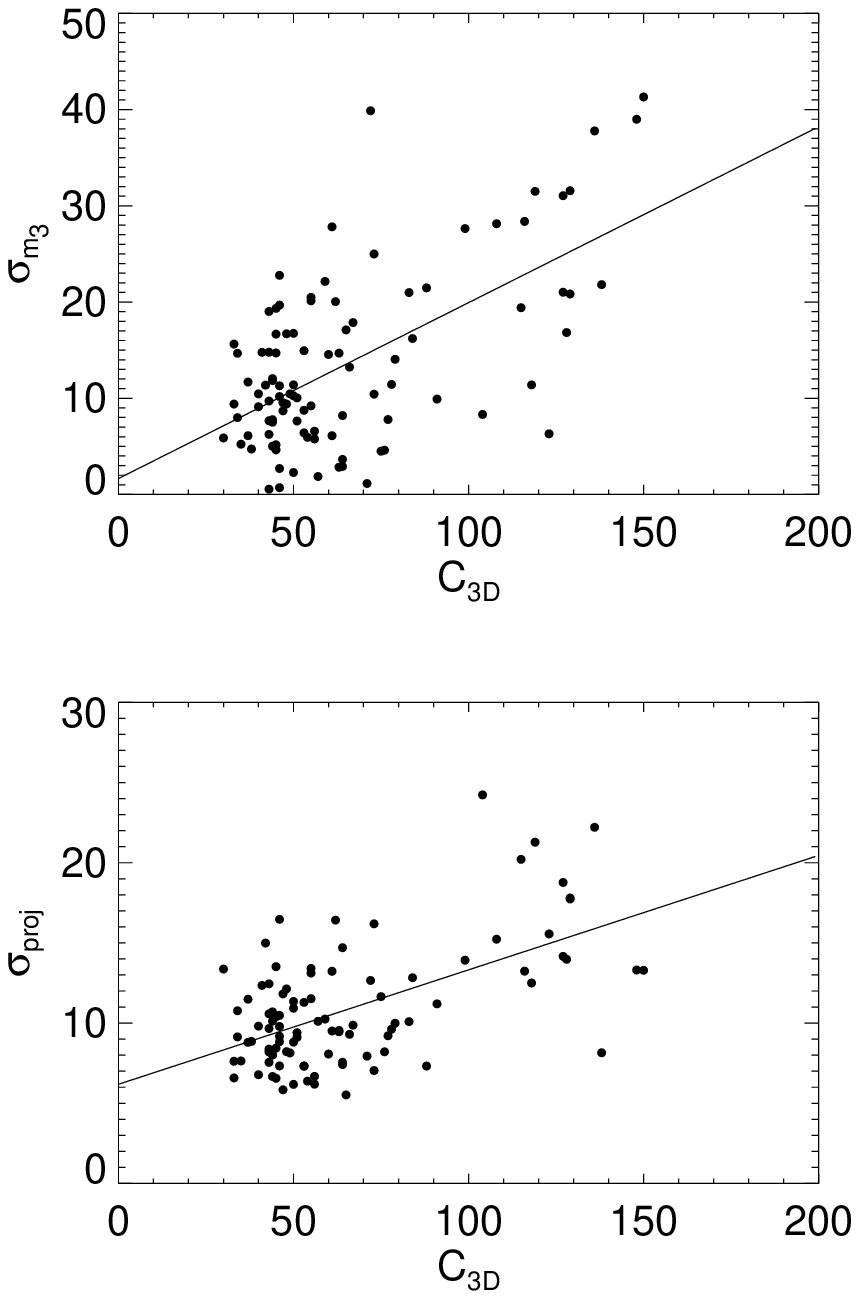,width=8.6cm,silent=}}
{{\bf Figure 10.} Noise in the 3D richness measure.
(a) Scatter in 3D richness due to $m_3$.
(b) Scatter in 3D richness due to projection.}

\subsection{Sources of incompleteness}

\tx In this section we estimate the {\it intrinsic}\ noise of the
richness measure due to its very definition and due to projection effects.
This allows us to find the scatter in the $M_{\rm G}$-$C_{\rm 3D}$
relation, as the total scatter found in Section 5.3 is the combination
of all these.

\subsubsection{Evolution of the third brightest galaxy}

\tx Recall that richness is defined as the number of galaxies
within the magnitude range $[m_3,m_3+2]$. Because the luminosity function
is not very well sampled at the high-end, $m_3$ can vary quite significantly
from cluster to cluster for equal total galaxy numbers and identical
luminosity functions. In other words, the
upper limit of the magnitude range, the magnitude of the third brightest
galaxy, is far more important for the richness measure than the lower limit.
A consequence of this is that the richness of a cluster can change
significantly with time if $m_3$ changes due to, for example, a galaxy
merger event.

We estimate the r.m.s.\ scatter due to the evolution of $m_3$ by taking
$m_i$ for $i=\{1,2,3,4,5\}$ and calculating the standard deviation of
the five corresponding richness measures. This gives us a measure for
the `jump' in richness due to a change of the third brightest galaxy.
We find a spread of around 18 per cent on average, with a minimum of 1
per cent (model 4), and a maximum of 52 per cent (model 21).
More quantitatively, the scatter due to $m_3$ as a function of $C_{\rm 3D}$
is plotted in Fig.\ \the\Frichkpgm(a), and a linear fit gives
$\sigma_{m_3}(C_{\rm 3D})=2+0.18C_{\rm 3D}$.

\subsubsection{Projection effects}

\tx Since richness is a `projected' quantity, measured within a cylinder,
it is strongly dependent on the orientation of the simulation volume
towards the observer. This projection effect is strongest for hierarchical
formation scenarios like CDM, which produce highly irregular,
clumpy galaxy distributions. We calculated the variation of richness
due to orientation by observing each model cluster from 200 random
directions. We find that the spread in $R$ due to projection
effects is on average 13 per cent, with a maximum of 39 per cent for
model 97 (a poor cluster), and a minimum of 7 per cent for model 76
(a fairly compact, $C_{\rm 3D}$=75 cluster).
We find from Fig.\ \the\Frichkpgm(b) that, on average,
$\sigma_{\rm proj}(C_{\rm 3D})=5.7+0.072C_{\rm 3D}$.

\subsubsection{Scatter in the $M_{\rm G}$-$C_{\rm 3D}$ relation}

\tx The noisyness in $m_3$ and the strong dependence on projection
make richness a measure with a rather large intrinsic spread.
Quadratically adding these two noise
sources, we estimate the total observational scatter in $C_{\rm 3D}$ as
$\sigma_{\rm int}(C_{\rm 3D})=6+0.2C_{\rm 3D}$.
For $C_{\rm ACO}$ we need to add the scatter in the
$C_{\rm ACO}$-$C_{\rm 3D}$ relation as well, yielding
$\sigma_{\rm int}(C_{\rm ACO})=6+0.27C_{\rm ACO}$.

Note that there are further sources of noise in richness. For example,
one might easily miss a few galaxies when observing a rich cluster in
projection due to overcrowding, especially if there is a large cD-galaxy
present. However, we consider these noise sources as insignifant as compared
to those discussed, as only 25 per cent of all clusters has a cD galaxy
(Leir \& van den Bergh 1977).

With $\sigma(C_{\rm 3D})$ known from the fit shown in Fig.\ \the\Fmassrich (b),
we can quadratically subtract $\sigma_{\rm int}(C_{\rm 3D})$ from that
to find a scatter of about $0.25C_{\rm 3D}$ in the
$M_{\rm G}$-$C_{\rm 3D}$ relation.

%% file: cat6.tex
\section{The final cluster catalogue}

\subsection{Timing of the models}

\tx The normalisation of the standard $\Omega_0=1$ CDM density fluctuation
is a free parameter that can be changed without
changing the characteristic scales of the CDM scenario. So a
numerical simulation adopting such a scenario can be run to an
arbitrary time and then renormalised to fix the present epoch.
Usually one assumes a linear bias between the galaxy and the dark
matter fluctuation spectra, and the observation that at present the
{\it r.m.s.}\ of the 8$h^{-1}$Mpc Top-Hat smoothed galaxy distribution
is about unity (Davis \& Peebles 1983).
A major advantage of actually forming galaxies during the
simulation is that it allows us to make a direct comparison to
observations without having to assume a value for the bias.

Initial conditions for the cluster models were set up such that
$\sigma_8$, the r.m.s.\ of the 8$h^{-1}$Mpc Top-Hat smoothed total
density field, is unity at the present epoch, i.e.\ there is no
linear bias. For the standard CDM scenario $\sigma_8$ was found
to be significantly smaller than unity in most earlier work
(e.g.\ Davis et al.\ 1985; Frenk et al.\ 1990; Bertschinger \& Gelb 1991).
We ran all models up to $\sigma_8=0.65$, and a subset to $\sigma_8=1$,
the unbiased normalisation.

\subsubsection{Galaxy autocorrelation function}

\tx A traditional measure for timing numerical simulations of the
large-scale structure of the universe is the galaxy autocorrelation
function. Interestingly, this function is claimed to be mainly
determined by galaxies within 10 $h^{-1}$Mpc of cluster centres
(Peebles 1974; McGill 1991). The galaxy autocorrelation function
obtained from our cluster models might therefore be representative
of the global one.
However, quite a few problems might spoil this idea in practice.
First of all, the dynamical evolution of both the galaxy and the dark
matter component inside the cluster is simulated only approximately,
because several physical processes are modelled only crudely, or not at all.
Secondly, we did not transform possible cD galaxies into single galaxy
particles. Since about 25 per cent of all observed clusters contain one or
more cD's (Leir \& van den Bergh 1977), we need to recognise these cases and
separate cluster dark matter particles and those belonging to the cD.

We therefore believe a determination of the galaxy autocorrelation
function from our cluster model will not be sufficiently unbiased.
For this reason we decided to use the normalisation found by van Kampen
(1996) from several unconstrained simulations, i.e.\ average patches
of universe for the same scenario.
These `field models' provide better estimates for the large-scale
autocorrelation function corresponding to the cosmology chosen.
The same galaxy formation parameters were used as for the cluster
models presented in this paper. The galaxy autocorrelation function
was calculated for the simulated galaxy distribution, and compared
to the observed one. It was found that both its slope and the amplitude
argue for a $\sigma_8$ in the range 0.46 to 0.56 (van Kampen 1996),
although its slow evolution does not rule out values somewhat outside
of this range.

Once we have chosen $\sigma_8$, i.e.\ defined the present epoch in the
simulations, we can test statistical properties of the catalogue.
Alternatively, we can use such statistics as an additional and
independent means of normalising the cosmological power spectrum.
Besides the autocorrelation function we use the distribution of
{\it richness}\ as an independent measure to constrain $\sigma_8$ by
matching it to that observed for Abell clusters.
If we have selected the correct scenario and if our modelling is
sufficiently realistic, both measures should yield consistent timing,
certainly if these measures are on roughly the same length-scale.
Note that the results of {\it COBE}\ demand $\sigma_8\approx1.3$ for
this scenario (Bunn, Scott \& White 1995), i.e.\ a linear {\it anti-}bias,
but remember that {\it COBE}\ measures power on a scale well outside the
dynamical range of our simulations (indicated by the two vertical lines in
Fig.\ \the\Fspectra). This will be discussed further in Section 8.

\newcount\Frichness
\Frichness=11
\figps{11}{S}{\psfig{file=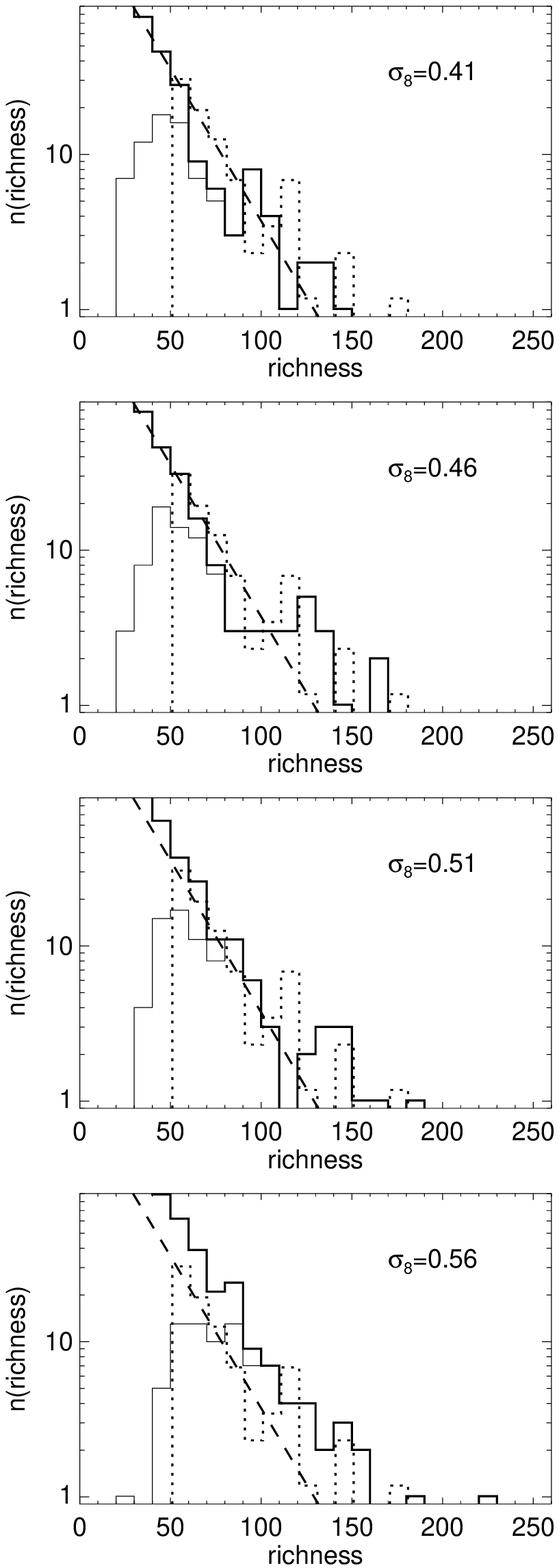,width=8.3cm,silent=}}
{{\bf Figure 11.} Uncorrected distributions of the richness measure
$C_{\rm 3D}$ obtained for our model clusters (thin solid lines),
and those corrected for incompleteness (thick solid lines),
for four different choices of $\sigma_8$. The dotted lines give the
observed distributions of $C_{\rm ACO}$ for the ENACS sample,
the dashed lines a fit to the distribution for the ACO sample,
normalised to the ENACS volume.}

\newcount\Frichevo
\Frichevo=12
\figps{12}{S}{\psfig{file=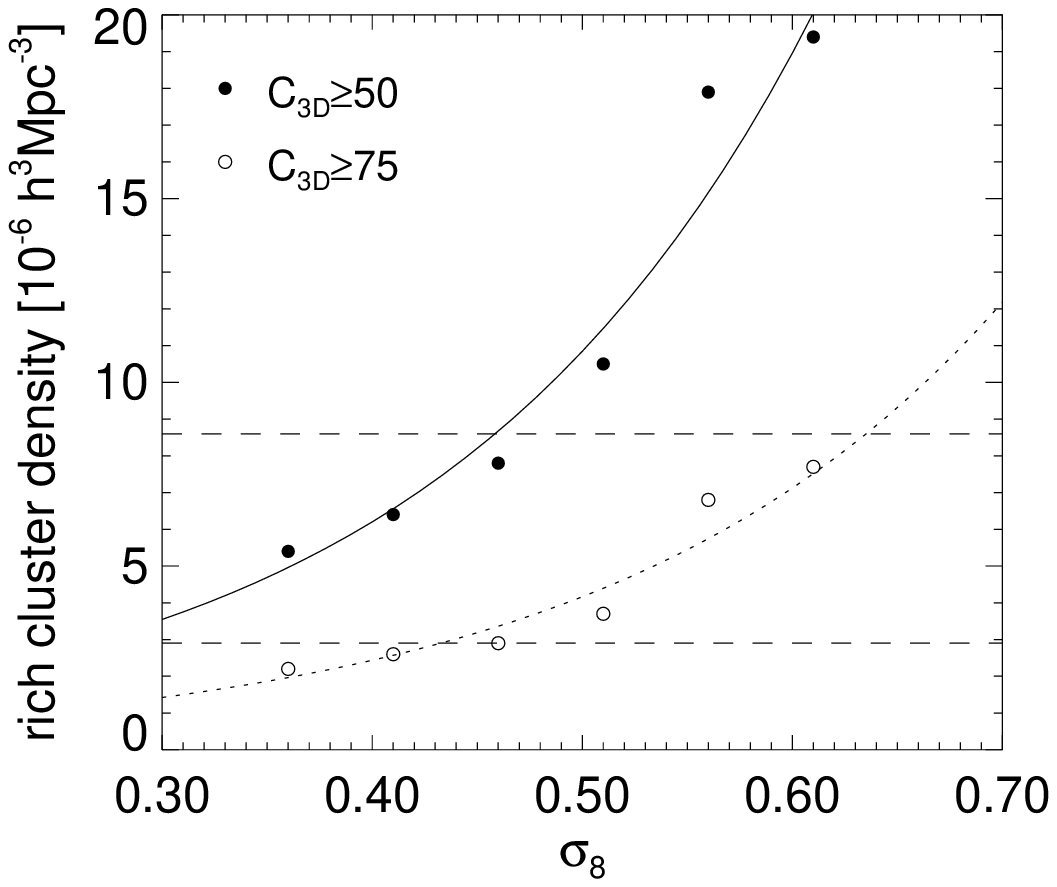,width=8.5cm,silent=}}
{{\bf Figure 12.} Number density of rich Abell clusters as a function
of $\sigma_8$ for the standard CDM scenario. Closed symbols are for
the corrected $C_{\rm 3D}\ge50$ sample, open ones are for the complete
$C_{\rm 3D}\ge75$ subsample. The horizontal dashed lines indicate the
observed densities for both samples, whereas the curves show exponential
fits to the plotted points. The preferred value for $\sigma_8$ is 0.46.}

\subsubsection{Richness distribution}

\tx We now combine all richness determinations for the individual
models to obtain a statistical distribution function. 
Fig.\ \the\Frichness\ shows histograms of the richness measure $C_{\rm 3D}$
found from our models (thin solid lines) for four values of $\sigma_8$.
Only the 86 clusters above the $M_{\rm G}$ threshold are used for
each histogram. We applied the method of Section 5.3 to correct for
incompleteness. The corrected distributions are plotted in
Fig.\ \the\Frichness\ as thick solid lines.

The dotted lines denote the distributions of richness measure
$C_{\rm ACO}$ for the ENACS sample, which is complete (see Section 5.1).
To guide the eye, we also included, as a dashed line, the fit to the
richness distribution for the complete ACO catalogue
(see Fig.\ \the\Frichaco), rescaled to the ENACS volume.
As the distributions for $C_{\rm ACO}$ and $C_{\rm 3D}$ are statistically
equivalent, we can directly compare our corrected $C_{\rm 3D}$
distribution to $C_{\rm ACO}$ distribution of the ENACS sample and
the rescaled ACO catalogue.

We clearly observe a monotonous evolution of our corrected $C_{\rm 3D}$
distribution. Those for $\sigma_8=0.46$ or $\sigma_8=0.51$ fit the
observed distribution best, although the values of 0.41 and 0.56
are not strongly ruled out. However, even smaller or higher values for
$\sigma_8$ are excluded (we did not plot these in Fig.\ \the\Frichness).
We can summarize the evolution by plotting the density
of rich clusters as a function of $\sigma_8$. We do this for both
the corrected $C_{\rm 3D}\ge50$ sample and the complete $C_{\rm 3D}\ge75$
subsample. These are shown in Fig.\ \the\Frichevo, along with exponential
fits to them. This functional form probably fits best for the range of
$\sigma_8$ we considered here, although one expects the curve to stop
rising exponentially for large $\sigma_8$.

From the exponential fits we find that $\sigma_8=0.46$ matches best.
White, Efstathiou \& Frenk (1993) find a value of $\approx0.52$ from
a comparison of cluster abundances calculated using the Press-Schechter
(1974) formalism and standard N-body modelling (i.e.\ no modelling of
galaxy formation) to X-ray data. A similar comparison to the optical
cluster abundance obtained from the median velocity dispersion of
$C_{\rm ACO}\ge50$ clusters gives $\sigma_8\approx0.62$.
Eke, Cole \& Frenk (1996) find 0.50$\pm$0.04 from similar
simulations (but at a higher resolution) as White et al.\ ({\it ibid.}),
when compared to just the X-ray data, where they used the same X-ray data
as White et al.\ but with a slightly different way of inferring the
X-ray cluster abundance.
Unfortunately it is not straightforward to compare these inferred values
for $\sigma_8$ to the one we infer from the richness distribution,
as there might be systematic differences between the richness distribution
and the X-ray temperature function.

\subsubsection{Choice for $\sigma_8$}

\tx The galaxy autocorrelation function and the richness distribution
for the models leave a reasonably narrow range of allowed values for
$\sigma_8$. In defining the catalogue using the expected final mass
$M_{\rm G}$, we have adopted $\sigma_8=0.46$ as the final epoch, as it
is the preferred value for the richness distribution, which is the most
restrictive of the two measures.

By adopting $\sigma_8=0.46$ we have a single catalogue that can be used
for subsequent study. This should include a study of the distribution of
intrinsic properties of clusters, and a more a detailed comparison to
the ENACS and other observed samples.

\subsection{Selection of catalogue entries}

\tx Now that we `calibrated' $\sigma_8$ to be 0.46, we select those cluster
models from the set of 99 simulated that have $C_{\rm 3D}\ge50$ in
order to construct a richness selected catalogue.
We find that we have simulated 60 rich clusters out of the
86 desired, i.e.\ we are 70 per cent complete in richness $C_{\rm 3D}$.
However, the density of rich clusters we find for $\sigma_8=0.46$ is only
$7.8\times10^{-6}h^3$Mpc$^{-3}$, while that should be 1.1 times larger
according to the fit of Fig.\ \the\Frichevo\ {\it and}\ the density
found for the ENACS sample (Mazure et al.\ 1996).
So we expect to find 66 rich clusters in our $M_{\rm G}$ selected catalogue,
i.e.\ we expect to be 77 per cent complete, and to miss 20 rich clusters
that will form from peaks below the $M_{\rm G}$ threshold for the
whole set of 99 models. The fraction of peaks that form a rich cluster at
the present epoch is quite small there, so it is not straightforward to
model those clusters (see also Fig.\ \the\Fmassrich (a)).

Fortunately, as discussed in Section 5.3, we do have a complete subsample
of $C_{\rm 3D}\ge 75$ clusters in our simulation set, i.e.\ a sample
for which no statistical corrections are needed and for which we have
each entry numerically modelled.

The resulting catalogue is presented in Table 1, which lists the
defining constraints as well as some global physical properties
(discussed below).
The cluster models that are not rich enough are listed in italics,
and those that are in the complete $C_{\rm 3D}\ge75$ are shown in
bold face.
Note that the selection of models that are part of the catalogue is
specific for the choice of the present epoch.
An extreme example of this is model no.\ 60.
Its mass within 6$h^{-1}$Mpc almost doubles from $\sigma_8=0.46$ to
$\sigma_8=0.51$ because another clump enters the Top-Hat window.
However, it is also the least rich model of all (because
its line-of-sight is perpendicular to the merging direction),
and therefore well below the richness limit.

\subsection{Global properties of the cluster models}

\tx The properties of all models are listed in Table 1.
Columns 2 to 5 give the peak parameters used to define
the cluster, which also determine the expected final cluster mass
$M_{\rm G}(\nu,x)$ listed in column 6. The mass within 6$h^{-1}$Mpc
found in the simulations, used to fit (\the\Mtheory), is given in
column 7. The Abell mass $M_{\rm A}$, i.e.\ the mass within a sphere
of 1.5$h^{-1}$ Mpc, can be obtained for most observed clusters
(although usually only approximately), and is therefore listed
in column 8. The 3D richness measure $C_{\rm 3D}$, which is
{\it on average}\ equal to the ACO richness measure that usually
defines an observational catalogue, is listed in column 9. Since
its value can be quite dependent on the chosen line-of-sight,
especially for low richness clusters, we show in column 10 the same
quantity averaged over 200 lines of sight.

The next four columns list various velocity dispersions, starting with
the intrinsic velocity dispersions $\sigma_{\rm dm}$ and $\sigma_{\rm gal}$,
for the dark matter and the galaxies respectively, which are obtained
within the 3D Abell radius and divided by $\sqrt{3}$ in order to compare
them to the line-of-sight velocity dispersion.
Column 12 gives the line-of-sight velocity dispersion
$\sigma_{\rm l.o.s.}$ obtained from the projected galaxy distribution.
Column 13 provides, for comparison, the same quantity averaged over
200 lines of sight.
The remaining three columns provide a few more useful cluster properties:
the masses of the third and the tenth ranked galaxies, and the cluster
turn-around radius.

\table{1}{D}{\bf Table 1. \rm Parameters and global properties of the
$\Omega_0=1$ CDM model cluster catalogue for $\sigma_8$=0.46,
with $H_0$=50 km s$^{-1}$ Mpc$^{-1}$.
Please refer to Section 6.3 for a detailed description of each column.}
{\halign{%
\hskip4pt\hfil#\hskip7pt&\hfil#\hskip7pt&\hfil#\hskip7pt&\hfil#\hskip7pt&\hfil#\hskip7pt&\hfil#\hskip7pt&\hfil#\hskip7pt&\hfil#\hskip7pt&\hfil#\hskip7pt&\hfil#\hskip7pt&\hfil#\hskip7pt&\hfil#\hskip7pt&\hfil#\hskip7pt&\hfil#\hskip7pt&\hfil#\hskip7pt&\hfil#\hskip7pt&\hfil#\hskip14pt \cr
\multispan{17}\hrulefill\cr
\noalign{\vfilneg\vskip -0.4cm}
\multispan{17}\hrulefill\cr
\noalign{\smallskip}
no.&$\nu$ & $x$ & $a_{12}$ & $a_{13}$ &\hfil$M_{\rm G}(\nu,x)$\hfil & $M_{\rm sim}$ & $M_{\rm A}$ & $C_{\rm 3D}$ & $\overline{C_{\rm 3D}}$ & $\sigma_{\rm dm}$ & $\sigma_{\rm gal}$ & $\sigma_{\rm los}$ & $\overline{\sigma_{\rm los}}$ & $m_3$ & $m_{10}$ & $r_{\rm turn}$ \hfil\cr
(a) & \multispan4 &\multispan3 \hrulefill\ [10$^{15}$M$_\odot$]\ \hrulefill \hskip14pt & & & \multispan4 \hrulefill\ [km s$^{-1}$] \hrulefill \hskip14pt & \multispan2 [10$^{12}$M$_\odot$] \hskip7pt & [Mpc] \hfil\cr
\noalign{\smallskip}
\multispan{17}\hrulefill\cr
\noalign{\smallskip}
\input table1a.tex
\noalign{\smallskip}
\multispan{17}\hrulefill\cr
\noalign{\vfilneg\vskip -0.4cm}
\multispan{17}\hrulefill\cr
\noalign{\smallskip}
}
notes: (a) bold face: $C_{\rm 3D}\ge 75$, italics: $C_{\rm 3D}< 50$.
}

\table{2}{D}{\bf Table 1. \rm continued}
{\halign{%
\hskip4pt\hfil#\hskip7pt&\hfil#\hskip7pt&\hfil#\hskip7pt&\hfil#\hskip7pt&\hfil#\hskip7pt&\hfil#\hskip7pt&\hfil#\hskip7pt&\hfil#\hskip7pt&\hfil#\hskip7pt&\hfil#\hskip7pt&\hfil#\hskip7pt&\hfil#\hskip7pt&\hfil#\hskip7pt&\hfil#\hskip7pt&\hfil#\hskip7pt&\hfil#\hskip7pt&\hfil#\hskip14pt \cr
\multispan{17}\hrulefill\cr
\noalign{\vfilneg\vskip -0.4cm}
\multispan{17}\hrulefill\cr
\noalign{\smallskip}
no.&$\nu$ & $x$ & $a_{12}$ & $a_{13}$ &\hfil$M_{\rm G}(\nu,x)$\hfil & $M_{\rm sim}$ & $M_{\rm A}$ & $C_{\rm 3D}$ & $\overline{C_{\rm 3D}}$ & $\sigma_{\rm dm}$ & $\sigma_{\rm gal}$ & $\sigma_{\rm los}$ & $\overline{\sigma_{\rm los}}$ & $m_3$ & $m_{10}$ & $r_{\rm turn}$ \hfil\cr
(a) & \multispan4 &\multispan3 \hrulefill\ [10$^{15}$M$_\odot$]\ \hrulefill \hskip14pt & & & \multispan4 \hrulefill\ [km s$^{-1}$] \hrulefill \hskip14pt &  \multispan2 [10$^{12}$M$_\odot$] \hskip7pt & [Mpc] \hfil\cr
\noalign{\smallskip}
\multispan{17}\hrulefill\cr
\noalign{\smallskip}
\input table1b.tex
\noalign{\smallskip}
\multispan{17}\hrulefill\cr
\noalign{\vfilneg\vskip -0.4cm}
\multispan{17}\hrulefill\cr
\noalign{\smallskip}
}
notes: (a) bold face: $C_{\rm 3D}\ge 75$, italics: $C_{\rm 3D}< 50$.
}

%% file: table1a.tex
{\bf   1} &  3.957 &  3.054 &  1.163 &  1.250 & 2.89 & 3.28 & 1.10 &  76 &  76 & 1265 & 1276 & 1189 & 1201 & 9.53 & 5.60 &12.73 \cr
  2 &  2.986 &  2.085 &  1.206 &  1.706 & 2.14 & 2.02 & 0.41 &  63 &  59 &  974 &  914 &  648 &  611 & 5.49 & 3.61 & 9.32 \cr
  3 &  2.781 &  2.037 &  1.355 &  1.608 & 1.94 & 2.26 & 0.31 &  51 &  43 &  532 &  507 &  518 &  540 & 7.23 & 5.26 & 8.61 \cr
{\bf   4} &  3.842 &  3.338 &  1.206 &  1.706 & 2.69 & 2.67 & 0.78 & 113 & 123 & 1235 & 1279 & 1351 & 1124 & 3.87 & 3.10 &11.38 \cr
  5 &  3.306 &  2.502 &  1.876 &  2.119 & 2.36 & 2.37 & 0.50 &  55 &  57 &  778 &  686 &  643 &  730 & 6.40 & 3.90 &10.31 \cr
{\it   6} &  2.874 &  2.484 &  2.358 &  3.300 & 1.91 & 1.91 & 0.36 &  28 &  33 &  653 &  683 &  688 &  675 & 4.76 & 2.80 & 7.89 \cr
{\bf   7} &  3.517 &  3.195 &  1.227 &  1.427 & 2.39 & 2.42 & 0.77 & 114 & 129 &  896 &  835 &  746 &  837 & 4.03 & 2.99 &11.39 \cr
{\bf   8} &  3.600 &  3.518 &  1.332 &  1.504 & 2.38 & 2.36 & 1.03 & 131 & 128 & 1049 & 1030 &  883 &  985 & 4.74 & 3.52 &11.95 \cr
{\bf   9} &  3.881 &  3.763 &  1.208 &  1.460 & 2.61 & 2.65 & 0.82 & 120 & 127 & 1164 & 1127 & 1095 &  962 & 3.61 & 2.78 &10.97 \cr
{\bf  10} &  3.470 &  3.540 &  1.672 &  1.898 & 2.24 & 2.13 & 0.83 & 139 & 138 & 1048 &  982 &  879 &  903 & 3.86 & 3.05 &10.33 \cr
{\bf  11} &  3.528 &  3.606 &  1.302 &  1.389 & 2.28 & 2.25 & 0.90 & 121 & 118 & 1159 & 1114 & 1084 & 1063 & 4.56 & 3.21 &11.25 \cr
{\bf  12} &  3.601 &  3.252 &  2.294 &  2.710 & 2.46 & 2.31 & 0.58 & 124 & 116 & 1141 & 1101 & 1028 &  973 & 4.20 & 3.47 &10.03 \cr
{\bf  13} &  4.507 &  4.061 &  1.120 &  1.302 & 3.19 & 3.15 & 1.33 & 161 & 150 & 1232 & 1240 & 1149 & 1299 & 4.66 & 3.44 &12.91 \cr
 14 &  3.206 &  2.376 &  2.451 &  2.857 & 2.29 & 2.38 & 0.45 &  58 &  51 &  851 &  797 &  701 &  668 & 5.93 & 4.14 & 9.94 \cr
{\bf  15} &  3.675 &  3.325 &  1.346 &  1.770 & 2.52 & 2.56 & 0.83 & 160 & 136 & 1256 & 1279 & 1274 & 1192 & 3.58 & 2.68 &11.14 \cr
 16 &  3.284 &  2.448 &  1.553 &  2.841 & 2.35 & 2.46 & 0.56 &  69 &  75 &  741 &  770 &  763 &  819 & 5.41 & 4.23 &10.94 \cr
{\it  17} &  2.560 &  1.580 &  1.595 &  3.322 & 1.83 & 1.68 & 0.36 &  46 &  40 &  814 &  843 &  771 &  690 & 7.63 & 4.08 & 8.67 \cr
{\bf  18} &  3.653 &  3.190 &  1.203 &  1.580 & 2.53 & 2.63 & 0.82 &  92 & 104 & 1013 & 1104 & 1054 & 1149 & 4.80 & 3.36 &11.59 \cr
{\it  19} &  3.218 &  2.677 &  1.527 &  2.058 & 2.22 & 2.11 & 0.49 &  44 &  44 &  683 &  573 &  636 &  640 & 9.58 & 4.73 &10.03 \cr
{\it  20} &  3.233 &  3.262 &  1.477 &  1.715 & 2.07 & 1.95 & 0.61 &  46 &  46 &  995 &  993 &  939 &  880 & 8.69 & 4.94 & 9.85 \cr
{\it  21} &  2.828 &  2.036 &  3.610 &  4.762 & 1.99 & 1.76 & 0.40 &  37 &  45 &  810 &  939 &  848 &  728 & 6.36 & 3.01 & 9.00 \cr
{\it  22} &  2.671 &  1.930 &  1.513 &  1.988 & 1.85 & 1.95 & 0.28 &  30 &  41 &  645 &  677 &  857 &  655 & 6.48 & 3.16 & 6.99 \cr
{\bf  23} &  3.629 &  2.829 &  1.083 &  1.359 & 2.61 & 2.65 & 0.63 &  86 &  84 &  981 &  970 &  876 &  828 & 6.21 & 4.57 &11.64 \cr
{\it  24} &  3.064 &  2.520 &  1.560 &  1.709 & 2.10 & 2.05 & 0.41 &  47 &  49 &  900 &  910 &  844 &  764 & 6.68 & 3.45 & 9.81 \cr
{\it  25} &  3.004 &  2.704 &  1.621 &  1.957 & 1.98 & 1.79 & 0.50 &  48 &  46 &  780 &  704 &  651 &  704 & 6.34 & 2.85 & 7.96 \cr
{\bf  26} &  3.538 &  4.184 &  1.110 &  1.272 & 2.13 & 2.11 & 0.69 & 135 & 127 & 1035 & 1042 &  992 &  909 & 3.63 & 2.69 & 9.79 \cr
 27 &  3.081 &  2.907 &  1.264 &  1.497 & 2.01 & 2.02 & 0.52 &  50 &  55 &  832 &  729 &  697 &  677 & 5.60 & 2.74 &10.18 \cr
 28 &  3.521 &  3.392 &  1.242 &  1.704 & 2.34 & 2.32 & 0.65 &  62 &  61 &  913 &  822 &  656 &  696 & 6.97 & 2.93 &11.03 \cr
{\bf  29} &  3.309 &  2.933 &  2.597 &  3.175 & 2.24 & 2.21 & 0.52 & 115 & 108 &  825 &  738 &  646 &  684 & 3.15 & 2.24 & 9.26 \cr
{\bf  30} &  3.768 &  3.838 &  1.052 &  1.368 & 2.47 & 2.10 & 0.92 &  76 &  72 & 1076 & 1015 & 1005 & 1085 & 6.22 & 3.44 &11.16 \cr
{\it  31} &  2.895 &  2.400 &  1.252 &  2.000 & 1.95 & 1.73 & 0.30 &  40 &  46 &  692 &  616 &  622 &  602 & 6.17 & 4.30 & 8.67 \cr
{\bf  32} &  3.929 &  4.345 &  1.312 &  1.570 & 2.50 & 2.48 & 1.20 & 126 & 119 & 1403 & 1353 & 1294 & 1245 & 4.54 & 2.83 &11.93 \cr
{\bf  33} &  3.412 &  3.022 &  1.387 &  1.953 & 2.32 & 2.47 & 0.48 & 105 &  99 &  878 &  792 &  846 &  829 & 4.86 & 3.45 & 9.24 \cr
{\it  34} &  2.751 &  1.489 &  2.193 &  2.933 & 2.06 & 1.77 & 0.25 &  43 &  44 &  672 &  634 &  515 &  586 & 8.89 & 5.24 & 8.85 \cr
 35 &  2.689 &  1.741 &  1.689 &  2.062 & 1.92 & 1.70 & 0.27 &  51 &  48 &  606 &  583 &  556 &  547 & 4.38 & 3.22 & 6.54 \cr
 36 &  2.938 &  2.016 &  1.364 &  1.845 & 2.11 & 1.63 & 0.56 &  53 &  56 &  764 &  746 &  723 &  739 & 7.66 & 4.63 & 8.89 \cr
{\bf  37} &  3.193 &  3.229 &  1.266 &  1.590 & 2.04 & 1.97 & 0.65 &  75 &  77 & 1055 & 1016 & 1003 &  966 & 6.39 & 4.05 & 9.94 \cr
{\it  38} &  2.902 &  2.373 &  1.342 &  1.859 & 1.97 & 1.77 & 0.37 &  44 &  45 &  789 &  938 &  885 &  795 & 5.14 & 2.94 & 9.44 \cr
{\bf  39} &  2.792 &  2.232 &  2.203 &  2.506 & 1.89 & 1.68 & 0.46 &  77 &  71 &  745 &  814 &  727 &  669 & 5.17 & 3.32 & 8.81 \cr
{\it  40} &  2.827 &  2.241 &  1.073 &  1.927 & 1.93 & 1.78 & 0.36 &  38 &  44 &  696 &  698 &  689 &  659 & 5.35 & 2.93 & 8.24 \cr
{\bf  41} &  4.294 &  3.682 &  1.198 &  1.328 & 3.07 & 3.12 & 1.27 & 140 & 148 & 1258 & 1268 & 1201 & 1186 & 5.10 & 3.63 &12.91 \cr
{\bf  42} &  3.442 &  3.195 &  1.085 &  1.479 & 2.31 & 2.20 & 0.65 & 121 & 129 &  849 &  821 &  753 &  765 & 4.20 & 2.51 &10.38 \cr
{\bf  43} &  3.302 &  3.067 &  1.276 &  1.376 & 2.20 & 2.30 & 0.85 &  83 &  73 & 1207 & 1181 & 1164 & 1074 & 7.14 & 4.49 &11.03 \cr
 44 &  2.860 &  2.407 &  1.736 &  2.114 & 1.91 & 1.76 & 0.44 &  57 &  54 &  756 &  669 &  627 &  624 & 7.48 & 4.57 & 8.69 \cr
{\it  45} &  3.345 &  2.957 &  2.506 &  3.096 & 2.27 & 1.99 & 0.43 &  37 &  46 &  694 &  701 &  484 &  579 & 5.16 & 2.49 & 8.95 \cr
{\bf  46} &  3.311 &  2.694 &  1.404 &  1.733 & 2.31 & 2.51 & 0.61 &  79 &  66 & 1213 & 1170 &  934 &  914 & 8.16 & 5.20 &10.27 \cr
 47 &  3.429 &  3.107 &  1.610 &  1.869 & 2.32 & 2.19 & 0.47 &  63 &  73 &  770 &  676 &  706 &  729 & 4.43 & 2.40 & 9.42 \cr
{\bf  48} &  3.431 &  3.331 &  1.145 &  1.250 & 2.26 & 2.15 & 0.87 &  93 &  88 & 1247 & 1171 & 1147 & 1058 & 5.41 & 3.04 &11.14 \cr
 49 &  2.778 &  2.177 &  1.828 &  2.037 & 1.89 & 2.09 & 0.36 &  57 &  63 &  786 &  724 &  730 &  629 & 4.45 & 3.07 & 8.74 \cr
{\it  50} &  2.946 &  1.916 &  2.755 &  3.636 & 2.14 & 2.34 & 0.30 &  47 &  43 &  653 &  572 &  622 &  595 & 5.58 & 4.14 & 9.00 \cr
{\bf  51} &  3.265 &  3.010 &  1.473 &  1.613 & 2.17 & 2.16 & 0.68 & 105 &  91 & 1163 & 1121 &  995 &  956 & 5.77 & 3.82 &10.68 \cr
 52 &  3.229 &  2.938 &  1.845 &  2.141 & 2.15 & 2.18 & 0.47 &  61 &  64 &  828 &  789 &  676 &  782 & 5.67 & 3.25 & 8.93 \cr
{\bf  53} &  3.193 &  2.862 &  1.166 &  1.550 & 2.14 & 2.15 & 0.51 & 102 & 115 &  965 &  965 &  845 &  824 & 4.22 & 2.59 & 9.52 \cr
{\it  54} &  2.980 &  2.239 &  1.055 &  1.490 & 2.09 & 2.11 & 0.38 &  38 &  43 &  636 &  658 &  679 &  616 & 8.18 & 4.02 & 9.43 \cr
 55 &  3.294 &  3.233 &  1.124 &  1.391 & 2.14 & 2.06 & 0.60 &  69 &  79 &  838 &  803 &  701 &  720 & 5.83 & 2.87 &10.42 \cr

%% file: table1b.tex
{\it  56} &  2.923 &  2.554 &  1.490 &  1.789 & 1.94 & 1.89 & 0.39 &  47 &  45 &  763 &  727 &  673 &  689 & 6.08 & 4.11 & 9.31 \cr
{\bf  57} &  3.128 &  2.863 &  1.353 &  1.460 & 2.07 & 2.08 & 0.75 &  90 &  83 & 1182 & 1259 & 1220 & 1045 & 5.37 & 3.62 &10.03 \cr
 58 &  3.145 &  2.362 &  1.389 &  1.637 & 2.23 & 2.46 & 0.63 &  60 &  64 &  956 &  960 &  915 &  892 & 7.65 & 5.76 &10.99 \cr
{\it  59} &  3.247 &  1.978 &  1.739 &  2.020 & 2.44 & 2.67 & 0.49 &  40 &  35 &  839 &  800 &  863 &  757 & 8.50 & 4.28 &10.40 \cr
{\it  60} &  2.980 &  2.564 &  2.558 &  3.030 & 2.00 & 2.31 & 0.36 &  22 &  34 &  632 &  682 &  594 &  614 & 7.14 & 2.24 & 9.02 \cr
 61 &  3.035 &  2.676 &  1.309 &  1.515 & 2.02 & 2.12 & 0.49 &  50 &  50 &  830 &  879 &  710 &  716 & 7.70 & 5.88 & 9.81 \cr
{\it  62} &  2.822 &  1.788 &  1.946 &  2.193 & 2.05 & 1.99 & 0.29 &  42 &  46 &  584 &  590 &  619 &  577 & 4.59 & 2.18 & 8.68 \cr
 63 &  3.348 &  3.592 &  1.151 &  1.323 & 2.10 & 2.12 & 0.70 &  58 &  67 &  933 &  987 &  878 &  950 & 5.13 & 2.38 &10.92 \cr
{\it  64} &  2.971 &  1.758 &  1.350 &  1.497 & 2.21 & 2.34 & 0.28 &  40 &  37 &  605 &  540 &  485 &  551 &10.45 & 4.57 & 9.94 \cr
{\bf  65} &  2.746 &  2.089 &  1.550 &  1.862 & 1.88 & 1.90 & 0.37 &  79 &  53 &  631 &  651 &  634 &  545 & 4.96 & 3.56 & 8.50 \cr
{\it  66} &  3.180 &  2.646 &  1.272 &  1.733 & 2.19 & 2.12 & 0.32 &  37 &  43 &  649 &  757 &  589 &  605 & 6.06 & 2.79 & 9.87 \cr
{\it  67} &  3.031 &  2.504 &  1.161 &  1.502 & 2.07 & 2.09 & 0.45 &  48 &  47 &  722 &  634 &  753 &  705 & 7.11 & 4.07 & 9.59 \cr
{\it  68} &  3.116 &  2.993 &  1.684 &  2.469 & 2.02 & 2.03 & 0.44 &  47 &  50 &  772 &  767 &  749 &  725 & 4.85 & 2.57 & 9.48 \cr
 69 &  2.957 &  2.432 &  1.453 &  1.718 & 2.01 & 1.93 & 0.40 &  57 &  53 &  919 &  979 &  900 &  792 & 5.71 & 4.65 & 9.55 \cr
 70 &  3.018 &  2.034 &  1.196 &  1.783 & 2.19 & 2.41 & 0.34 &  63 &  47 &  735 &  663 &  690 &  687 & 7.52 & 5.38 & 9.39 \cr
{\it  71} &  2.883 &  2.221 &  1.447 &  2.137 & 1.99 & 2.05 & 0.31 &  48 &  50 &  648 &  695 &  542 &  561 & 6.05 & 4.33 & 8.48 \cr
 72 &  3.621 &  2.663 &  2.494 &  3.195 & 2.65 & 2.22 & 0.49 &  51 &  55 &  817 &  688 &  693 &  728 & 6.19 & 3.42 & 8.50 \cr
 73 &  3.153 &  2.657 &  1.148 &  1.560 & 2.15 & 2.21 & 0.49 &  73 &  63 &  751 &  695 &  597 &  653 & 6.91 & 5.34 &10.24 \cr
{\it  74} &  3.157 &  3.253 &  1.403 &  1.616 & 1.99 & 2.05 & 0.49 &  45 &  40 &  735 &  749 &  682 &  737 & 6.68 & 4.61 & 9.75 \cr
 75 &  3.037 &  2.901 &  1.290 &  1.706 & 1.96 & 2.11 & 0.49 &  60 &  48 &  875 &  815 &  795 &  764 & 5.78 & 4.02 & 9.83 \cr
 76 &  3.350 &  4.140 &  1.050 &  1.323 & 1.95 & 2.08 & 0.82 &  66 &  65 &  999 &  932 &  794 &  876 & 7.60 & 4.83 &10.42 \cr
{\it  77} &  3.218 &  3.718 &  1.220 &  1.497 & 1.92 & 2.07 & 0.60 &  28 &  62 &  914 &  903 &  860 &  917 &10.88 & 3.50 &10.27 \cr
 78 &  3.074 &  3.417 &  1.441 &  1.656 & 1.86 & 1.97 & 0.50 &  58 &  60 &  899 &  855 &  895 &  817 & 5.41 & 3.15 & 9.57 \cr
{\bf  79} &  3.149 &  3.416 &  1.337 &  1.842 & 1.94 & 1.95 & 0.58 &  84 &  78 &  722 &  676 &  773 &  833 & 4.18 & 3.34 & 9.55 \cr
 80 &  3.040 &  3.143 &  1.185 &  1.433 & 1.90 & 2.02 & 0.61 &  55 &  56 &  810 &  802 &  812 &  809 & 6.66 & 3.71 &10.11 \cr
{\it  81} &  3.041 &  3.213 &  1.252 &  1.475 & 1.88 & 2.00 & 0.56 &  38 &  38 &  846 &  802 &  727 &  834 & 8.29 & 4.09 & 9.59 \cr
{\it  82} &  3.037 &  3.247 &  1.927 &  2.604 & 1.87 & 1.77 & 0.52 &  43 &  45 &  924 & 1045 &  983 &  883 & 6.77 & 3.79 & 9.44 \cr
{\it  83} &  3.005 &  3.219 &  1.534 &  2.110 & 1.84 & 1.74 & 0.48 &  38 &  43 &  808 &  773 &  692 &  761 & 7.53 & 3.43 & 9.17 \cr
{\it  84} &  2.972 &  3.166 &  2.041 &  2.278 & 1.82 & 1.69 & 0.55 &  45 &  51 &  736 &  662 &  667 &  757 & 6.46 & 4.13 & 8.85 \cr
{\it  85} &  2.949 &  2.853 &  1.582 &  2.242 & 1.88 & 1.99 & 0.26 &  34 &  42 &  610 &  607 &  655 &  614 & 5.69 & 3.12 & 8.63 \cr
{\it  86} &  2.969 &  2.951 &  1.621 &  1.799 & 1.88 & 1.77 & 0.40 &  49 &  55 &  890 &  858 &  812 &  817 & 7.39 & 4.32 & 8.32 \cr
{\it  87} &  2.983 &  2.998 &  1.202 &  1.370 & 1.88 & 1.85 & 0.45 &  37 &  34 &  746 &  781 &  657 &  646 & 7.25 & 4.27 & 9.07 \cr
 88 &  2.992 &  2.941 &  1.513 &  1.653 & 1.90 & 2.05 & 0.64 &  60 &  64 & 1044 & 1078 & 1092 &  932 & 6.68 & 3.81 & 9.85 \cr
{\it  89} &  2.918 &  2.681 &  1.661 &  2.506 & 1.90 & 1.85 & 0.34 &  48 &  50 &  642 &  667 &  633 &  667 & 5.62 & 2.62 & 8.56 \cr
 90 &  3.003 &  2.863 &  1.838 &  2.028 & 1.94 & 1.98 & 0.49 &  61 &  53 &  829 &  728 &  725 &  736 & 7.49 & 5.32 & 9.59 \cr
 91 &  2.899 &  2.691 &  1.364 &  1.931 & 1.88 & 2.03 & 0.27 &  51 &  37 &  528 &  508 &  494 &  503 & 8.69 & 6.33 & 7.06 \cr
 92 &  2.859 &  2.692 &  1.499 &  1.980 & 1.83 & 1.80 & 0.49 &  71 &  44 &  851 &  839 &  809 &  731 & 8.34 & 5.33 & 8.37 \cr
{\it  93} &  2.823 &  2.645 &  1.608 &  1.776 & 1.81 & 1.93 & 0.39 &  28 &  45 &  752 &  736 &  629 &  680 & 5.94 & 2.34 & 8.65 \cr
 94 &  2.823 &  2.523 &  1.754 &  2.066 & 1.84 & 1.95 & 0.32 &  69 &  61 &  640 &  676 &  505 &  553 & 4.89 & 3.65 & 7.26 \cr
{\it  95} &  2.794 &  2.401 &  1.267 &  1.715 & 1.85 & 1.90 & 0.41 &  46 &  46 &  882 &  866 &  773 &  722 & 6.24 & 3.96 & 8.93 \cr
{\it  96} &  2.660 &  1.955 &  1.416 &  2.747 & 1.83 & 1.71 & 0.18 &  43 &  43 &  498 &  463 &  473 &  474 & 4.93 & 2.52 & 6.18 \cr
{\it  97} &  2.646 &  2.076 &  1.475 &  1.912 & 1.78 & 1.77 & 0.15 &  24 &  30 &  561 &  556 &  639 &  545 & 7.67 & 3.45 & 9.13 \cr
 98 &  2.611 &  1.957 &  2.532 &  2.933 & 1.78 & 1.85 & 0.22 &  51 &  44 &  474 &  447 &  422 &  499 & 8.06 & 4.92 & 6.66 \cr
{\it  99} &  2.540 &  1.643 &  1.634 &  1.832 & 1.79 & 2.05 & 0.23 &  33 &  33 &  645 &  682 &  630 &  553 & 5.63 & 2.13 & 8.02 \cr

%% file: cat7.tex
\section{Some testable properties of the $\Omega_0=1$ CDM catalogue}

\newcount\Fvlosfunc
\Fvlosfunc=13
\figps{13}{S}{\psfig{file=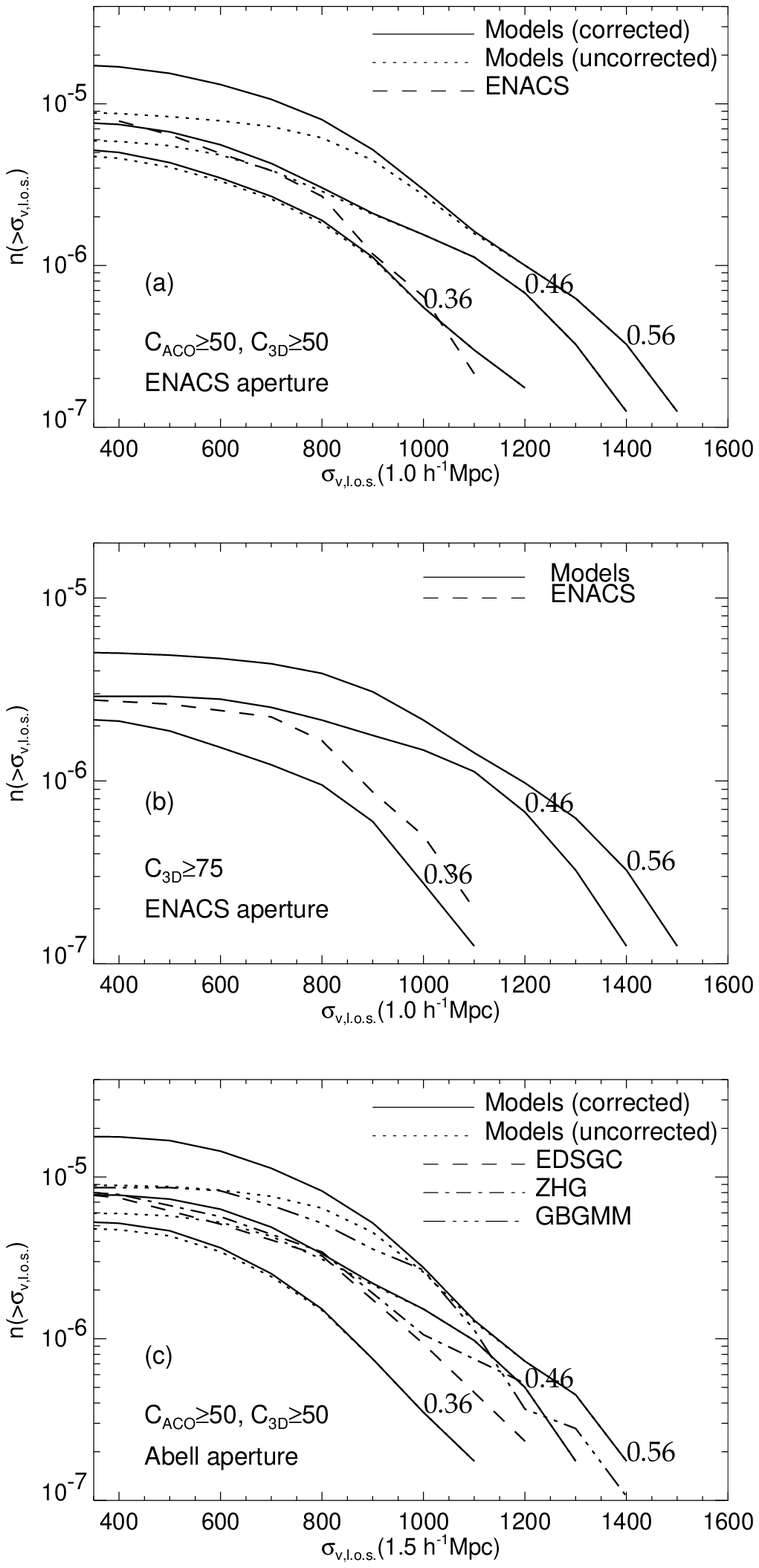,width=8.5cm,silent=}}
{{\bf Figure 13.} (a) Cumulative distributions of the line-of-sight velocity
dispersion within 1.0 $h^{-1}$Mpc, normalised to the average cluster density.
The dashed line denotes the observed distribution for the ENACS sample.
Numbers denote the value of $\sigma_8$ for each model curve.
(b) same as (a), but for dispersion within 1.5 $h^{-1}$Mpc.
The non-solid linestyles represent the distribution for various observed
samples.
(c) same as (a), but for the complete subsample of $C_{\rm 3D}\ge75$ clusters.}

\subsection{Cumulative distribution of line-of-sight velocity dispersions}

\tx The velocity dispersion of a galaxy cluster generally increases
with time for most cosmological scenarios. This makes it a useful
quantity for testing the timing of our CDM $\Omega=1$ model catalogue.
We compare to the ENACS data (Mazure et al. 1996)
and observations by Collins et al.\ (1995), taken in the context of the
Edinburgh-Milano Cluster Redshift Survey (EDSGC for short).
Both these samples are relatively complete
and should provide the best comparison, especially the ENACS sample
because we used exactly the same data analysis procedures.
We also consider the samples of Zabludoff, Huchra \& Geller (1990,
ZHG from here on) and Girardi et al. (1993, GBGMM from here on), although
both of these are not drawn from a complete sample. The ZHG sample might
be complete but is rather small (25 clusters), which makes completeness
hard to establish, while the GBGMM sample is biased towards richer
clusters (although an attempt has been made to correct for this).

Clearly we have to use the line-of-sight velocity dispersion, denoted by
$\sigma_{\rm l.o.s.}$, for a comparison to observations.
There are quite a few methods for obtaining line-of-sight
velocity dispersions from observational data, all dealing with the
removal of foreground and background galaxies, and `interlopers',
unbound galaxies near the cluster turn-around radius. 
With these galaxies removed, the velocity dispersions found are
generally smaller than obtained from the contaminated velocity
distribution (e.g.\ Lucey 1983; Frenk et al.\ 1990;
den Hartog \& Katgert 1996; Mazure et al.\ 1996).

However, as discussed in Section 5.1.1, in our models we do not have any
foreground or background galaxies due to the limited simulation volume.
With respect to removing interlopers, for the most conservative schemes,
like `$3\sigma$-clipping', we find very few interlopers in our models.
For more strict (but physically motivated) interloper removers,
like those proposed by den Hartog \& Katgert (1996), we indeed find
quite a few more, and more importantly, in roughly the same numbers as
for observed clusters. Furthermore, the interlopers that are removed
are indeed outside the turn-around radius, justifying the method
of den Hartog \& Katgert ({\it ibid.}).

Because EDSGC, ZHG and GBGMM use fairly conservative interloper removers, 
we do not need to remove any galaxies from our models for a comparison
to these samples. For a match to the ENACS clusters, we
apply the interloper remover of den Hartog \& Katgert ({\it ibid.}),
and also use the smaller aperture of 1.0$h^{-1}$Mpc instead of the Abell
radius used for the other samples. This smaller aperture results in
slightly larger line-of-sight velocity dispersions, while the interloper
remover used for the ENACS sample produces significantly smaller
dispersions. 

\subsubsection{Correction for incompleteness}

\tx Before we can compare our line-of-sight velocity dispersions to 
those observed, we again need to correct for incompleteness in $C_{\rm 3D}$.
We use a similar method as that used to correct the richness distributions
(see Section 5.3), but now using the $\sigma_{\rm l.o.s.}$-$M_{\rm G}$
relation. We fit an exponential function to the
($M_{\rm G}$,$\sigma_{\rm l.o.s.}$) pairs
obtained from the simulations, and calculate the scatter of the residuals
for several bins in $M_{\rm G}$. 
We then sample the velocity dispersions from a normal distributed with
mean and variance given by these two fits. As we use the same $M_{\rm G}$'s
as in the correction procedure for richness $C_{\rm 3D}$, we can link
them and obtain ($C_{\rm 3D}$,$\sigma_{\rm l.o.s.}$) pairs. Adding these to
the ones taken directly from the simulations gives us a complete distribution
over line-of-sight velocity dispersions.

\subsubsection{Comparison to the ENACS catalogue}

\tx The corrected $\sigma_{\rm l.o.s.}$ distributions are plotted as
solid lines in Fig.\ \the\Fvlosfunc(a) for several $\sigma_8$'s, along
with the uncorrected ones (dotted lines), and the
observed distribution found for the ENACS sample (dashed line).
We find that the observed distribution matches the modelled one for
$\sigma_8\approx 0.35-0.45$, but has a slightly different slope.

The second comparison we make is that of the complete subsample of
$C_{\rm 3D}\ge75$ clusters to a similar complete subsample of the
ENACS sample. Note that we do not have to apply any incompleteness
corrections here, as both sample are complete for this richness limit.
The model distributions for $\sigma_8\approx0.4$ matches
the observed one best, as shown in Fig.\ \the\Fvlosfunc(b). The shapes of
the curves compare better than for the $C_{\rm 3D}\ge50$ sample.

\subsubsection{Comparison to other observed samples}

\tx We can do the same comparison for the EDSGC, GBGMM, and ZHG samples,
although only the first of these is really complete in richness.
The line-of-sight velocity dispersions within the Abell radius for
several values for $\sigma_8$ are plotted in Fig.\ \the\Fvlosfunc(c),
along with the observed distributions. Please note that the interloper
removal method of den Hartog \& Katgert (1996) was {\it not}\ used here,
as it was not used for any of these observed distributions either,
and also note that the Abell aperture was adopted.

For the EDSGC sample we find that the model curve fits best for
$\sigma_8\approx0.4$. The other two samples prefer much higher values, but
are not complete and biased to the richest clusters, and therefore
to higher velocity dispersions.

\subsection{Other observables}

\tx To test the chosen scenario one can use many other observables besides
the one considered above. For example, the line-of-sight velocity dispersion
{\it profiles} might prove useful, although unfortunately at present these
can only be obtained for a few bins.
However, a classification derived from this binned profile can be very
instructive (den Hartog \& Katgert 1996).

Another distribution function is that of the projected shapes of clusters.
In a preliminary comparison presented elsewhere (de Theije, Katgert \& van
Kampen 1995) it was found that the predicted distribution of
shape parameters of model clusters differs significantly from the observed
one. Although the observed sample used in the study of de Theije et al.\
({\it ibid.}) is not complete, the deviations are unlikely to be due to
incompleteness.
For a subset of our model catalogue that was rerun for the $\Omega=0.2$ CDM
scenario (in future work we will build complete catalogues for this and
other scenarios) there seems to be better agreement between models and data.

%% file: cat8.tex
\section{Summary and discussion}

\tx The aim of this paper was to present a technique that produces a
fair catalogue of individually simulated high-resolution cluster models.
We used this technique to build one specific model catalogue that we
compared to observed catalogues.

Under the assumption that rich clusters of galaxies originate from
peaks in the initial density field Gaussian smoothed at $4h^{-1}$Mpc,
we select a catalogue on the basis of characteristic parameters of
these peaks. Given the fact that we need to use the constrained
random field method to generate initial conditions for the individual
numerical models, we need to select the catalogue on the basis of
linear functionals of the initial density field. With this in mind,
we argued that cluster mass is the best catalogue defining quantity,
because it can be predicted reasonably well from a linear function of
the amplitude {\it and}\ the curvature of the initial density peak
which is the progenitor of the cluster. Besides this practical argument,
total cluster mass {\it is}\ a basic property of a galaxy cluster.

For the individual models a numerical code was used that contains a
prescription to form galaxies, and also allows galaxies already formed
to grow (i.e.\ accrete particles) and merge with other galaxies.
Not only does this allow us to directly compare to optical catalogues of
galaxy clusters, but the properties of the galaxy population also directly
set the present epoch for the cosmological scenario in which the model
catalogue is embedded, i.e.\ they set the amplitude of the initial
cosmological fluctuation spectrum.
Not all cosmological scenarios allow this, but the scenario we chose,
viz.\ $\Omega_0=1$ CDM, does permit this {\it a posteriori}\
normalisation and also makes it possible to compare our results to
published numerical work.

We used a traditional, single large-scale simulation in order to test
several issues related to the construction of the catalogue and the
reliability of the individual models. We traced peaks in these
simulations, and found that they do not merge during the entire run.
So we indeed have a one-to-one mapping of initial peaks to final rich
Abell clusters.
Note that this may not be so in other cosmological scenarios.
We also found that both the distribution and the evolution of the
peak parameters in the single large-scale low-resolution simulation
are similar to those of the set of individual high-resolution simulations.
Finally we found that the expected mass relation, a linear function
of peak amplitude and peak curvature, is almost equal to that obtained
from the clusters extracted from the single simulation.

Having built a catalogue selected on expected final cluster mass, we
found that is was 70 per cent complete for $C_{\rm 3D}\ge 50$. 
This incompleteness mostly concerns the low-mass end and is mainly due
to the noise still present in the relation between final cluster mass and
initial peak parameters. But note that the richness measure itself has
a rather large intrinsic noise originating from its very definition. 
To allow a comparison of statistical distribution functions of model
clusters properties to those observed we devised a method to corrected
for this incompleteness. However, we do have a complete subsample for
$C_{\rm 3D}\ge 75$, which permits a fully unbiased comparison to
observations.

We use the richness distribution of both samples to set the present
epoch in the models, i.e.\ fix $\sigma_8$. A value of 0.46 for
$\sigma_8$ matches best; this is also in the range of allowed values
derived from a match of the galaxy autocorrelation function for a set
of field models similar to the clusters models discussed here to
the observed autocorrelation function (van Kampen 1996).
The elimination of the free parameter $\sigma_8$ of the $\Omega_0=1$
CDM scenario allows us to test it with other measures.

We find that the cumulative distribution of line-of-sight velocity
dispersions obtained for the ENACS sample (Katgert et al.\ 1996),
and for other samples as well, fits the model
distribution best for $\sigma_8\approx 0.4$. This is fairly
consistent with both the auto-correlation function and the richness
distribution. We can obtain a slightly larger $\sigma_8$ if we invoke a 
{\it velocity}\ bias, but this is not seen in our models
if we compare the intrinsic velocity dispersions of the dark matter
to those of the galaxies (van Kampen 1994). Other studies, like those
of Cen \& Ostriker (1992) and Katz {\it et al.}\ (1992) do not find
a significant velocity bias either.

However, there are also problems with the standard CDM scenario. We mention
two major ones here (see Ostriker 1993 for a more extensive discussion).
De Theije et al.\ (1995) compared the distribution of cluster shapes for our
catalogue to an observed sample, and found that these differ significantly.
The results from {\it COBE}\ imply that $\sigma_8\approx 1.3$ for our
chosen scenario (Bunn et al.\ 1995). Note however that {\it COBE} traces
length-scales well outside the dynamical range of the individual models.
We will need to investigate whether these inconsistencies remain for
other tests before we completely discard the $\Omega_0=1$ CDM scenario
{\it on the scales we studied here}.
It is still useful to test more of the statistical properties
of this scenario, and use successes and failures of such tests as a
guide for a new choice for the most likely cosmological scenario. 
Of the scenarios plotted in Fig.\ \the\Fspectra, we should take a closer
look at the $k^{-2}$ spectrum, or low-$\Omega_0$ CDM, as both of these have
more power on larger scales and therefore are in better agreement with the
{\it COBE}\ results.

\section*{ACKNOWLEDGEMENTS}

\tx We would like to thank Roland den Hartog, Rien van de Weygaert and
Tim de Zeeuw for many useful suggestions and discussion. 
Joshua Barnes and Piet Hut are gratefully acknowledged for allowing use of
their treecode, Edmund Bertschinger and Rien van de Weygaert for providing
their constrained random field code and Edmund Bertschinger for his P$^3$M 
code, Erik Deul for allowing us to use the computer systems that are
part of the DENIS project, and the ENACS collaboration for allowing us
to use their data before publication.
EvK acknowledges EelcoSoft Software Services for partial financial support,
and an European Community Research Fellowship as part of the HCM programme.

%% file: catrefs.tex
\section*{REFERENCES}

\bibitem Abell G.O., 1958, ApJS, 3, 211
\bibitem Abell G.O., Corwin H.G., Olowin R.P., 1990, ApJS, 70, 1
\bibitem Adler R.J., 1981, The Geometry of Random Fields. Wiley, Chichester.
\bibitem Aquilar L.A., White S.D.M., 1985, ApJ, 295, 374
\bibitem Bardeen J.M., Bond J.R., Kaiser N., Szalay A.S., 1986, ApJ, 304, 15
\bibitem Barnes J.E., Hut P., 1986, Nat, 324, 446
\bibitem Barnes J.E., Hernquist L., 1992, ARAA, 30, 705
\bibitem Beers T.C., 1992, in Feigelson E.D, Babu G.J., eds., Statistical
	Challenges In Modern Cosmology. Springer-Verlag, New York, Inc., pp. 111-135
\bibitem Bertschinger E., 1987, ApJ, 323, L103
\bibitem Bertschinger E., Jain B., 1994, ApJ, 431, 495
\bibitem Bertschinger E., 1993, in Akerlof C.W., Srednicki M.A., eds.,
	Texas/PASCOS 92: Relativistic Astrophysics and Particle Cosmology.
	The New York Academy of Sciences, New York, p. 297
\bibitem Bertschinger E., Gelb, J.M., 1991, Computers in Physics, 5, 164
\bibitem Binggeli B., 1982, A\&A, 107, 338
\bibitem Bunn E.F., Scott D., White M., 1995, ApJ, 441, L9
\bibitem Carlberg R.G., 1994, ApJ, 433, 468
\bibitem Cavaliere A., Santangelo P., Tarquini G., Vittorio N., 1986, ApJ, 305, 651
\bibitem Cen R., Ostriker J.P., 1992, ApJL, 399, L113
\bibitem Collins C.A., Guzzo L., Nichol R.C., Lumsden S.L., 1995, MNRAS, 274, 1071
\bibitem Dalton G.B., Efstathiou G., Maddox S.J., Sutherland W.J., 1992, ApJ,
390, L1
\bibitem Davis M., Efstathiou G., Frenk C.S., White S.D.M., 1985, ApJ, 292, 371
\bibitem Davis M., Efstathiou G., Frenk C.S., White S.D.M., 1992, Nat, 356, 489
\bibitem Davis M., Peebles P.J.E., 1983, ApJ, 267, 465
\bibitem den Hartog R., 1995, Ph.D. thesis, Rijksuniversiteit Leiden
\bibitem den Hartog R., Katgert P., 1996, MNRAS, 279, 349
\bibitem de Theije P.A.M., Katgert P., van Kampen E., 1995, MNRAS, 273, 30
\bibitem de Theije P.A.M., van Kampen E., Slijkhuis R., submitted to MNRAS
\bibitem de Zeeuw P.T., Franx M., 1991, ARAA, 29, 239 
\bibitem Doroshkevich A.G., 1970, Astrophysica, 6, 320
\bibitem Efstathiou G., 1990, in Peacock J.A., Heavens A.F., Davies A.T., eds.,
	Physics of the Early Universe. SUSSP publications, Edinburgh, p. 361
\bibitem Eke V.R., Cole S., Frenk C.S., 1996, MNRAS, 282, 263
\bibitem Evrard A.E., 1989, ApJ, 341, L71
\bibitem Evrard A.E., Mohr J.J., Fabricant D.G., Geller M.J., 1993, ApJ, 419, L9
\bibitem Franx M., 1988, MNRAS, 231, 285
\bibitem Frenk C.S., White S.D.M., Davis, M., Efstathiou G., 1988, ApJ, 327, 507
\bibitem Frenk C.S., White S.D.M., Efstathiou G., Davis, M., 1990, ApJ, 351, 10
\bibitem Forman W., Jones C., 1990, in Oegerle W.R., Fitchett M.J., Danly L.,
eds., Clusters of Galaxies. Cambridge University Press, Cambridge, p. 257
\bibitem Fort B., Mellier Y., 1994, A\&AR, 5, 239
\bibitem Girardi M., Biviano A., Guiricin G., Mardirossian F., Mezzetti M., 1993,
ApJ, 404, 38
\bibitem Giovanelli R., Haynes M.P., 1991, ARA\&A, 29, 499
\bibitem Hoffman Y., Ribak E., 1991, ApJ, 380, L5
\bibitem Kaiser N., 1984, ApJ, 284, L9
\bibitem Kaiser N., 1987, MNRAS, 227, 1
\bibitem Katgert P., Mazure A., Jones B., den Hartog R., Biviano A., Dubath P.,
Escalera E., Focardi P., Gerbal D., Giuricin G., Le F\`evre O., Moles O.,
Perea J., Rhee G., 1996, A\&A, 310, 8
\bibitem Katz N., Hernquist L., Weinberg D.W., 1992, ApJL, 399, L109
\bibitem Kent S.M., Gunn J.E., 1982, AJ, 87, 945
\bibitem Kolb E.W., Turner M.S., 1990, The Early Universe. Addison-Wesley,
  Reading CA
\bibitem Leir A.A., van den Bergh S., 1977, ApJS, 34, 381
\bibitem Lucey J.R., 1983, MNRAS, 204, 33
\bibitem Lumsden S.L., Nichol R.C., Collins C.A., Guzzo L., 1992, MNRAS, 258, 1
\bibitem Mazure A., Katgert P., den Hartog R., Biviano A., Dubath P.,
Escalera E., Focardi P., Gerbal D., Giuricin G., Jones B., Le F\`evre O., Moles O.,
Perea J., Rhee G., 1996, A\&A, 310, 31
\bibitem McGill C., 1991, MNRAS, 250, 340
\bibitem Mo H.J., Jing Y.P., B\"orner G., 1993, MNRAS, 264, 825
\bibitem Ostriker J.P., 1993, ARA\&A, 31, 689
\bibitem Peacock J.A., Heavens A.F., 1985, MNRAS, 217, 805
\bibitem Peebles P.J.E., 1974, A\&A. 32, 197
\bibitem Peebles P.J.E., 1990, ApJ, 365, 27
\bibitem Peebles P.J.E., 1993, Principles of Physical Cosmology.
Princeton University Press, Princeton
\bibitem Press W.H., Flannery B.P., Teukolsky S.A., Vetterling W.T., 1988,
	Numerical Recipes: The Art of Scientific Computing.
	Cambridge University Press, Cambridge.
\bibitem Press W.H., Schechter P., 1974, ApJ, 187, 425
\bibitem Rhee G.F.R.N., van Haarlem M.P., Katgert P., 1989, A\&AS, 91, 513
\bibitem Rice, S.O., 1954, in Wax N., ed., Selected papers on Noise and
	Stochastic Processes. Dover.
\bibitem Sarazin C.L., 1986, Rev Mod Phys, 58, 1
\bibitem Spitzer Jr.\ L., 1969, ApJ, 158, L139
\bibitem Suginohara T., Suto Y., 1992, ApJ, 396, 395
\bibitem van de Weygaert R., Bertschinger E., 1996, MNRAS, 281, 84
\bibitem van Haarlem M.P., Frenk C.S., White S.D.M., 1996, MNRAS, in press
\bibitem van Haarlem M.P., van de Weygaert R., 1993, ApJ, 418, 544
\bibitem van Kampen E., 1994, Ph.D. thesis, Rijksuniversiteit Leiden
\bibitem van Kampen E., 1995, MNRAS, 273, 295
\bibitem van Kampen E., 1996, submitted to MNRAS
\bibitem West M.J,, Oemler Jr. A., Dekel A., 1988, ApJ, 327, 1
\bibitem White S.D.M., Davis M., Efstathiou G., Frenk C.S., 1987, Nat, 330, 451
\bibitem White S.D.M., Efstathiou G., Frenk C.S., 1993, MNRAS, 262, 1023
\bibitem Zabludoff A.I., Huchra J.P., Geller M.J., 1990, ApJS, 74, 1
\bibitem Zurek W.H., Warren M.S., Quinn P.J., Salmon J.K., 1993, in Bouchet F.R.,
Lachi\`eze-Rey M., eds., Cosmic Velocity Fields. Edition Fronti\`eres,
Gif-sur-Yvette, p. 465

%% file: catapps.tex
\section*{Appendix A: Expressions from the theory of Gaussian random fields}

\eqnumber=1

\tx In this Appendix we summarize some functions taken from Bardeen
et al.\ (1986; BBKS from here on), slightly rewritten for our purposes,
as used in Section 2. The first function is a factor in the probability
distribution for the peak height $\nu$, which depends
on the spectral parameters $\gamma$ and $R_{\rm G}$.
It depends on $\gamma$ and $\nu$ only:
$${\cal F}_1 (\gamma, \nu) = 
   {\gamma^3\nu^3-3\gamma^3\nu+[B(\gamma)\nu^2+C_1(\gamma)] e^{-A(\gamma)\nu^2}
   \over 1+C_2(\gamma)e^{-C_3(\gamma)\nu}} \eqno(A\neweq)$$
(from equation (4.4) of BBKS), with 
$$A(\gamma)={5\gamma^2\over(18-10\gamma^2)}\ ,\ \
  B(\gamma)={432\gamma^2\over \sqrt{10\pi}(9-5\gamma^2)^{5/2}} \eqno(A\neweq)$$
and the fits
$$\eqalign{C_1(\gamma)=&1.84+1.13(1-\gamma^2)^{5.72}\cr
           C_2(\gamma)=&8.91+1.27e^{6.51\gamma^2}\cr
	   C_3(\gamma)=&2.58\gamma e^{1.05\gamma^2}\ .\cr }\eqno(A\neweq)$$
The second function is used for the curvature distribution function:
$$\eqalign{{\cal F}_2 & (x) = {(x^3-3x)\over 2} \Bigl[
  {\rm Erf}\Bigl(\sqrt{5\over 2}\ x\Bigr)+
  {\rm Erf}\Bigl(\sqrt{5\over 8}\ x\Bigr)\Bigr]+\cr
  &{\sqrt{2\over 5\pi}}\Bigl[\ \Bigl({31x^2\over4}+{8\over5}\Bigr)e^{-{5x^2\over8}}+
  \Bigl({x^2\over 2}-{8\over 5}\Bigr)e^{-{5x^2\over2}} \Bigr].\cr}\eqno(A\neweq)$$ 
The next two functions are used for the asphericity probabilities:
$${\cal F}_3 (a_{12},a_{13}) = 
 {27 (a_{12}^2-1) (a_{13}^2-1)(a_{13}^2-a_{12}^2) \over
  2 a_{12}^{-2} a_{13}^{-2} (1+a_{12}^2+a_{13}^2)^6 } \eqno(A\neweq) $$
and
$${\cal F}_4 (a_{12},a_{13}) =
  {a_{13}^4+a_{12}^4-a_{12}^2-a_{13}^2-a_{12}^2a_{13}^2+1 \over
  (1+a_{12}^2+a_{13}^2)^2}\ . \eqno(A\neweq)$$
Finally we list the function used to describe the asphericity of peaks in the
second order Taylor expansion, which is a function of the spherical coordinate
angles $\theta$ and $\phi$ and axial ratios $a_{12}$ and $a_{13}$:
$$\eqalign{{\cal F}_5 & (a_{12},a_{13},\theta,\phi) =
	2(2a_{13}^2-1) - (2a_{12}^2-1) + \cr
 & 3\sin^2\theta\{ \sin^2\phi - (2a_{13}^2 - 1) + (2a_{12}^2 -1) \cos^2\phi \}\ .\cr}
  \eqno(A\neweq)$$
This is an adaptation from eq.\ (7.4) of BBKS.
It is equal to unity for spherical peaks.

We draw from the probability distributions (\the\Npeaks) and (\the\Psize) by
using the rejection method. The comparison function needed is found by
substituting $\nu^3$ for ${\cal F}_1(\gamma, \nu)$ in (\the\Npeaks).
Its integral is
$$ {V \over (2\pi)^2 R_*^3} \Bigl[2-(2+\nu^2)e^{-\nu^2/2}\Bigr] \eqno(A\neweq)$$
The comparison function for drawing $x$, $f_{\rm c}(x,\mu,\sigma)$, is obtained
by replacing ${\cal F}_2(x)$ with $x^3$, where we have defined $\mu=\gamma\nu$ and
$\sigma=\sqrt{1-\gamma^2}$. The integral $F_{\rm c}(x,\mu,\sigma)$ of the comparison
function is
$$\eqalign{F_{\rm c}(x,\mu,\sigma)=&-{{{\sigma^2}\,\left({\mu^2}+2\,{\sigma^2}+\mu\,x+
    {x^2}\right)}{{e^{-{{{{\left(x-\mu\right) }^2}}\over {2\,{\sigma^2}}}}}}} \cr
  &+ \mu\,\sigma{\sqrt{{{\pi}\over 2}}}\,\left( {\mu^2} + 3\,{\sigma^2} \right) \,
    {\rm Erf}\bigl({{x-\mu}\over {{\sqrt{2}\,\sigma}}}\bigr)\ .\cr}\eqno(A\neweq)$$
An approximate formula for the expectation value of $x$ for a given $\nu$ is
given by eq.\ (6.14) of BBKS:
$$\langle x \rangle = \gamma \nu + 
	{3(1-\gamma^2)+(1.216-0.9\gamma^4)e^{-\gamma^3\nu^2/8} \over
	[3(1-\gamma^2)+0.45+(\gamma\nu/2)^2]^{1/2}+\gamma\nu/2} \eqno(A\neweq)$$
The expectation values for $a_{12}$ and $a_{13}$ are functions of
$y\equiv(6+5x^2)^{-1/2}$:
$$\eqalign{\langle a_{12} \rangle =& \sqrt{1-60y^4\over 1-3y+30y^4} \cr
	   \langle a_{13} \rangle =& \sqrt{1+3y-30y^4\over 1-3y+30y^4}\ .\cr }
	\eqno(A\neweq)$$


\section*{Appendix B: Derivation of the predicted mass relation}

\eqnumber=1

\tx In this appendix we show how the curvature of the initial peak, $x$, comes
into play in the non-linear evolution of total cluster mass. We start from the
initial density profile, given by (\the\BBKSprofile):
$$\delta_{\rm G}(r)={\nu-\gamma x\over\sigma_0(1-\gamma^2)}\xi_{\rm G}(r)
   +{\nu-x/\gamma\over3\sigma_0(1-\gamma^2)}\nabla^2\xi_{\rm G}(r)\ .
   \eqno(B\neweq)$$
We neglect non-linear evolution of $\sigma_0$, so that $\gamma$ remains
constant ($\sigma_1$ and $\sigma_2$ remain well below unity). We
assume that the autocorrelation function is a power law with index $n$,
which evolves as $n(t)=n(t_{\rm i})\epsilon(t)$, where $\epsilon(t)$ is
unity at $t=t_{\rm i}$ and increases slowly. This gives us approximately
$$\eqalign{\nabla^2\xi_{\rm G}(0,t)
	\approx & {\epsilon(t)(n(t_{\rm i})\epsilon(t)+1)
	\over n(t_{\rm i})+1}a^2(t) \nabla^2\xi_{\rm G}(0,t_{\rm i}) \cr
     = & -3{\epsilon(t)(n(t_{\rm i})\epsilon(t)+1)\over n(t_{\rm i})+1}
    \gamma^2\sigma_0^2(t)\ ,\cr} \eqno(B\neweq)$$
where we have used eq.\ (\the\initnabla) for the second part, and thus
\newcount\evoprofile
\evoprofile=\eqnumber
$$\delta_{\rm G}(0,t)=\Bigl[{\nu(t)-\gamma x(t)\over (1-\gamma^2)}
	+ \eta(t)\gamma{x(t)-\gamma\nu(t)\over(1-\gamma^2)}\Bigr]
	\sigma_0(t) \ \eqno(B\neweq)$$
with
$$\eta(t)\equiv{\epsilon(t)[n(t_{\rm i})\epsilon(t)+1]\over n(t_{\rm i})+1} \ ,
  \eqno(B\neweq)$$
which monotonically increases (slowly) from its initial value
$\eta(t_{\rm i})=1$. The second term in eq.\ ($B$\the\evoprofile)\
evolves somewhat faster than the first because of the
increasing slope $n(t)$ of the autocorrelation function.
This might seem unimportant, but the point is that the $x$ terms,
which grow relatively fast (as clearly seen in Fig.\ \the\Fnux),
do not cancel anymore because of the different growth rates of the
two terms in eq.\ ($B$\the\evoprofile).

Rearranging the terms, we find
$$\delta_{\rm G}(0,t)={\nu(t)\bigl[1-{\gamma^2\eta(t)}\bigr]
	-\gamma x(t)\bigl[1-\eta(t)\bigr] \over(1-\gamma^2)}
	\sigma_0(t)\ . \eqno(B\neweq)$$
We can rewrite this in terms of growth factors for $\nu$ and $x$:
$$\delta_{\rm G}(0,t)=\bigl\{g_\nu(t)\nu_{\rm i} + g_x(t)x_{\rm i} \bigr\}
	\sigma_0(t_0)a(t)\ . \eqno(B\neweq)$$
with $\nu_{\rm i}\equiv \nu(t_{\rm i})$, $x_i\equiv x(t_{\rm i})$, and
$$g_\nu(t) \equiv {\nu(t)\over\nu_{\rm i}}
	\Bigl[{1-\gamma^2\eta(t)\over 1-\gamma^2}\Bigr] \eqno(B\neweq)$$
and
$$g_x(t) \equiv {x(t)\over x_{\rm i}}
	\Bigl[{\gamma\eta(t)-\gamma\over 1-\gamma^2}\Bigr] \ . \eqno(B\neweq)$$
The CDM spectrum smoothed on 4$h^{-1}$Mpc has an autocorrelation function
with an initial slope of $n\approx-1.5$ for the range of wavenumbers in our
simulations (Efstathiou 1990). As an example of what the final relation
will look like we assume that $\eta(t_0)\approx 1.2$.
Because $\gamma(4h^{-1}{\rm Mpc})=0.729$ for CDM, we get
$$\delta_{\rm G}(t_0) \approx 0.8 {\nu(t)\over \nu_{\rm i}} + 
	0.2 {x(t)\over x_{\rm i}}\ . \eqno(B\neweq)$$
The peak amplitude $\nu(t)$ just grows, but the evolution of the curvature
$x(t)$ is more complicated. Van Haarlem \& van de Weygaert (1993) performed
cluster simulations with constraints set on the same smoothing scale, and
found that a small initial $x_{\rm i}$ results in a larger final $x(t_0)$ than a
larger initial $x_{\rm i}$. For three simulations with $\nu_{\rm i}=3$ and
$x_i$'s corresponding to the 5, 50 an 95 percentiles of
$P(x|3\sigma_0,4h^{-1}{\rm Mpc})$, a linear relation provides a good fit to
their results:
\newcount\vdWvH
\vdWvH=\eqnumber
$$x(t_0) \approx 27 - 4.6 x_{\rm i}\ . \eqno(B\neweq)$$
We should note that both constants are present epoch values of unknown functions
of time, but here we just look for a relation between initial and final values.
This relation is obtained for $\nu_{\rm i}=3$ only, but since that is
roughly the average value for our model cluster sample, a linear relation
for $x(t_0)$ seems to be a good approximation for all $\nu_{\rm i}$.

We now assume a linear relation for $\nu(t_0)$ as well, i.e.\
$\nu(t_0) \approx a\nu_{\rm i} + b$, where again the fitting parameters
are functions of $t_0$. This linear relation is a good
approximation because we only consider a relatively small range of
$\nu_{\rm i}$, so that the non-linear relation between $\nu(t_0)$ and
$\nu_{\rm i}$ can locally be well approximated by a linear function.

Combining both linear relations we can write the expected final total mass as
$$M_{\rm G}(\nu_{\rm i},x_{\rm i},t_0) = (2\pi R_{\rm G}^2)^{3/2}\bar\rho
	\bigl[c_0 + c_1\nu_{\rm i} + c_2 x_{\rm i}\bigl]\ . \eqno(B\neweq)$$
The constant $c_1$ will be larger than one, as $\nu(t)$ is a monotonically
growing function. On the other hand, we expect $c_2$ to be negative on
the basis of eq.\ (B\the\vdWvH).

\eqnumber=1

\section*{Appendix C: Reliability of the cluster models}


\tx Our models could be susceptible to various systematic effects and errors.
The first source of error that comes to mind is the N-body code itself,
since the integration of the particle trajectories is performed with a
certain accuracy. However, the energy is conserved sufficiently well for
all models: about one procent change in $E/U(0)$ (the total energy divided
by the initial total potential energy). The shape of the initially spherical
distribution of particles changes little and its borders do not move inwards.
This means that we have indeed included the complete region of influence of
the cluster, i.e.\ the turnaround radius is well within the simulation volume.
The autocorrelation function lacks
some of the large-scale power at late times, but that is only the case for
$\sigma_8$ larger than the value of 0.46 that we adopt. However, we will need
to take larger simulation volumes for scenarios with more power on large
scales, like the CDM $\Omega_0=0.2$ scenario.

Systematic errors are most likely due to the approximate way in which we model
the formation and evolution of galaxies. For example, mass loss of galaxies is
not modelled, and dynamical friction may therefore be unrealistically large.
However, the galaxy formation parameters were chosen such that only the cores
of particle groups that were found to be virialised were tranformed into a
single galaxy particle (see also van Kampen 1995).
This allows the remaining outer particles to be stripped, even though
this need not necessarily happen. The crude way in which merging is
modeled is probably the biggest defect. Merging may therefore be
underestimated in some regions and overestimated in others.
In future work we will try to add a separate merger criterion,
even though this will introduce more parameters to the modelling.


